\documentstyle[11pt]{article}
\begin{document}
\begin{titlepage}
\vspace*{-60pt}
\begin{flushright}
SUSSEX-AST 95/8-3 \\
FERMILAB-Pub-95/280-A \\
astro-ph/9508078\\
\end{flushright}
\vspace*{10pt}
\begin{center}
\Large
{\bf Reconstructing the Inflaton Potential --- an Overview}\\
\vspace{.8cm}
\normalsize
\large{James E.~Lidsey$^1$, Andrew R.~Liddle$^2$, Edward W.~Kolb$^{3,4}$, 
Edmund J.~Copeland$^2$, Tiago Barreiro$^2$ and Mark Abney$^4$} \\
\normalsize
\vspace{.6 cm}
{\em $^1$Astronomy Unit,\\ School of Mathematical Sciences,\\
Queen Mary and Westfield College,\\ Mile End Road, London E1 4NS,~~~U.~K.}\\
\vspace{.4 cm}
{\em $^2$Division of Physics and Astronomy, \\
University of Sussex, \\ Falmer, Brighton BN1 9QH,~~~U.~K.}\\
\vspace{.4 cm}
{\em $^3$NASA/Fermilab Astrophysics Center,\\ Fermi National Accelerator 
Laboratory,\\ Batavia, Illinois 60510,~~~U.~S.~A.}\\
\vspace{.4 cm}
{\em $^4$Department of Astronomy and Astrophysics,\\
Enrico Fermi Institute\\
University of Chicago,\\ Chicago, Illinois 60637,~~~U.~S.~A.}
\end{center}

\vspace{.4 cm}
\begin{abstract}
\noindent
We review the relation between the inflationary potential and the
spectra of density (scalar) perturbations and gravitational waves
(tensor perturbations) produced, with particular emphasis on the
possibility of reconstructing the inflaton potential from
observations. The spectra provide a potentially powerful test of the
inflationary hypothesis; they are not independent but instead are
linked by consistency relations reflecting their origin from a single
inflationary potential. To lowest-order in a perturbation expansion
there is a single, now familiar, relation between the tensor spectral
index and the relative amplitude of the spectra. We demonstrate that
there is an infinite hierarchy of such consistency equations, though
observational difficulties suggest only the first is ever likely to be
useful. We also note that since observations are expected to yield
much better information on the scalars than on the tensors, it is
likely to be the next-order version of this consistency equation which
will be appropriate, not the lowest-order one. If inflation passes the
consistency test, one can then confidently use the remaining
observational information to constrain the inflationary potential, and
we survey the general perturbative scheme for carrying out this
procedure. Explicit expressions valid to next-lowest order in the
expansion are presented. We then briefly assess the prospects for
future observations reaching the quality required, and consider 
simulated data sets motivated by this outlook.
\end{abstract}

\end{titlepage}


\tableofcontents

\newpage

\section{~~~Introduction}

\def\theequation{1.\arabic{equation}}
\setcounter{equation}{0}

Observational cosmology is entering a new era where it is becoming
possible to make detailed quantitative tests of models of the early
universe for the first time. Such observations are presently the most
plausible route towards learning some of the details of physics at
extremely high energies, and the possibility of testing some of the
speculative ideas of recent years has generated much excitement.

One of the most important paradigms in early universe cosmology is
that of cosmological inflation, which postulates a period of
accelerated expansion in the universe's distant past (Starobinsky,
1980; Guth, 1981; Sato, 1981; Albrecht and Steinhardt, 1982; Hawking and
Moss, 1982; Linde, 1982a, 1983). Although originally introduced as a
possible solution to a host of cosmological conundrums such as the
horizon, flatness and monopole problems, by far the most useful
property of inflation is that it generates spectra of both density
perturbations (Guth and Pi, 1982; Hawking, 1982; Linde, 1982b; Starobinsky,
1982; Bardeen, Steinhardt, and Turner, 1983) and gravitational waves
(Starobinsky, 1979; Abbott and Wise, 1984a). These extend from extremely
short scales to scales considerably in excess of the size of the
observable universe. During inflation the scale factor grows
quasi-exponentially, while the Hubble radius remains almost constant.
Consequently the wavelength of a quantum fluctuation -- either in the
scalar field whose potential energy drives inflation or in the
graviton field -- soon exceeds the Hubble radius.  The amplitude of
the fluctuation therefore becomes `frozen'. Once inflation has ended,
however, the Hubble radius increases faster than the scale factor, so
the fluctuations eventually reenter the Hubble radius during the
radiation- or matter-dominated eras. The fluctuations that exit around
60 $e$-foldings or so before reheating reenter with physical
wavelengths in the range accessible to cosmological observations.
These spectra provide a distinctive signature of inflation. They can
be measured in a variety of different ways including the analysis of
microwave background anisotropies, velocity flows in the universe,
clustering of galaxies and the abundances of gravitationally bound
objects of various types (for reviews, see Efstathiou (1990); Liddle and 
Lyth (1993a)).

Until the measurement of large angle microwave background anisotropies
by the COsmic Background Explorer (COBE) satellite (Smoot et al., 1992; 
Wright et al., 1992; Bennett et al., 1994, 1996; see White, Scott, and Silk 
(1994) for a general discussion of the microwave background), such 
observations covered a fairly limited range of scales, and it was 
satisfactory to treat the prediction of a generic inflationary scenario as 
giving rise to a scale-invariant (Harrison--Zel'dovich) spectrum of density 
perturbations (Harrison, 1970; Zel'dovich, 1972) and a negligible amplitude 
of gravitational waves (though even then, it was recognized that the 
scale-invariance was only approximate (Bardeen et al., 1983)). Since the 
detection by COBE, however, the spectra are now constrained over a range of 
scales covering some four orders of magnitude from one megaparsec up to 
perhaps ten thousand megaparsecs. Moreover, shortly after the COBE 
detection, a number of authors reexamined the possibility that a significant 
fraction of the signal could be due to gravitational waves (Krauss and 
White, 1992; Davis et al., 1992; Salopek, 1992; Liddle and Lyth, 1992; 
Lidsey and Coles, 1992; Lucchin, Matarrese, and Mollerach, 1992; Souradeep 
and Sahni, 1992; Adams et al., 1993; Dolgov and Silk, 1993).

Thus, the inflationary prediction must now be considered with much
greater care, even in order to deal with {\em present} observations.
At the next level of accuracy, one finds that different inflation
models make different predictions for the spectra, which can be viewed
as differing magnitudes of variation from the scale-invariant result.
In the simplest approximation the spectra are taken to be power-laws.
Hence, modern observations discriminate between different inflationary
models, and are already sufficient to rule out some models completely
(see e.~g.~Liddle and Lyth, 1992) and substantially constrain the
parameter space of others (Liddle and Lyth, 1993a). Future observations
will make even stronger demands on theoretical precision, and will
certainly tightly constrain inflation.

These deviations from highly symmetric situations such as a
scale-invariant spectrum provide an extremely distinctive way of
probing inflation. This is considerably more powerful than employing
historically emphasised predictions such as a spatially flat universe.
Although a spatially flat universe is indeed a typical (but not
inevitable, see e.g., Sasaki et al. (1993); Bucher, Goldhaber, and Turok
(1995)) outcome of inflation, it appears unlikely that this feature
will be unique to inflation. Moreover, the power that observations
such as microwave background anisotropies provides may be sufficient
to override the rather subjective arguments often made against
inflation models because of their apparent `unnaturalness'. Regardless
of whether a model appears natural or otherwise, it should be the
observations which decide whether it is correct or not.

In a wide range of inflationary models, the underlying dynamics is
simply that of a single scalar field --- the {\em inflaton} ---
rolling in some underlying potential. This scenario is generically referred 
to as {\em chaotic inflation} (Linde, 1983, 1990b) in reference to its 
choice of initial conditions. This picture is widely favored because of its 
simplicity and has received by far the most attention to date. Furthermore, 
many superficially more complicated models can be rewritten in this 
framework. In view of this we shall concentrate on such a type of model 
here.

The generation of spectra of density perturbations and gravitational
waves has been extensively investigated in these theories. The usual
strategy is an expansion in the deviation from scale-invariance,
formally expressed as the {\em slow-roll expansion} (Steinhardt and
Turner, 1984; Salopek and Bond, 1990; Liddle, Parsons, and Barrow, 1994). At
the simplest level of approximation, the spectra can be expressed as
power-laws in wavenumber; further accuracy entails calculation of the
deviations from this power-law approximation.

A crucial aspect of the two spectra is that they are not independent.
In a general sense, this is clear since they correspond at the formal
level to two continuous functions that both have an origin in the
single continuous function expressing the scalar field potential. Such
a link was noted in the simplest situation, where the spectra are
approximated by power-laws, by Liddle and Lyth (1992); the general
situation where the two are linked by a {\em consistency equation} was
expounded in Copeland et al. (1993b, henceforth CKLL1), and an
explicit higher-order version of the simplest equation was found by
Copeland et al. (1994a, henceforth CKLL2). If one had complete
expressions for the entire problem, the consistency relation would be
represented as a differential equation relating the two spectra.
However, we shall argue that it is preferable to express the spectra
via an order-by-order expansion. In this case one obtains a finite set
of {\em algebraic} expressions which represent the coefficients of an
expansion of the full differential equation. The familiar situation is
a single consistency equation that relates the gravitational wave
spectral index to the relative amplitudes of the spectra. This is a
result of the lowest-order expansion. The general situation of
multiple consistency equations does not seem to have been expounded
before, though a second consistency equation did appear in Kosowsky and
Turner (1995). In practice, the observational difficulties associated
with measurements of the details of the gravitational wave spectrum
make it extremely unlikely that any but the first consistency equation
shall ever be needed.

Given a particular set of observations of some accuracy, one can
attempt the bold task of reconstructing the inflaton potential from
the observations. In fact, the situation one hopes for is stronger
than a simple reconstruction, the language of which suggests the
possibility of finding a suitable potential regardless of the
observations. With sufficiently good observations, one can first test
whether the consistency equation is satisfied; in situations where
observations make this test non-trivial it provides a very convincing
vindication of the inflationary scenario. Thus emboldened, one could
then go on to use the remaining, nondegenerate, information to
constrain features of the inflaton potential. Figure 1 illustrates
this procedure schematically.

The main obstacle in reconstruction is the limited range of scales
accessible. Although the observations may span up to four orders of
magnitude, the expansion of the universe is usually so fast during
inflation that this typically translates into only a brief range of
scalar field values. One should therefore not overexaggerate the
usefulness of this approach in determining the detailed structure of
physics at high energy, but one should bear in mind that this may be
the only observational information available of any kind at such
energies.

A second obstacle is that one doesn't observe the primordial spectra 
directly, but rather after they have evolved considerably. Although this is 
a linear problem (except on the shortest scales) and hence computationally 
tractable, the evolution necessarily depends on the various cosmological 
parameters, such as the expansion rate and the nature of any dark matter. 
The form of the initial spectra must be untangled from their influence. 
We shall discuss this in some detail in Section \ref{obs}.

Earlier papers discuss two possible ways of treating observational
data. The bolder strategy is to use estimates of the spectra as
functions of scale (Hodges and Blumenthal, 1990; Grishchuk and Solokhin,
1991; CKLL1). In practice, however, this approach founders through the
lack of theoretically derived exact expressions for the spectra
produced by an arbitrary potential. We shall therefore argue in this
review in favor of the alternative approach, which is usually called
perturbative reconstruction (Turner, 1993a; Copeland et al., 1993a;
CKLL1; Turner, 1993b; CKLL2; Liddle and Turner, 1994). In this approach,
the consistency equation and scalar potential are determined as an
expansion about a given point (regarded either as a single scale in
the spectra or as a single point on the potential), allowing
reconstruction of a region of the potential about that point. This has
the considerable advantage that one can terminate the series when
either theoretical or observational knowledge runs out.

The outline of this review is as follows. We devote two Sections to a
review of the inflation driven by a (slowly) 
rolling scalar field. We begin by considering the
classical scalar field dynamics and then proceed to discuss the
generation of the spectra of density perturbations and gravitational
waves. Because an accurate derivation of the predicted spectra is
crucial to this programme, we provide a detailed account of the most
accurate calculation presently available, due to Stewart and Lyth
(1993). In Section \ref{first} we consider the simplest possible
scenario allowing reconstruction, and introduce the notion of the
consistency equation.  Section \ref{second} reviews the present
state-of-the-art, where next--order corrections are incorporated into
all expressions. One hopes that observational accuracy will justify
this more detailed analysis, though this depends upon which (if any)
inflation model proves correct. Section \ref{genfram} then expands on
this by describing the full perturbative reconstruction framework,
illustrating how much information can be obtained from which
measurements and demonstrating that one can write a hierarchy of
consistency equations. We then briefly illustrate worked examples on
simulated data in Section \ref{obs}. Before concluding, we devote a section 
to an examination of other proposals for constraining the inflaton 
potential, without using large-scale structure observations.

\section{~~~Inflationary Cosmology and Scalar Fields}
\label{dyn}

\def\theequation{2.\arabic{equation}}
\setcounter{equation}{0}

\subsection{The fundamentals of inflationary cosmology}

Observations indicate that the density distribution in the universe is 
nearly smooth on large scales, but contains significant irregularities on 
small scales. These correspond to a hierarchy of structures including 
galaxies, clusters and superclusters of galaxies. 
One of the most important questions that modern cosmology must address is 
why the observable universe is almost, but not quite exactly, homogeneous 
and isotropic on sufficiently large scales. 

The hot big bang model is able to explain the current expansion of the 
universe, the primordial abundances of the light elements and 
the origin of the cosmic microwave background radiation; for a review of all 
these successes see Kolb and Turner (1990). However, this model as it stands 
is unable to explain the origin of structure in the universe. This problem 
is related to the well known flatness problem (Peebles and Dicke, 1979) and 
is essentially a problem of initial data. It arises because the entropy in 
the universe is so large, $S \approx 10^{88}$ (Barrow and Matzner, 1977). 
One expects this quantity to be of order unity since it is a dimensionless 
constant. 

This paradox can  be made more quantitative in the following way. The 
dynamics of a Friedmann--Robertson--Walker (FRW) universe containing 
matter with density $\rho$ and pressure $p$ is determined by the Einstein 
acceleration equation
\begin{equation}
\label{acc}
\frac{\ddot{a}}{a} = - \frac{4\pi}{3m_{\rm Pl}^2} 
(\rho +3p) ,
\end{equation}
the Friedmann equation
\begin{equation}
\label{Fequation}
H^2 = \frac{8\pi}{3m_{\rm Pl}^2} \rho - \frac{k}{a^2},
\end{equation}
and the mass conservation equation
\begin{equation}
\label{mass}
\dot{\rho} +3H(\rho +p) =0  ,
\end{equation}
where $a(t)$ is the scale factor of the universe, $H\equiv \dot{a}/{a}$ is 
the Hubble expansion parameter, a dot denotes differentiation with respect 
to cosmic time $t$, $m_{\rm Pl}$ is the Planck mass and $k=0,-1, +1$ for 
spatially flat, open, or closed cosmologies, respectively.  Units are chosen 
such that $c=\hbar =1$. 

The Friedmann equation (\ref{Fequation}) may be expressed in terms of the  
$\Omega$--parameter. This parameter is defined as the ratio of the energy 
density of the universe to the critical energy density $\rho_{\rm c}$ that 
is just sufficient to halt the current expansion:
\begin{equation}
\Omega \equiv \frac{\rho}{{\rho}_{\rm c}}, \qquad
{\rho}_{\rm c} \equiv \frac{3 m^2_{\rm Pl} H^2}{8\pi}.
\end{equation}
The current observational values for these parameters are ${\rho}_{\rm c} = 
1.88h^2 \times 10^{-29} \quad {\rm g} \quad {\rm cm}^{-3}$ and $H_0 =100 h$ 
km ${\rm s}^{-1}$ ${\rm Mpc}^{-1}$ where conservatively we have 
$0.4 \le h \le 0.8$. Eq. (\ref{Fequation}) simplifies to 
\begin{equation}
\label{Omegafriedmann}
\Omega -1 =\frac{k}{a^2 H^2},
\end{equation}
and this implies that 
\begin{equation}
\label{of}
\frac{\Omega -1}{\Omega} =\frac{3 m^2_{\rm Pl}}{8\pi} 
\frac{k}{\rho a^2} .
\end{equation}

Now, for a radiation--dominated universe, the 
equation of state is given by $\rho =3p = \pi^2 g_{\rho} T^4/30$ 
at some temperature $T$, where $g_{\rho} = {\cal{O}} (10^2)$ represents 
the total number of  relativistic  degrees of freedom in the 
matter sector at that time. Thus the  scale factor grows as $a 
(t) \propto t^{1/2}$ when $k=0$ and the expansion rate is given by
\begin{equation}
\label{hub}
H=1.66 g_{\rho}^{1/2} \left( \frac{T^2}{m_{\rm Pl}} \right) = 
\frac{1}{2t}  .
\end{equation}
Eq. (\ref{hub}) yields  the useful expression 
\begin{equation}
\label{veryuseful}
\left( \frac{t}{\rm sec} \right) \approx \left( \frac{T}{\rm MeV} 
\right)^{-2}
\end{equation}
and substituting Eqs. (\ref{hub}) and (\ref{veryuseful}) 
into Eq. (\ref{of}) implies that 
\begin{equation}
\label{Omega}
\left| \frac{\Omega -1}{\Omega} \right| \approx \frac{10^{43}}{S^{2/3}} 
\left( \frac{t}{\rm sec} \right) \approx \frac{10^{37}}{S^{2/3}} 
\left( \frac{{\rm GeV}}{T} \right)^2  \,,
\end{equation}
where $S \approx 10^{88}$ is the entropy contained within the present 
horizon. The large amount of entropy in the universe therefore implies that 
$\Omega$ must have been very close to unity at early times. 
Indeed, we find that  $\Omega =1 \pm 10^{-16}$ just one second after the
big bang, the time of nucleosynthesis. 

The flatness problem is therefore a problem of understanding why the 
(classical) initial conditions corresponded to a universe that was so close 
to spatial flatness. In a sense, the problem is one of fine--tuning and 
although such a balance is possible in principle, one nevertheless feels 
that it is unlikely. On the other hand, the flatness problem arises because 
the entropy in a comoving volume is conserved. It is possible, therefore,  
that the problem could be resolved if the cosmic expansion was 
non--adiabatic for some finite time interval $t \in [t_{\rm i},t_{\rm f}]$ 
during the early history of the universe. 

This point was made explicitly by Guth in his seminal paper of 1981. He 
postulated that the entropy changed by an amount
\begin{equation}
S_{\rm f}=Z^3S_{\rm i}
\end{equation}
during this time interval, where $Z$ is a numerical factor. In Guth's 
original model, this entropy production occurred at, or just below, the 
energy scale $T_{\rm GUT} = {\cal{O}} (10^{17})$ GeV associated with the 
Grand Unified (GUT) phase transition. This corresponds to a timescale $t 
\approx 10^{-40}$ s. Eq. (\ref{Omega}) then implies that the flatness 
problem is solved, in the sense that $|{\Omega}_{\rm i}^{-1}-1| = {\cal{O}} 
(1)$, if $Z \ge 10^{28}$. It can be shown that the other problems of the big 
bang model, such as the horizon and monopole problems are also solved if $Z$ 
satisfies this lower bound (Guth, 1981). 

Guth called this process of entropy production {\em inflation}, 
because the volume of the universe also grows by the factor $Z^3$
between $t = t_{\rm i}$ and  $t = t_{\rm f}$. Indeed, the expansion of the 
universe during the inflationary epoch is very rapid. Further insight into 
the nature of this expansion may be gained by considering Eq. (\ref{of}). 
This expression  implies that the quantity $({\Omega}^{-1}-1){\rho}a^2$ is 
conserved for an arbitrary equation of state. It follows, therefore, that 
\begin{equation}
({\Omega}^{-1}_{\rm i}-1)a_{\rm i}^2{\rho}_{\rm i}=({\Omega}^{-1}_{\rm f}-1)
a_{\rm f}^2{\rho}_{\rm f}
\end{equation}
and, if we assume that the standard, big bang 
model is valid for $t>t_{\rm f}$, we may deduce that (Lucchin and 
Matarrese, 1985b)
\begin{equation}
{\rho}_{\rm i}a_{\rm i}^2|{\Omega}_{\rm i}^{-1}-1| \approx 10^{-56}
 {\rho}_{\rm f}a_{\rm f}^2 |{\Omega}_0^{-1}-1 |  .
\end{equation}
Since our current observations imply that $|{\Omega}_0^{-1}-1| =  {\cal{O}} 
(1)$, the flatness problem is solved if ${\rho}_{\rm f}a_{\rm f}^2 \gg 
{\rho}_{\rm i}a_{\rm i}^2$. However, Eq. (\ref{Fequation}) implies that 
the quantity $3\dot{a}^2-(8\pi /m^2_{\rm Pl}) \rho a^2$ is also conserved.
Consequently, this inequality is satisfied if $\dot{a}_{\rm f}
>{\dot{a}_{\rm i}}$. Thus, a necessary condition for inflation to proceed is 
that the scale factor of the universe {\it accelerates} with respect to 
cosmic time:
\begin{equation}
\label{acceleration}
{\ddot a}(t) >0 \,.
\end{equation}
This is in contrast to the decelerating expansion that arises in the big 
bang model. 

The question now arises as to the nature of the energy source that drives 
this accelerated expansion. It follows from Eq. (\ref{acc}) that Eq. 
(\ref{acceleration}) is satisfied if $\rho +3p <0$ and this is equivalent to 
violating the strong energy condition (Hawking and Ellis, 1973). The 
simplest way to achieve such an antigravitational effect is by the presence 
of a homogeneous scalar field, $\phi$, with some self--interaction 
potential $V(\phi ) \ge 0$. In the FRW universe, such a field is equivalent 
to a perfect fluid with energy density and pressure given by 
\begin{equation}
\rho =\frac{1}{2} \dot{\phi}^2 + V(\phi) 
\end{equation}
and 
\begin{equation}
p=\frac{1}{2} \dot{\phi}^2 - V(\phi ) \,,
\end{equation}
respectively. Other matter fields play a negligible role in the evolution 
during the inflation, so their presence will be ignored. In this case, 
Eqs. (\ref{Fequation}) and (\ref{mass}) are given by 
\begin{equation}
\label{F}
H^2 = \frac{8\pi}{3m_{\rm Pl}^2} \left( \frac{1}{2} \dot{\phi}^2 
+V(\phi) \right) -\frac{k}{a^2}
\end{equation}
and 
\begin{equation}
\label{phieqn}
\ddot{\phi} +3H\dot{\phi} =-V' (\phi)   ,
\end{equation}
where here 
and throughout a prime denotes differentiation with respect to $\phi$. 
Hence, $-\rho \le p \le \rho$ and we have the inflationary requirement 
$\ddot{a}>0$ as long as $\dot{\phi}^2 <V$. Inflation is thus achieved 
when the matter sector of the theory applicable at some stage in the early 
universe is dominated by vacuum energy. 

Recently, an alternative inflationary scenario --- the pre--big bang
cosmology --- has been developed whereby the accelerated expansion is
driven by the kinetic energy of a scalar field rather than its
potential energy (Gasperini and Veneziano, 1993a,b, 1994). If the field is 
non--minimally coupled to gravity in an appropriate fashion, this kinetic 
energy can produce a sufficiently negative pressure and a violation of the 
strong energy condition (Pollock and Sahdev, 1989; Levin, 1995a). Such
couplings arise naturally within the context of the string effective
action. However, models of this sort inherently suffer from a `graceful 
exit' problem due to the existence of singularities in both the curvature 
and the scalar field motion (Brustein and Veneziano, 1994; Kaloper, Madden 
and Olive, 1995, 1996; Levin, 1995b; Easther, Maeda, and Wands, 1996). 
Moreover, a satisfactory mechanism for generating structure formation and 
microwave background anisotropies in these models has yet to be developed, 
although it is possible that such inhomogeneities may be generated by 
quantum fluctuations in the electromagnetic field (Gasperini, Giovannini and 
Veneziano, 1995). 

In view of this, we shall restrict our discussion to potential-driven 
models. We will focus in this work on some of the general features 
of the chaotic inflation scenario (Linde, 1983, 1990b). Although Linde's 
original paper considered a specific potential (a quartic one), the theme 
was much more general. We adopt the modern usage of {\em chaotic inflation} 
to refer to any model where inflation is driven by a single scalar field 
slow-rolling from a regime of extremely high potential energy. The phrase 
does not imply any particular choice of potential. Most, though not quite 
all, modern inflationary models fall under the umbrella of this definition. 
Since the precise identity of the scalar field driving the inflation is 
unknown, it is usually referred to as the {\em inflaton field}. 

In the chaotic inflation scenario, it is assumed that the universe emerged 
from a quantum gravitational state with an energy density comparable to that 
of the Planck density. This implies that $V (\phi ) \approx m^4_{\rm Pl}$ 
and results in a large friction term in the Friedmann equation (\ref{F}). 
Consequently, the inflaton will slowly roll  down its potential, i.e., $| 
\ddot{\phi} |\ll H |\dot{\phi} |$ and $\dot{\phi}^2 \ll V$. The condition 
for inflation is therefore satisfied and the scale factor grows as 
\begin{equation}
a(t) =a_{\rm i} \exp \left( \int^t_{t_{\rm i}} dt' H(t') \right)  .
\end{equation}
The expansion is quasi--exponential in nature, since $H(\phi) \approx 8\pi 
V(\phi) / 3m_{\rm Pl}^2$ is almost constant, and the curvature term $k/a^2$ 
in Eq. (\ref{F}) is therefore rapidly redshifted away. The kinetic energy of 
the inflaton gradually increases as it rolls down the potential towards the 
global minimum. Eventually, its kinetic energy dominates over the potential 
energy and inflation comes to an end when $\dot{\phi}^2 \approx V(\phi )$. 
The field then oscillates rapidly about the minimum and the couplings of 
$\phi$ to other matter fields then become important. It is these 
oscillations that result in  particle production and a reheating of the 
universe. 

The simplest chaotic inflation model is that of a free field with a 
quadratic potential, $V(\phi) =m^2 \phi^2/2$, where $m$ represents the mass 
of the inflaton. During inflation the scale factor grows as 
\begin{equation}
a(t) = a_{\rm i} e^{2\pi (\phi^2_{\rm i} - \phi^2 (t))}
\end{equation}
and inflation ends when $\phi = {\cal{O}} (1)$ $ m_{\rm Pl}$. If inflation 
begins when $V(\phi_{\rm i} ) \approx m_{\rm Pl}^4$, the scale factor grows 
by a factor $\exp( 4\pi m_{\rm Pl}^2/m^2)$ before the inflaton reaches the 
minimum of its potential (Linde, 1990b). One can further show that the mass 
of the field should be $m \approx 10^{-6}m_{\rm Pl}$ if the microwave 
background constraints are to be satisfied. This implies that the volume of 
the universe will increase by a factor of $Z^3 \approx 10^{3 \times 
10^{12}}$ and this is more than enough inflation to solve the problems of 
the hot big bang model.

It is important to emphasize that in this scenario the initial value of the 
scalar field is randomly distributed in different regions of the universe. 
On the other hand, one need only assume that a small, causally connected, 
region of the pre--inflationary universe becomes dominated by the potential 
energy of the inflaton field. Indeed, if the original domain is only one 
Planck length in extent, its final size will be of the order $10^{10^{12}}$ 
cm; for comparison, the size of the observable universe is approximately 
$10^{28}$ cm. 

In conclusion, therefore, the chaotic inflationary scenario represents a 
powerful framework within which specific inflationary models can be 
discussed. The essential features of each model --- such as the final reheat 
temperature and the amplitude of scalar and tensor fluctuations --- 
are determined  by the specific form of the potential function $V(\phi)$. 
This in turn is determined by the particle physics sector of the theory. 

Unfortunately, however, there is currently much theoretical
uncertainty in the correct form  of the unified field theory above
the electroweak scale.  This has resulted in the development of a
large number of different inflationary scenarios and the identity of
the inflaton field is therefore somewhat uncertain.  Possible
candidates include the Higgs bosons of grand unified theories, the
extra degrees of freedom associated with higher metric derivatives in
extensions to general relativity, the dilaton field of string theory
and, more generally, the time--varying gravitational coupling that
arises in scalar--tensor theories of gravity.

It is not the purpose of this review to discuss the relative merits of 
different models, since this has been done elsewhere (Kolb and Turner, 1990; 
Linde, 1990b; Olive, 1990; Liddle and Lyth, 1993a). Traditionally, a 
specific potential with a given set of coupling constants is chosen. The 
theoretical predictions of the model are then compared with large--scale 
structure observations. The region of parameter space consistent with such 
observations may then be identified (Liddle and Lyth, 1993a). However, it is  
difficult to select a unique inflationary model by this procedure due to the 
large number of plausible  models available. 

In view of the above uncertainties and motivated by recent and forthcoming 
advances in observational cosmology, our aim will be to address the question 
of whether {\em direct} insight into the nature of the inflaton potential 
may be gained by studying the large--scale structure of the universe. We 
therefore assume nothing about the potential except that it leads to 
an epoch of inflationary expansion.  

We will proceed in the remainder of this Section by reviewing a
formalism that allows the classical dynamics of the scalar field
during inflation to be studied in full generality.  This formalism may
then be employed to discuss the generation of quantum fluctuations in
the inflaton and gravitational fields.  

\subsection{Scalar field dynamics in inflationary cosmology}

In view of the discussion in the previous Subsection, we will assume 
throughout this work that the universe was dominated during inflation by a 
single scalar field $\phi$ with a self-interaction potential $V(\phi)$, the
form of which it is our aim to determine. We shall further assume that
gravity is adequately described by Einstein's theory of general
relativity. We shall therefore employ  the four-dimensional action
\begin{equation}
\label{action}
S= -\int d^4 x \sqrt{-g} \left[ \frac{m_{{\rm Pl}}^2R}{16\pi} 
	-\frac{1}{2} \left( \nabla {\phi} \right)^2 +V( {\phi}) 
	\right] \,,
\end{equation}
where $R$ is the Ricci curvature scalar of the space--time with metric 
$g_{\mu\nu}$ and $g\equiv {\rm det} g_{\mu\nu}$. 

Actually, these restrictions are not as strong as they seem. For
example, even theories such as hybrid inflation, which feature
multiple scalar fields, are usually dynamically dominated by only one
degree of freedom (Linde, 1990a, 1991, 1994; Copeland et al., 1994b;
Mollerach, Matarrese, and Lucchin, 1994).  Many other models invoke
extensions to general relativity, and much effort has been devoted to
studying inflation in the Bergmann-Wagoner class of generalized
scalar-tensor theories (Bergmann, 1968; Wagoner, 1970) and higher-order
pure gravity theories in which the Einstein-Hilbert lagrangian is
replaced with some analytic function $f(R)$ of the Ricci curvature.
Such theories can normally be rewritten via a conformal transformation
as general relativity plus one or more scalar fields, again with the
possibility that only one such field is dynamically relevant (Higgs,
1959; Whitt, 1984; Barrow and Cotsakis, 1988; Maeda, 1989; Kalara, Kaloper,
and Olive, 1990; Lidsey, 1992; Wands, 1994). 

We are unable to discuss models where more than one field is
dynamically important in the reconstruction context. While
considerable progress has been made recently in understanding the
perturbation spectra from these models (Starobinsky and Yokoyama, 1995;
Garc\'{\i}a-Bellido and Wands, 1995, 1996; Sasaki and Stewart, 1996; 
Nakamura and Stewart, 1996), the extra freedom of the second field thwarts 
any attempt at finding a unique reconstruction, though it is possible to
find some general inequalities relating the spectra (Sasaki and Stewart,
1996). These problems arise both because there is no longer a unique
trajectory, independent of initial conditions, into the minimum of the
potential, and because with a second field one can generate
isocurvature perturbations as well as adiabatic ones. Fortunately, it
appears that it is hard, though not impossible, to keep models of this kind 
consistent with observation, as the density perturbations tend to be large 
whatever the energy scale of inflation (Garc\'{\i}a-Bellido, Linde, and 
Wands, 1996). A completely different way of using two fields is to drive
successive periods of inflation, as in the double inflation scenario
(Polarski and Starobinsky, 1995 and refs therein). This can impose very
sharp features in the spectra which, although rather distinctive, are
not amenable to the perturbative approach that reconstruction
requires.

As we saw above, the accelerated expansion during inflation causes the 
spatial hypersurfaces to rapidly tend towards flatness. Moreover, any 
initial anisotropies and inhomogeneities in the universe are washed away 
beyond currently observable scales by the rapid expansion. Since only the 
final stages of the accelerated expansion are important from an 
observational point of view, we can assume that the space-time metric may be 
described as a spatially flat FRW metric, given by
\begin{equation}
\label{background}
ds^2=L^2(t) dt^2 -e^{2\alpha (t)} [dx^2+dy^2+dz^2] \,,
\end{equation}
where $L(t)$ represents the lapse function and $a(t) =e^{\alpha (t)}$
is the scale factor of the universe. 

By taking this metric, we prevent ourselves from studying reconstruction in 
the recently discovered versions of inflation giving an open universe (Gott, 
1982; Gott and Statler, 1984; Sasaki et al., 1993; Bucher et al., 1995; 
Linde, 1995; Linde and Mezhlumian, 1995). In fact these models have not yet 
been developed sufficiently to provide the information we need --- in 
particular the gravitational wave spectrum has not been predicted --- and 
the generalization of the reconstruction program to these models must await 
further developments.

Our analysis will however apply in full to low-density cosmological
models where the spatial geometry is kept flat by the introduction of
a cosmological constant (or similar mechanism). Our discussion is entirely
focussed on the initial spectra, which are independent of the material
composition of the universe at late times. Of course, in such a
cosmology the details of going between these spectra and actual
observables will be changed, and the impact of this on reconstruction
has been studied by Turner and White (1995).

Substitution of the metric ansatz Eq.~(\ref{background}) into the
theory given by Eq.~(\ref{action}) leads to an Arnowitt, Deser and
Misner (ADM) (1962) action of the form
\begin{equation}
\label{ADM}
S=\int dt \, U L e^{3\alpha} \left[ -\frac{3m_{{\rm Pl}}^2}{8\pi}
\frac{\dot{\alpha}^2}{L^2} + \frac{1}{2} \frac{\dot{\phi}^2}{L^2}
-V(\phi) \right] \,,
\end{equation}
where $U\equiv \int d^3 {\bf x}$ is the comoving volume of the
universe and a dot denotes differentiation with respect to $t$.
Without loss of generality we may normalize the comoving volume to
unity.

In recent years, considerable progress in the treatment of scalar
fields within the environment of the very early universe has been
made. The approach we adopt in this work is to view the scalar field
itself as the dynamical variable of the system (Grishchuk and Sidorav,
1988; Muslimov, 1990; Salopek and Bond, 1990, 1991; Lidsey, 1991b). This
allows the Einstein-scalar field equations to be written as a set of
first-order, non-linear differential equations.

The Hamiltonian constraint ${\cal{H}}=0$ is derived by functionally
differentiating the action Eq.~(\ref{ADM}) with respect to the
non-dynamical lapse function. One arrives at the Hamilton-Jacobi
equation
\begin{equation}
\label{HJ}
 -\frac{4\pi}{3m_{{\rm Pl}}^2} \left( \frac{\partial S}{\partial \alpha}
\right)^2 + \left( \frac{\partial S}{\partial \phi} \right)^2 +2
e^{6\alpha} V(\phi ) =0 \,,
\end{equation}
where the momenta conjugate to $\alpha$ and $\phi$ are $p_{\alpha} =
\partial S/\partial \alpha = -3m_{{\rm Pl}}^2 e^{3\alpha} 
\dot{\alpha}/4\pi L$ and $p_{\phi}= \partial S/\partial \phi = e^{3\alpha}
\dot{\phi}/L$, respectively. This equation follows from the invariance
of the theory under reparametrizations of time. The classical dynamics
of this model is determined by the real, separable solution
\begin{equation}
\label{sepact}
S=-\frac{m_{{\rm Pl}}^2}{4\pi} e^{3\alpha} H(\phi) \,,
\end{equation}
where $H(\phi)$ satisfies the differential equation (Grishchuk and
Sidorav, 1988; Muslimov, 1990; Salopek and Bond, 1990, 1991)
\begin{equation}
\label{H} 
\left( \frac{dH}{d\phi} \right)^2 -\frac{12\pi}{m_{{\rm Pl}}^2} H^2 (\phi) 
=-\frac{32\pi^2}{m_{{\rm Pl}}^4} V(\phi) \,.
\end{equation}
In the gauge $L=1$, substitution of ansatz Eq.~(\ref{sepact}) into the
expressions for the conjugate momenta implies that
\begin{equation}
\label{field}
 H(\phi)=\dot{\alpha} \quad ; \quad -\frac{m_{{\rm Pl}}^2}{4\pi}
	\frac{dH}{d\phi} = \dot{\phi} \,.
\end{equation}
Thus, $H(\phi)$ represents the Hubble expansion parameter expressed as
a function of the scalar field $\phi$. It follows immediately from the
second of these expressions that $\dot{H}<0$. Consequently, the
physical Hubble radius $H^{-1}$ increases with time as the inflaton
field rolls down its potential. The Hubble radius can only remain
constant if the inflaton field is trapped in a meta-stable false
vacuum state; this is forbidden in the context of `old' inflation as
it can never successfully escape this state, but may be possible in
the context of single-bubble open inflationary models which are
outside the scope of this paper (see, for example, Sasaki et al.,
1993; Bucher et al., 1994; Bucher, Goldhaber, and Turok, 1995; Linde,
1995).

The solution to Eq.~(\ref{H}) depends on an initial condition, the
value of $H$ at some initial $\phi$ (Salopek and Bond, 1990, 1991). If
we are to obtain unique results, the late-time evolution (that is, the
evolution during which the perturbations we see are generated) of $H$
in terms of the scalar field must be independent of the initial
condition chosen, and fortunately one can easily show that this is the
case (Salopek and Bond, 1990; Liddle et al., 1994); the late-time
behavior is governed by an inflationary `attractor' solution, which
is approached exponentially quickly during inflation.

The Hamilton--Jacobi formalism we have outlined is equivalent to the
more familiar version of the equations of motion given by Eqs. (\ref{F}) and 
(\ref{phieqn}) (for $k=0$). Eq. (\ref{H}) is equivalent to the time--time 
component of the Einstein field equations and therefore represents the 
Friedmann equation (\ref{F}). In the form given by Eqs. (\ref{F}) and 
(\ref{phieqn}), $\dot{\phi}$ is an initial condition at some value of $t$; 
in the Hamilton-Jacobi formalism the equivalent freedom allows one to 
specify $H$ at some initial value of $\phi$. 

The above analysis of the Hamilton--Jacobi formalism assumes
implicitly that the value of the scalar field is a monotonically
varying function of cosmic time. In particular, it breaks down if the
field undergoes oscillations (though one can attempt to patch together
separate solutions). As a result, this formalism is not directly
suitable for investigating the dynamics of a field undergoing
oscillations in a minimum of the potential, for example.  However, the
scalar and tensor fluctuations relevant to large-scale structure
observations are generated when the field is still some distance away
from the potential minimum. Moreover, the piece of the potential
corresponding to these scales is relatively small, so it is reasonable
to assume that the potential is a smoothly decreasing function in this
regime. The scalar field will therefore roll down this part of the
potential in an unambiguous fashion. In the following, we will assume, 
without loss of generality, that $\dot{\phi} > 0$, so that $H' (\phi) <0$. 
This choice allows us to fix the sign of any prefactors that arise when 
square roots appear.

In principle, the Hamilton--Jacobi formalism enables us to treat the
dynamical evolution of the scalar field exactly, at least at the
classical level. In practice, however, the separated Hamilton-Jacobi
equation, Eq.~(\ref{H}), is rather difficult to solve. On the other
hand, the analysis can proceed straightforwardly once the functional
form of the expansion parameter $H(\phi)$ has been determined. This
suggests that one should view $H(\phi)$ as the fundamental quantity in
the analysis (Lidsey, 1991a, 1993).  This is in contrast to the more
traditional approaches to inflationary cosmology, whereby the particle
physics sector of the model --- as defined by the specific form of the
inflaton potential $V(\phi)$ --- is regarded as the input parameter.
In the reconstruction procedure, however, the aim is to determine this
quantity from observations, so one is free to choose other quantities
instead. It proves convenient to express the scalar and tensor
perturbation spectra in terms of $H(\phi)$ and its derivatives.

Unfortunately, exact expressions for these perturbations have not yet
been derived in full generality. All calculations to date have
employed some variation of the so-called `slow-roll' approximation
(Steinhardt and Turner, 1984; Salopek and Bond, 1990; Liddle and Lyth, 1992;
Liddle et al., 1994). It is important to emphasize that there are two 
different versions of the slow-roll approximation, with their attendant 
slow-roll parameters $\epsilon$, $\eta$, etc, depending on whether one is 
taking the potential or the Hubble parameter as the fundamental quantity --- 
the differences are described in considerable detail in Liddle et al.
(1994). Here we are defining them in terms of the Hubble parameter.

We represent the slow-roll approximation as an expansion in terms of
quantities derived from appropriate derivatives of the Hubble
expansion parameter. Since {\em at a given point} each derivative is
independent, there are in general an infinite number of these terms.
Typically, however, only the first few enter into any expressions
of interest. We define the first three as\footnote{Note that the definition 
of the third parameter is different to that made in CKLL2, $\xi_{\rm CKLL2} 
= (m_{{\rm Pl}}^2/4\pi) H'''/H'$. The two are related by $\xi^2
=\epsilon \xi_{\rm CKLL2}$. The former definition has proven awkward;
because of the derivative on the denominator it need not be small in
the scale-invariant limit (though the combination $\sqrt{\epsilon
\xi_{{\rm CKLL2}}}$ must be). We choose to use this better definition,
as introduced by Liddle et al. (1994) who give further details and a
collection of useful formulae.}:
\begin{equation}
\label{epsilon}
\epsilon (\phi) \equiv \frac{3\dot{\phi}^2}{2} \left[ V+\frac{1}{2} 
	\dot{\phi}^2 \right]^{-1} =\frac{m_{{\rm Pl}}^2}{4\pi} \left( 
	\frac{H' (\phi) }{H(\phi)} \right)^2 \,,
\end{equation}
\begin{equation}
\label{eta}
\eta (\phi)
\equiv -\frac{\ddot{\phi}}{H\dot{\phi}} = \frac{m_{{\rm Pl}}^2}{4\pi} 
	\frac{H''(\phi)}{H(\phi)} = \epsilon -\frac{m_{{\rm Pl}} \, 
	\epsilon'}{\sqrt{16\pi \epsilon}} \,,
\end{equation}
\begin{equation}
\label{xi}
\xi (\phi) \equiv \frac{m_{{\rm Pl}}^2}{4\pi} \left( \frac{H'(\phi) H''' 
	(\phi)}{H^2(\phi)} \right)^{1/2} = \left( \epsilon \eta -
	\left( \frac{m_{{\rm Pl}}^2 \, \epsilon}{4\pi} \right)^{1/2} 
	\eta' \right)^{1/2} \,.
\end{equation}
One need not be concerned as to the sign of the square root in the
definition of $\xi$; it turns out that only $\xi^2$, and not $\xi$
itself, will appear in our formulae (Liddle et al., 1994). We emphasize
that the choice $\dot{\phi} > 0$ implies that $\sqrt{\epsilon} = -
\sqrt{m_{{\rm Pl}}^2/4\pi} \, H'/H$.

Modulo a constant of proportionality, $\epsilon$ measures the relative
contribution of the field's kinetic energy to its total energy
density. The quantity $\eta$, on the other hand, measures the ratio of
the field's acceleration relative to the friction acting on it due to
the expansion of the universe. The slow-roll approximation applies
when these parameters are small in comparison to unity, i.~e.~~$\{
\epsilon, |\eta |, \xi \} \ll 1$; this corresponds to being able to
neglect the first term in Eq.~(\ref{H}) and its first few derivatives.
Inflation proceeds when the scale factor accelerates, $\ddot{a}>0$,
and this is precisely equivalent to the condition $\epsilon <1$.
Inflation ends once $\epsilon$ exceeds unity. It is interesting that
the conditions leading to a violation of the strong energy condition
are uniquely determined by the magnitude of $\epsilon$ alone. In
principle, inflation can still proceed if $|\eta |$ or $|\xi |$ are
much larger than unity, though normally such values would drive a
rapid variation of $\epsilon$ and bring about a swift end to
inflation.

For specific results, we shall not go beyond these three parameters. 
However, in general one can define a full hierarchy of slow-roll parameters 
(Liddle et al., 1994): 
\begin{eqnarray}
\label{generalsro}
\beta_n & \equiv & \left\{\, \prod_{i=1}^{n}\left[- \frac{d \ln
	H^{\left(i\right)}}{d \ln a}\right] \right\}^{\frac{1}{n}} 
	\,, \nonumber \\
& = & \frac{m^{2}_{{\rm Pl}}}{4\pi} \left(\frac{ 
	\left(H'\right)^{n-1}H^{ \left( n+1\right)}}{ 
	H^{n}}\right)^{\frac{1}{n}} \,,
\end{eqnarray}
where $\beta_1 \equiv \eta$, $\beta_2 \equiv \xi$, etc, and a
superscript $(m)$ indicates the $m$-th derivative with respect to
$\phi$. The $\epsilon$ parameter has to be defined separately, though
it may be referred to as $\beta_0$.

These slow-roll parameters, along with analogues defined in terms of
the potential, can be used as the basis for a slow-roll expansion to
derive arbitrarily accurate solutions given a particular choice of
potential.  However, this formalism is not necessary when making
general statements about inflation without demanding a specific
potential.

The amount of inflationary expansion within a given timescale is most
easily para\-met\-rized in terms of the number of $e$-foldings that
occur as the scalar field rolls from a particular value $\phi$ to its
value $\phi_e$ when inflation ends:
\begin{equation}
\label{efolds}
N(\phi, \phi_e ) \equiv \int_t^{t_e} H(t) dt =-\frac{4\pi}{m_{{\rm Pl}}^2}
 \int^{\phi_e}_{\phi} d\phi \frac{H(\phi)}{H'(\phi)}  \,.
\end{equation}

Thus, with the help of Eq.~(\ref{efolds}), we may relate the value of
the scale factor $a(\phi) =e^{\alpha (\phi)}$ at any given epoch
during inflation directly to the value of the scale factor at the end
of inflation, $a_e$:
\begin{equation}
a(\phi) = a_e \exp [-N(\phi)] \,.
\end{equation}

An extremely useful formula is that which connects the two epochs at
which a given scale equals the Hubble radius, the first during
inflation when the scale crosses outside and the second much nearer
the present when the scale crosses inside again. A comoving scale $k$
crosses outside the Hubble radius at a time which is $N(k)$
$e$-foldings from the end of inflation, where
\begin{equation}
\label{Ncross}
N(k) = 62 - \ln \frac{k}{a_0 H_0} - \ln \frac{10^{16} 
        {\rm GeV}}{V_k^{1/4}}
	+ \ln \frac{V_k^{1/4}}{V_{{\rm end}}^{1/4}} - \frac{1}{3} \ln
	\frac{V_{{\rm end}}^{1/4}}{\rho_{{\rm reh}}^{1/4}} \,.
\end{equation}
The subscript `0' indicates present values; the subscript `$k$'
specifies the value when the wave number $k$ crosses the Hubble radius
during inflation ($k=aH$); the subscript `end' specifies the value at
the end of inflation; and $\rho_{{\rm reh}}$ is the energy density of
the universe after reheating to the standard hot big bang evolution.
This calculation assumes that instantaneous transitions occur between
regimes, and that during reheating the universe behaves as if
matter-dominated.

It is fairly standard to make a generic assumption about the number of
$e$-foldings before the end of inflation at which the scale presently
equal to the Hubble radius crossed outside during inflation; most
commonly one sees this number taken as either 50 or 60. Within the
context of making predictions from a given potential this can have a
slight effect on results, but it is completely unimportant as regards
reconstruction.

What we do need for reconstruction is a measure of how rapidly scales
pass outside the Hubble-radius as compared to the evolution of the
scalar field; this is essential for calculating such quantities as the
spectral indices of scalar and tensor perturbations. The formal
definition we take of a scale matching the Hubble radius is that $k =
aH$. Then one can write
\begin{equation}
\label{scalescalar}
k(\phi) = a_e H(\phi) \exp[ -N(\phi) ] \,,
\end{equation}
where $N(\phi)$ is given by Eq.~(\ref{efolds}).  Differentiating with
respect to $\phi$ therefore yields
\begin{equation}
\label{kphi} 
\frac{d \ln k}{d \phi} =\frac{4\pi}{m_{{\rm Pl}}^2} \frac{H}{H'} 
	(\epsilon - 1) \,.
\end{equation}

This concludes our discussion on the classical dynamics of the scalar
field during inflation. In the following Section, we will proceed to
discuss the consequences of quantum fluctuations that arise in both
the inflaton and graviton fields.

\section{~~~The Quantum Generation of Perturbations}
\label{pert}

\def\theequation{3.\arabic{equation}}
\setcounter{equation}{0}

During inflation, the inflaton and graviton fields undergo
quantum-mechanical fluctuations. The most important observational
consequences of the inflationary scenario derive from the significant
effects these perturbations may have on the large-scale structure of
the universe at the present epoch.  In this Section we shall discuss
how these fluctuations arise and present expressions for their
expected amplitudes. Since the inflaton and gravitational
perturbations are produced in a similar fashion, we shall begin with a
qualitative description of the effects of the former. We shall then
proceed with an extensive account of the calculation of both spectra
by Stewart and Lyth (1993), which is the most accurate analytic
treatment presently available.

\subsection{Qualitative discussion}

Fluctuations in the inflaton field lead to a stochastic spectrum of
density (scalar) perturbations (Guth and Pi, 1982; Hawking, 1982; Linde,
1982b; Starobinsky, 1982; Bardeen et al., 1983; Lyth, 1985; Mukhanov,
1985; Sasaki, 1986; Mukhanov, 1989; Salopek, Bond, and Bardeen, 1989).
Physically, these arise because the inflaton field reaches the global
minimum of its potential at different times in different places in the
universe. This results in a time shift in how quickly the rollover
occurs. Thus, constant $\delta\rho$ does not correspond to a
constant-time hypersurface; in other words, there is a density
distribution produced by the kinetic energy of the inflaton field for
a given constant-time hypersurface. It is widely thought that these
density perturbations result in the formation of large-scale structure
in the universe via the process of gravitational instability. They may
also be responsible for anisotropic structure in the temperature
distribution of the cosmic microwave background radiation.

Typically, the inflationary scenario predicts that the spectrum of
density perturbations should be gaussian and scale-dependent. This is
certainly true for the class of models that we shall be considering
here, in which it is assumed that the inflaton field is weakly
coupled. However, one should bear in mind that the prediction of
gaussianity is not generic to all inflationary models; it is possible
to contrive models with nongaussian perturbations by introducing
features in just the right part of the inflationary potential (Allen,
Grinstein, and Wise, 1987; Salopek and Bond, 1990).

The historical viewpoint on the scale-dependence of the fluctuations
was that they were of scale-invariant (Harrison--Zel'dovich) form, though 
it had been recognized that the scale-invariance was only approximate 
(Bardeen et al., 1983; Lucchin and Matarrese, 1985a). This is because the 
scalar field must be undergoing some kind of evolution if inflation is to 
end eventually, and this injects a scale-dependence into the spectra. As we 
shall see, this effect should be easy to measure.

To take advantage of accurate observations, it is imperative that the
spectra be calculated as accurately as possible. However, let us first
make a qualitative discussion of the generation mechanism.

In a spatially flat, isotropic and homogeneous universe, the Hubble
radius, $H^{-1}(t)$, represents the scale beyond which causal
processes cannot operate. The relative size of a given scale to this
quantity is of crucial importance for understanding how the primordial
spectrum of fluctuations is generated. Quantities such as the power
spectrum are defined via a Fourier expansion as functions of {\em
comoving} wavenumber $k$, and the combination $k/aH$ appears in many
equations. Different physical behavior occurs depending on whether
this quantity is much greater or smaller than unity.

Inflation is defined as an epoch during which the scale factor
accelerates, and so the comoving Hubble radius, $(aH)^{-1}$, must
necessarily decrease.  This is an important feature of the
inflationary scenario, because it means that physical scales will grow
more rapidly than the Hubble radius. As a result, a given mode will
start within the Hubble radius.  In this regime the expansion is
negligible and the microphysics in operation at that epoch will
therefore be relevant. This is determined by the usual flat-space
quantum field theory for which the vacuum state of the scalar field
fluctuations is well understood. As the inflationary expansion
proceeds, however, the mode grows much more rapidly than the Hubble
radius (in physical coordinates) and soon passes outside it. One can
utilize a Heisenberg picture of quantum theory to say that the
operators obey the classical equations of motion, and so the evolution
of the vacuum state can be followed until it crosses outside the
Hubble radius. At this point the microphysics effectively becomes
`frozen'. It turns out that the asymptotic state is not a
zero-particle state --- particles are created by the gravitational
field. Corresponding perturbations in the gravitational field itself
are also generated, so a spectrum of gravitational wave (tensor)
fluctuations is independently produced by the same mechanism.

Once inflation is over, the comoving Hubble radius begins to grow.
Eventually, therefore, the mode in question is able to come back
inside the Hubble radius some time after inflation. The overall result
is that perturbations arising from fluctuations in the inflaton field
can be imprinted onto a given length scale during the inflationary
epoch when that scale first leaves the Hubble radius. These will be
preserved whilst the mode is beyond the Hubble radius and will
therefore be present when the scale re-enters during the
radiation-dominated or matter-dominated eras.

\subsection{Quantitative analysis}

If one is to take full advantage of the observations to the extent one
hopes, it is crucial to have extremely accurate predictions for the
spectra induced by different inflationary models. For example,
microwave background theorists have set themselves the stringent goal
of calculating the radiation angular power spectrum (the $C_l$
discussed later in this paper) to within one percent (Hu et al., 1995),
in the hope that satellite observations may one day provide
extremely accurate measurements of the anisotropies across a wide
range of angular scales (Tegmark and Efstathiou, 1996). This involves a
detailed treatment with a host of subtle physical effects. If
inflationary models are to capitalize on this sort of accuracy, it is
essential to have as accurate a determination as possible of the
initial spectra which are to be input into such calculations. Given
that the slow-roll parameters are typically at least a few percent,
that implies that a determination of the spectra to at least one order
beyond leading order in the slow-roll expansion is desired.

The calculations we make are based on linear perturbation theory.
Since the observed anisotropies are small, this approximation is
considerably more accurate than the slow-roll approximation, and we
need not attempt to go beyond it, though it is possible to extend
calculations beyond linear perturbation theory (Durrer and
Sakellariadou, 1994).

Before proceeding, however, let us clarify a notational point. In
earlier literature, especially CKLL2 and Liddle and Turner (1994),
orders were referred to as first-order, second-order etc. However, we
feel this can be misleading, because it might suggest that all terms
containing say two slow-roll parameters in any given expression are
supposed to be neglected.  This is not the intention, because in many
expressions the lowest-order term already contains one or more powers
of the slow-roll parameters. Because differentiation respects the
order-by-order expansion, while multiplying each term by a slow-roll
parameter, it is always valid to take terms to the same number of
orders, however many slow-roll parameters the actual terms possess.
Therefore, in order to clarify the meaning, we choose to always employ
the phrase \underline{lowest-order} to indicate the term containing
the least number of powers of the slow-roll parameters, however many
that may be for a specific expression. The phrase next-to-lowest
order, abbreviated to \underline{next-order}, then indicates
correction terms to this which contain one further power of the
slow-roll parameters than the lowest-order terms.

The calculation of the spectra to next-order has been provided by
Stewart and Lyth (1993). Because of its crucial importance, we shall
devote quite some time to describing it. The basic principle is to
start with the one known situation where the spectra can be calculated
exactly, that of power-law inflation. This corresponds to each of the
slow-roll parameters having the same constant value. To next-order, a
general inflationary potential can be considered via an expansion in
$(\epsilon - \eta)$ about a power-law inflation model with the same
$\epsilon$; as we shall see, it is an adequate approximation to treat
$\epsilon$ and $\eta$ as different constant values.

In fact, the logic we develop is slightly different to that of Stewart
and Lyth (1993). They computed an exact solution for the situation
where $\epsilon$ and $\eta$ are treated as exactly constant with
different values.  Formally, this situation does not exist as
$\epsilon$ precisely constant implies $\epsilon = \eta$. They then
treated power-law inflation as an exact special case of this
situation, and a general inflation model to next-order as an expansion
about their more general result. Logically, it is more accurate to
expand directly about the exact power-law inflation result, but
nevertheless the final answer is guaranteed to be the same.

\subsubsection{Scalar perturbations}

Throughout the calculations to derive the spectra of scalar and tensor
fluctuations, the space-time representing our universe is decoupled
into two components, representing the background and perturbation
contributions. The background part is taken to be the homogeneous and
isotropic FRW metric. This is a reasonable assumption to make in view
of the high degree of spatial uniformity in the temperature of the
cosmic microwave background. In this paper we assume the background is
also spatially flat with a line element given by
Eq.~(\ref{background}). The perturbed sector of the metric then
determines by how much the actual universe deviates from this
idealization.

Four quantities are required to specify the general nature of a scalar
perturbation. These may be denoted by $A$, $B$, $\Psi$ and $E$ and
these are functions of the space and time coordinates. It has been
shown by Bardeen (1980) and by Kodama and Sasaki (1984) that the most
general form of the line element for the background and scalar metric
perturbations is given by
\begin{equation}
\label{metric}
ds^2 =a^2(\tau) \left[ (1+2A)d\tau^2 -2\partial_i B dx^i d\tau -
\left[ (1-2\Psi ) \delta_{ij} +2 \partial_i \partial_j E \right] dx^i dx^j 
\right] \,,
\end{equation}
where $\tau \equiv \int dt/a(t)$ is conformal time.

The perturbations can be measured by the intrinsic curvature perturbation of 
the comoving hypersurfaces, which has the form
\begin{equation}
\label{intrinsic}
{\cal{R}} = -\Psi -\frac{H}{\dot{\phi}} \delta \phi\,,
\end{equation}
during inflation, where $\delta\phi$ represents the fluctuation of the
inflaton field and $\dot{\phi}$ and $H$ are calculated from the
background field equations Eqs.~(\ref{H})-(\ref{field}). To proceed,
we follow Mukhanov, Feldman and Brandenberger (1992) and introduce the
gauge-invariant potential
\begin{equation}
\label{v}
u \equiv a\left[ \delta \phi +\frac{\dot{\phi}}{H} \Psi \right] \,.
\end{equation}
It also proves convenient to introduce the variable
\begin{equation}
\label{z}
z\equiv \frac{a\dot{\phi}}{H} \,,
\end{equation}
and it follows immediately that
\begin{equation}
\label{vR}
u=-z{\cal{R}} \,.
\end{equation}

The evolution of the perturbations is determined by the Einstein
action. The first-order perturbation equations of motion are given by
a second-order action. Hence, the gravitational and matter sectors are
separated and each expanded to second-order in the perturbations. The
result for the gravitational component is simplified by employing the
ADM form of the action (Arnowitt, Deser, and Misner, 1962; Misner et al., 
1973). The action for the matter perturbations, on the other hand, can be 
calculated by expanding the Lagrangian as a Taylor series about a fixed 
value of the scalar field, applying the background field equations and 
integrating by parts. Mukhanov et al. (1992) show that the full action for 
linear scalar perturbations is given by
\begin{equation}
\label{pertact}
S=\int d^4x{\cal{L}} = \frac{1}{2} \int d\tau d^3{\bf x} \left[ 
	\left( \partial_{\tau} u \right)^2 -\delta^{ij} \partial_i
	u\partial_j u + \frac{z_{\tau\tau}}{z} u^2 \right] \,,
\end{equation}
where a subscript $\tau$ denotes partial differentiation with respect
to conformal time. For further details the reader is referred to
Mukhanov (1989) and Makino and Sasaki (1991).

Formally, this is equivalent to the action for a scalar field in flat
space-time with a time-dependent effective mass $m^2 =-
z_{\tau\tau}/z$. This equivalence implies that one can consider the
quantum theory in an analogous fashion to that of a scalar field
propagating on Minkowski space-time in the presence of a time-varying
external field (Grib, Mamaev, and Mostepanenko, 1980). The
time-dependence has its origin in the variation of the background
space-time (Birrell and Davies, 1982).

The momentum canonical to $u$ is given by
\begin{equation}
\label{canmom}
\pi (\tau ,{\bf x}) =\frac{\partial {\cal{L}}}{\partial (u_{\tau})}
=u_{\tau} (\tau , {\bf x}) \,,
\end{equation}
and the theory is then quantized by promoting $u$ and its conjugate
momentum to operators that satisfy the following commutation relations
on the $\tau ={\rm constant}$ hypersurfaces:
\begin{eqnarray}
\label{ETCR}
\left[ \hat{u} (\tau ,{\bf x}) ,\hat{u} (\tau ,{\bf y}) \right] =
\left[ \hat{\pi} (\tau ,{\bf x}) ,\hat{\pi} (\tau ,{\bf y}) \right] =0 \,, 
\\
\left[ \hat{u} (\tau ,{\bf x}) ,\hat{\pi} (\tau ,{\bf y}) 
\right] = i\delta^{(3)} ({\bf x} -{\bf y}) \,.
\end{eqnarray}

We expand the operator $\hat{u} (\tau ,{\bf x})$ in terms of plane waves
\begin{equation}
\label{vhat} 
\hat{u} (\tau ,{\bf x}) = \int \frac{d^3{\bf k}}{(2\pi)^{3/2}} \left[ 
 u_k (\tau) \hat{a}_{\bf k} e^{i{\bf k.x}} +u_k^* (\tau) 
\hat{a}^{\dagger}_{\bf k} e^{-i{\bf k.x}} \right] \,,
\end{equation}
and the field equation for the coefficients $u_k$ is derived by
setting the variation of the action Eq.~(\ref{pertact}) with respect
to $u$ equal to zero. It is given by (Mukhanov, 1985, 1988; Stewart and 
Lyth, 1993)
\begin{equation}
\label{vfield}
\frac{d^2u_k}{d\tau^2} +\left( k^2 - \frac{1}{z} \frac{d^2 z}{d\tau^2}
\right) u_k =0 \,.
\end{equation}
These modes are normalized so that they satisfy the Wronskian condition 
\begin{equation}
\label{norm}
u^*_k \frac{du_k}{d\tau} - u_k \frac{du^*_k}{d\tau} =-i \,,
\end{equation}
and this condition ensures that the creation and annihilation
operators $\hat{a}^{\dagger}_{\bf k}$ and $\hat{a}_{\bf k}$ satisfy
the usual commutation relations for bosons:
\begin{equation}
\label{usual}
[\hat{a}_{\bf k}, \hat{a}_{\bf l}]=[\hat{a}^{\dagger}_{\bf k},
\hat{a}^{\dagger}_{\bf l}]=0, \qquad [\hat{a}_{\bf k}, 
\hat{a}^{\dagger}_{\bf l}]=\delta^{(3)} ({\bf k}-{\bf l}) \,.
\end{equation}
The vacuum is therefore defined as the state that is annihilated by
all the $\hat{a}_{\bf k}$, i.~e., $\hat{a}_{\bf k} | 0 \rangle =0$.

The modes $u_k(\tau)$ must have the correct form at very short
distances so that ordinary flat space-time quantum field theory is
reproduced.  Thus, in the limit that $k/aH \rightarrow \infty$, the
modes should approach plane waves of the form
\begin{equation}
\label{short}
u_k(\tau) \rightarrow \frac{1}{\sqrt{2 k}} e^{-ik\tau}\,.
\end{equation}
In the opposite (long wavelength) regime where $k$ can be neglected in
Eq.~(\ref{vfield}), we see immediately that the growing mode solution
is
\begin{equation}
\label{long}
u_k \propto z \,,
\end{equation}
with no dependence on the behavior of the scale factor (except
insofar as implicitly through the definition of $z$).

Ultimately, the quantity in which we are interested is the curvature
perturbation ${\cal R}$. We expand this in a Fourier series
\begin{equation}
\label{fourierR}
{\cal{R}} =\int \frac{d^3{\bf k}}{(2\pi)^{3/2}} {\cal{R}}_{\bf k}
(\tau) e^{i{\bf k.x}} \,.
\end{equation}
The power spectrum ${\cal P}_{\cal{R}} (k)$ can then be defined in terms 
of the vacuum expectation value
\begin{equation}
\label{scalarpower}
\langle {\cal{R}}_{\bf k} {\cal{R}}^*_{\bf l} \rangle =
\frac{2\pi^2}{k^3} {\cal P}_{\cal{R}} \delta^{(3)} ({\bf k}-{\bf l}) \,,
\end{equation}
where the prefactor is in a sense arbitrary but is chosen to obey the
usual Fourier conventions. The left-hand side of this expression may
be evaluated by combining Eqs.~(\ref{vR}), (\ref{usual}) and
(\ref{fourierR}):
\begin{equation}
\label{vev}
\langle {\cal{R}}_{\bf k}{\cal{R}}^*_{\bf l} \rangle =\frac{1}{z^2}
|u_k|^2 \delta^{(3)} ({\bf k}-{\bf l}) \,,
\end{equation}
yielding
\begin{equation}
\label{pspec}
{\cal P}_{\cal R}^{1/2}(k) = \sqrt{\frac{k^3}{2\pi^2}} \, \left| 
\frac{u_k}{z} \right| \,.
\end{equation}
For modes well outside the horizon, the growing mode of $u_k$ will
dominate and so the spectrum will approach a constant value. It is
this value that we are aiming to calculate.

In order to provide a solution, we need an expression for
$z_{\tau\tau}/z$.  This can be straightforwardly obtained as
\begin{equation}
\label{zderiv}
\frac{1}{z} \frac{d^2z}{d\tau^2} =2a^2 H^2 \left[ 1+\epsilon -\frac{3}{2} 
\eta + \epsilon^2 -2 \epsilon \eta + \frac{1}{2} \eta^2 + \frac{1}{2} \xi^2
\right] \,,
\end{equation}
and despite its appearance as an expansion in slow-roll parameters,
this expression is exact.

\vspace*{12pt}
\noindent
{\bf Exact solution for power-law inflation}
\vspace*{12pt}

So far, all the expressions we have written down have been exact.
However, we have reached the limit of analytic progress for general
circumstances. The desired situation then is to obtain an exact
solution for some special case, about which a general expansion can be
applied in terms of the slow-roll parameters. Such an exact solution
is the case of power-law inflation, which we now derive\footnote{It is
at this point that our construction of the expansion begins to differ
in logical construction from Stewart and Lyth (1993), though the final
result will agree.}.

Power-law inflation, where the scale factor expands as $a(t) \propto
t^p$, corresponds to the particularly simple case where the Hubble
parameter is exponential in $\phi$ (Lucchin and Matarrese, 1985a,
1985b):
\begin{equation}
H(\phi) \propto \exp \left( \sqrt{\frac{4\pi}{p}} \, 
	\frac{\phi}{m_{{\rm Pl}}} \right) \,.
\end{equation}
It follows that the slow-roll parameters are not only constant but
equal; we are primarily interested in
\begin{equation}
\epsilon = \eta = \xi = \frac{1}{p} \,.
\end{equation}
With a constant $\epsilon$, an integration by parts 
\begin{equation}
\label{parts}
\tau = \int \frac{da}{a^2H} = -\frac{1}{aH} + \int \frac{\epsilon \, 
da}{a^2H} \,,
\end{equation}
supplies the conformal time as
\begin{equation}
\label{simply}
\tau =-\frac{1}{aH}\frac{1}{1-\epsilon} \,.
\end{equation}
Thus, $\tau$ is negative during inflation, with $\tau = 0$
corresponding to the infinite future.

Since the slow-roll parameters in Eq.~(\ref{zderiv}) are constant,
Eq.~(\ref{vfield}) simplifies to a Bessel equation of the form
\begin{equation}
\label{bessel}
\left[ \frac{d^2}{d\tau^2} +k^2 -\frac{(\nu^2 -\frac{1}{4})}{\tau^2} 
	\right] u_k =0 \,,
\end{equation}
where
\begin{equation}
\nu \equiv \frac{3}{2} + \frac{1}{p-1} \,.
\end{equation}
The appropriately normalized solution with the correct asymptotic
behavior at small scales is therefore given by\footnote{The choice of
phase factor ensures that the behavior described by Eq.~(\ref{short})
is reproduced at short scales, and the factor of $\sqrt{\pi}/2$
implies that condition Eq.~(\ref{norm}) is satisfied.}
\begin{equation}
\label{correctform}
u_k(\tau) =\frac{\sqrt{\pi}}{2} e^{i(\nu +1/2)\pi/2} (-\tau)^{1/2}
	H_{\nu}^{(1)} (-k\tau) \,,
\end{equation}
where $H_{\nu}^{(1)}$ is the Hankel function of the first kind of order
$\nu$. 

Ultimately, we are interested in the asymptotic form of the solution
once the mode is well outside the horizon. Taking the limit $k/aH
\rightarrow 0$ yields the asymptotic form
\begin{equation}
\label{modelimit}
u_k \rightarrow e^{i(\nu -1/2)\pi /2} 2^{\nu -3/2} 
\frac{ \Gamma (\nu)}{\Gamma (3/2)} \frac{1}{\sqrt{2k}} (-k \tau )^{-\nu 
+1/2} \,,
\end{equation}
and substituting this into Eq.~(\ref{pspec}) gives the asymptotic form
of the power spectrum
\begin{equation}
\label{scalaramp}
{\cal P}_{\cal{R}}^{1/2}(k) =2^{\nu -1/2} \frac{\Gamma(\nu)}{\Gamma(3/2)} 
	(\nu-1/2)^{1/2 - \nu} \frac{1}{m_{{\rm Pl}}^2} \left. 
	\frac{H^2}{|H'|} \right|_{k=aH} \,,
\end{equation}
where we have employed Eq.~(\ref{simply}) to substitute for $k\tau$.
A subtle point is that, despite the appearance of this equation, the
calculated value for the spectrum is {\em not} the value when the
scale crosses outside the Hubble radius. Rather, it is the asymptotic
value as $k/aH \rightarrow 0$, but rewritten in terms of the values
which quantities had at Hubble radius crossing.

This exact expression for the asymptotic power spectrum was first
derived in an earlier paper by Lyth and Stewart (1992). It is one of
only two known exact solutions, and is the only one for a realistic
inflationary scenario. The other known exact solution, found by
Easther (1996), arises in an artificial model designed to permit exact
solution, and while of theoretical interest is excluded by observations.

\vspace*{12pt}
\noindent
{\bf Slow-roll expansion for general potentials}
\vspace*{12pt}

Having obtained an exact solution, we can now make an expansion about
it.  The power-law inflation case corresponded to the slow-roll
parameters being equal, and hence exactly constant; we now wish to
allow them to be different which means they will pick up a time
dependence.

At this stage, there is no need to require that the parameter
$\epsilon$ be small, for the exact solution exists for all $\epsilon <
1$. However, the deviation of all higher slow-roll parameters from
$\epsilon$ must indeed be small, since the differences vanish for the
exact solution. Let us label the first of these as $\zeta = \epsilon -
\eta$. There are in general an infinite number of such small
parameters in the expansion but we shall only need this one.

The first step is to find a more general equation for $\tau$. By
integrating by parts in the manner of Eq.~(\ref{parts}) an infinite
number of times, one can obtain
\begin{equation}
\tau =-\frac{1}{aH}\frac{1}{1-\epsilon} - \frac{2\epsilon \zeta}{aH} + 
\mbox{expansion in slow-roll parameters $\zeta$ etc.} \,,
\end{equation}
where $\epsilon$ can now have arbitrary time dependence. This is all
very well, but even via an expansion in small $\zeta$ one cannot
analytically solve Eq.~(\ref{vfield}) for a general time-dependent
$\epsilon$; we must resort to a situation where $aH\tau$ can be taken
as constant for each $k$-mode (though not necessarily the same
constant for different $k$). The relevant equation to study is the
exact relation
\begin{equation}
\label{vareps}
\dot{\epsilon}/H = 2 \epsilon \zeta \,.
\end{equation}
What we are aiming to do is to shift the time dependence of $\epsilon$
to next-order in the expansion, so that it can be neglected. This is
achieved by assuming that $\epsilon$ is a small parameter as well as
$\zeta$ (that is, that both $\epsilon$ and $\eta$ are small), in which
case one can expand to lowest-order to get
\begin{equation}
\tau = -\frac{1}{aH} (1+\epsilon) \,.
\end{equation}
We will return to the question of the error in assuming constant
$\epsilon$ shortly.

Having this expression for $\tau$, we can now immediately use
Eq.~(\ref{zderiv}), which must also be truncated to first-order. This
gives the same Bessel equation Eq.~(\ref{bessel}), but now with $\nu$
given by
\begin{equation}
\nu = \frac{3}{2} + 2\epsilon - \eta \,.
\end{equation}
The assumption that treats $\epsilon$ as constant also allows $\eta$
to be taken as constant, but crucially, $\epsilon$ and $\eta$ need no
longer be the same since we are consistent to first-order in their
difference. The differences between further slow-roll parameters and
$\epsilon$ lead to higher order effects, and so incorporating
$\epsilon$ and $\eta$ in this manner is applicable to an arbitrary
inflaton potential to next-order.  The same solution
Eq.~(\ref{scalaramp}) can be used with the new form of $\nu$, but for
consistency it should be expanded to the same order. This gives the
final answer, which is true for general inflation potentials to this
order, of (Stewart and Lyth, 1993)
\begin{equation}
\label{2ndscal}
P_{\cal{R}}^{1/2}(k) = \left[ 1 - (2C+1) \epsilon + C \eta
	\right] \; \frac{2}{m_{{\rm Pl}}^2} \left. 
	\frac{H^2}{|H'|} \right|_{k=aH} \,,
\end{equation}
where $C = -2 +\ln 2 +\gamma \simeq -0.73$ is a numerical constant,
$\gamma$ being the Euler constant originating in the expansion of the
Gamma function. Since the slow-roll parameters are to be treated as
constant, they can also be evaluated at horizon crossing.

Let us now return to the question of the error in assuming $\epsilon$
is constant. The crucial aspect is that the variation of $\epsilon$ is
only important around $k = aH$. In either of the two extreme regimes
the evolution of $u_k$ (in relation to $z$) is independent of it
(Eqs.~(\ref{short}) and (\ref{long})). Assuming the variation of
$\epsilon$ is only important for some unspecified but finite number of
$e$-foldings, Eq.~(\ref{vareps}) measures that change (per
$e$-folding). As long as we are assuming $\epsilon$ small as well as
$\zeta$, that change is next-order and can be neglected along with all
the other next-order terms we did not attempt to include.

Finally, one can see from the complexity of this calculation the
obstacles to obtaining general expressions which go to yet another
higher order. This would involve finding some way of solving the
Bessel-like equation in the situation where its coefficients could not
be treated as constant.

This concludes our discussion on the generation of scalar
perturbations during inflation. In the remainder of this Section we
will present the analogous result for the tensor fluctuations.

\subsubsection{Gravitational waves}

The propagation of weak gravitational waves on the FRW background was
investigated by Lifshitz (1946). Quantum fluctuations in the
gravitational field are generated in a similar fashion to that of the
scalar perturbations discussed above. A gravitational wave may be
viewed as a ripple in the background space-time metric
Eq.~(\ref{background}) and in general the linear tensor perturbations
may be written as $g_{\mu\nu} =a^2 (\tau)[ \eta_{\mu\nu}
+h_{\mu\nu}]$, where $|h_{\mu\nu}| \ll 1$ denotes the metric
perturbation and $\eta_{\mu\nu}$ is the flat space-time metric
(Bardeen, 1980; Kodama and Sasaki, 1984). In the transverse-traceless
gauge, we have $h_{00} = h_{0i} = \partial^i h_{ij} =\delta^{ij}
h_{ij} =0$, and there are two independent polarization states (Misner et 
al., 1973). These are usually denoted as $\lambda = +,\times$.

The gravitons are the propagating modes associated with these two
states. The classical dynamics of the gravitational waves is
determined by expanding the Einstein-Hilbert action to quadratic order
in $h_{\mu\nu}$ and it can be shown that this action takes the form
(Grishchuk, 1974, 1977)
\begin{equation}
\label{gravitonaction}
S_g= \frac{m_{{\rm Pl}}^2}{64\pi} \int d\tau d^3 {\bf x} a^2 (\tau) 
\partial_{\mu} {h^i}_j \partial^{\mu}{h_i}^j \,.
\end{equation}
It proves convenient to introduce the rescaled variable
\begin{equation}
{P^i}_j (x) \equiv (m_{{\rm Pl}}^2 /32 \pi)^{1/2} a(\tau) {h^i}_j (x) \,,
\end{equation}
and substitution of this expression into the action 
Eq.~(\ref{gravitonaction}) implies that 
\begin{equation}
\label{newact}
S_g =\frac{1}{2} \int d\tau d^3{\bf x} \left[ \left( \partial_{\tau} 
{P_i}^j \right) \left( \partial^{\tau} {P^i}_j \right) 
- \delta^{rs} \left( \partial_r {P_i}^j \right) \left( \partial_s 
{P^i}_j \right) + \frac{a_{\tau\tau}}{a} {P_i}^j {P^i}_j 
\right] \,,
\end{equation}
where we have ignored a total derivative. This expression resembles
the equivalent action Eq.~(\ref{pertact}) for the scalar
perturbations. Indeed, we may interpret Eq.~(\ref{newact}) as the
action for two scalar fields in Minkowski space-time each with an
effective mass squared given by $a_{\tau\tau}/a $. This equivalence
between the two actions implies that the procedure for quantizing the
tensor fluctuations is essentially the same as in the scalar case.

We perform a Fourier decomposition of the gravitational waves by
expanding ${P^i}_j$:
\begin{equation}
{P^i}_j = \sum_{\lambda =+,\times} \int \frac{d^3 {\bf k}}{(2\pi )^{3/2}} 
 v_{{\bf k}, \lambda} (\tau) {\epsilon^i}_j ({\bf k};\lambda ) 
e^{i {\bf k.x}} \,.
\end{equation}
In this expression ${\epsilon^i}_j ({\bf k}; \lambda )$ is the
polarization tensor and satisfies the conditions $\epsilon_{ij}
=\epsilon_{ji}$, $\epsilon_{ii}=0$, $k^i\epsilon_{ij}=0$ and
${\epsilon^i}_j ({\bf k},\lambda ) {\epsilon_i}^{j*} ({\bf k},
\lambda' ) =\delta_{\lambda\lambda'}$. The analysis is further
simplified if we choose $\epsilon_{ij}(-{\bf k}, \lambda
)=\epsilon_{ij}^* ({\bf k} ,\lambda )$, since this ensures that
$v_{{\bf k} ,\lambda} =v^*_{-{\bf k} ,\lambda}$. We may consider each
polarization state separately. The effective graviton action during
inflation therefore takes the form
\begin{equation}
\label{rewrite}
S_g=\frac{1}{2} \sum_{\lambda =+,\times} \int d\tau d^3{\bf k} 
\left[ \left( \partial_{\tau} \left| v_{{\bf k},\lambda} \right| \right)^2 
-\left( k^2 -\frac{a_{\tau\tau}}{a} \right) \left| v_{{\bf k},\lambda} 
\right|^2 \right] \,.
\end{equation}

We quantize by interpreting $v_{{\bf k},\lambda} (\tau)$ as the operator
\begin{equation}
\label{quantize}
\hat{v}_{{\bf k},\lambda} (\eta) =v_k(\eta) \hat{a}_{{\bf k},\lambda} 
+v^*_k(\eta) \hat{a}^{\dagger}_{-{\bf k},\lambda} \,,
\end{equation}
where the modes $v_k$ satisfy the normalization condition
Eq.~(\ref{norm}) and have the form given by Eq.~(\ref{short}) as $aH/k
\rightarrow 0$.  This ensures that the creation and annihilation
operators satisfy
\begin{equation}
\label{gravityETCR}
[\hat{a}_{{\bf k},\lambda} \hat{a}^{\dagger}_{{\bf l},\sigma} ] 
=\delta_{\lambda\sigma} \delta^{(3)} ({\bf k}-{\bf l}), 
\qquad \hat{a}_{{\bf k},\lambda} |0\rangle \,,
\end{equation}
and the spectrum of gravitational waves ${\cal P}_g(k)$ is then defined by 
\begin{equation}
\langle \hat{v}_{{\bf k},\lambda } \hat{v}^*_{{\bf l},\lambda} \rangle 
=\frac{m_{{\rm Pl}}^2 a^2}{32\pi} 
\frac{2\pi^2}{k^3} {\cal P}_g \delta^{(3)} ({\bf k}-{\bf l}) \,.
\end{equation}

The field equation for $u_k$, derived by varying the action 
Eq.~(\ref{rewrite}), is 
\begin{equation}
\label{fieldeqn}
\frac{d^2v_k}{d\tau^2} +\left( k^2 -\frac{1}{a}\frac{d^2a}{d\tau^2} \right) 
v_k =0 \,,
\end{equation}
and the scale factor term can be written as
\begin{equation}
\frac{1}{a}\frac{d^2a}{d\tau^2} = 2a^2H^2 \left( 1 - \frac{1}{2} \epsilon
\right) \,.
\end{equation}
This puts us in a very similar situation to that for the density
perturbations. The situation is simplified since $a$ appears directly
in the equation of motion rather than $z$, but the strategy is exactly
the same.

For power-law inflation we can again solve exactly by writing
\begin{equation}
\frac{a_{\tau\tau}}{a} =\frac{1}{\tau^2} \left( \mu^2 -\frac{1}{4} 
\right)  \,,
\end{equation}
where
\begin{equation}
\mu \equiv \frac{3}{2} +\frac{1}{p-1} \,.
\end{equation}
For power-law inflation $\nu$ and $\mu$ coincide, though in general
they do not. The appropriate solution for $v_k$ is given by
Eq.~(\ref{correctform}), as before, after replacing $\nu$ with $\mu$.
It follows, therefore, that
\begin{equation}
\label{tensoramp}
{\cal P}^{1/2}_g (k) =\frac{2}{\sqrt{\pi}} \, 2^{\mu-1/2} 
	\frac{\Gamma (\mu)}{\Gamma (3/2)} (\mu-1/2 )^{1/2 -\mu} 
	\left. \frac{H}{m_{{\rm Pl}}} \right|_{k=aH} \,,
\end{equation}
where $P_g$ has been multiplied by a factor of $2$ to account for the
two polarization states. This exact solution was first obtained by
Abbott and Wise (1984a) and we note that for power-law inflation
\begin{equation}
\frac{{\cal P}_g^{1/2}}{{\cal P}_{\cal R}^{1/2}} = \frac{4}{\sqrt{p}} = 
4\sqrt{\epsilon} \,.
\end{equation}

The final step is to carry out the expansion in the same way as in the
scalar case to yield the slow-roll expression for the tensor spectrum.
This gives
\begin{equation}
\mu = \frac{3}{2} + \epsilon \,,
\end{equation}
and hence
\begin{equation}
\label{2ndtens}
{\cal P}^{1/2}_g (k) = \left[ 1 - (C+1) \epsilon \right] 
\frac{4}{\sqrt{\pi}} 
	\left. \frac{H}{m_{{\rm Pl}}} \right|_{k=aH} \,.
\end{equation}

\section{~~~Lowest-Order Reconstruction}
\label{first}

\def\theequation{4.\arabic{equation}}
\setcounter{equation}{0}

In the previous section we discussed the derivation of expressions for
the two {\em initial} spectra ${\cal P}^{1/2}_{{\cal{R}}}$ and ${\cal
P}^{1/2}_g$, which were accurate to next-order in the slow-roll
parameters. Before proceeding, let's relate our notation to other
notations that the reader may be familiar with, which concern the
present-day spectra. In order to derive these, one needs the transfer
functions $T(k)$ and $T_g(k)$ for both scalars (Efstathiou, 1990) and 
tensors (Turner, White, and Lidsey, 1993) respectively, which describe the 
suppression of growth on scale $k$ relative to the infinite wavelength mode. 
The transfer functions in general depend on a whole range of cosmological 
parameters, as discussed later. The present-day spectrum of density 
perturbations, denoted $P(k)$, is given by
\begin{equation}
\frac{k^3}{2\pi^2} P(k) = \left( \frac{k}{aH} \right)^4 \, T^2(k) \, 
	{\cal P}_{{\cal R}}(k) \,,
\end{equation}
whle the energy density (per octave) in gravitational waves is
\begin{equation}
\Omega_g(k) = \frac{1}{24} \, T_g^2(k) \, {\cal P}_g(k) \,.
\end{equation}
These expressions apply to a critical density universe; for models with a 
cosmological constant they require generalization (see Turner and White, 
1996). Note though that in the following Sections, we shall always be 
working with (rescaled versions of) the initial spectra, and not with the
present-day spectra.

In this Section, we shall concentrate on the lowest-order situation,
where all expressions are truncated at the lowest-order. This is not
equivalent to assuming that $\epsilon$ and $\eta$ are zero, for in
some expressions, such as the spectral indices, the lowest-order terms
contain $\epsilon$ and $\eta$, as we shall see.  This approximation
can be regarded as being extremely useful for the present state of
observations. However, optimistically one hopes that future
observations, particularly satellite-based high resolution microwave
background anisotropy observations, will require a higher degree of
accuracy as discussed in Section \ref{second}.

\subsection{The consistency equation and generic predictions of inflation}

In the forthcoming analysis it will prove convenient to work with
rescaled expressions for the spectra 
${\cal P}^{1/2}_{{\cal{R}}}$ and ${\cal P}^{1/2}_g$ which 
we will use throughout the rest of the paper. 
To lowest-order we obtain
\begin{eqnarray}
\label{firstscalar}
A_S(k) & \equiv & 2 {\cal P}^{1/2}_{{\cal{R}}}/5 = 
\left. \frac{4}{5} \frac{H^2}{m_{{\rm Pl}}^2|H'|} 
	\right|_{k=aH} \,, \\
\label{firsttensor}
A_T(k) & \equiv & {\cal
P}^{1/2}_g/10 = \left. \frac{2}{5\sqrt{\pi}} \frac{H}{m_{{\rm Pl}}}
	\right|_{k=aH} \,.
\end{eqnarray}
The specific choice of normalizations is arbitrary\footnote{We remark
that these expressions have different prefactors to those contained in
our original papers, CKLL1 and CKLL2; while one normalization is as
valid as any other, the normalizations chosen in those papers were
atypical of the literature. Those used here conform more readily with
the conventions employed in the existing literature and in particular
with the Stewart and Lyth (1993) calculation. In fact, the numerical
difference is only 0.3\%. The ratio of the tensor and scalar
amplitudes is unaffected by this change.}. The above choice ensures
that $A_S$ coincides precisely with the quantity $\delta_H$ as defined
by Liddle and Lyth (1993a, Eq.~(3.6)). This parameter may be viewed as
the density contrast at Hubble-radius-crossing. The normalization for
the tensor spectrum is then chosen so that to lowest-order $\epsilon =
A_T^2/A_S^2$.

During inflation the scalar field slowly rolls down its
self-interaction potential. This causes the Hubble parameter to vary
as a function of cosmic time and therefore with respect to the scale
at Hubble-radius crossing. The expressions for the perturbations
therefore acquire a dependence on scale and it is conventional to
express this variation in terms of {\em spectral indices}. In general,
these indices are themselves functions of scale and there appear to be
two ways in which they may be defined. In the first case, one may
simply write the power spectra as
\begin{equation}
A_S^2(k) = A_S^2(k_0) \left(\frac{k}{k_0}\right)^{\tilde{n}(k)-1} \quad ; 
\quad A_T^2(k) =A_T^2(k_0) \left(\frac{k}{k_0}\right)^{\tilde{n}_T(k)} \,.
\end{equation}
Although these definitions are completely general, they do require a
specific choice of $k_0$ to be made. This feature implies that the
definitions are non-local, a considerable drawback. A more suitable
alternative is to define the spectral indices differentially via
\begin{eqnarray}
n(k)-1 & \equiv & \frac{d \ln A_S^2}{d \ln k} \,,\\
n_T(k) & \equiv & \frac{d \ln A_T^2}{d \ln k} \,.
\end{eqnarray}
We shall adopt this second choice in this work. The two definitions
coincide for power-law spectra, where the indices are constant. In
general, however, they are inequivalent.

At the level of approximation we are considering in this Section, the
spectral indices may be expressed directly in terms of the slow-roll
parameters $\epsilon$ and $\eta$. One calculates the first derivatives
of the amplitudes from Eqs.~(\ref{firstscalar}) and
(\ref{firsttensor}) with respect to $\phi$ and converts to derivatives
with respect to wavenumber with the help of Eq.~(\ref{kphi}). It is
straightforward to show that
\begin{eqnarray}
\label{scalind1}
n(k)-1 & = & 2 \eta - 4 \epsilon \,, \\
\label{tensind1}
n_T(k) & = & - 2 \epsilon \,.
\end{eqnarray}

The conventional statement attached to these expressions is that
inflation predicts spectra which to the presently desired accuracy can
be approximated as power-laws; that is, that the slow-roll parameters
can be treated as constants. While this statement is formally correct,
it requires some discussion. In particular it is important to realize
that the power-law approximation has no direct connection to the
slow-roll approximation, but rather is a statement that the relevant
observations cover only a limited range of scale and do so with
limited accuracy. As far as the derivations of the spectra are
concerned, the approximation is that for each scale the parameters can
be treated as constant while that scale crosses outside the Hubble
radius. However, in this `adiabatic' approximation, there is no need
for those constant values to remain the same from scale to scale.
Thus, the expressions for the spectra can be applied across the
complete range of scales. Although they are an approximation at each
scale, the approximation does not deteriorate when one attempts to
study a wider range of scales. The feature that dictates whether the
spectra can be treated as power-laws is that the range of scales over
which observations can be made is quite small, in terms of the range
of $\phi$ values, and taking additional derivatives of the spectra
introduces into the lowest-order result an extra power in the
slow-roll parameters. For example, although differentiating
Eq.~(\ref{scalind1}) gives the correct lowest-order expression for
$dn/d \ln k$, this will be of order $\epsilon^2$ and hence a small
effect over the short range of scales large-scale structure samples.
Were large-scale structure able to sample, for example, scales
encompassing twenty orders of magnitude rather than four, the
approximation by power-law would be liable to break down for typical
inflation models. With high accuracy observations, the power-law
approximation represented by these lowest-order expressions may prove
inadequate even over the short range of accessible scales.

We emphasize that the spectral indices do not have to satisfy the
exact power-law result $n-1 = n_T$ at this level of approximation.
Each spectrum is uniquely specified by its amplitude and spectral
index. The overall amplitude is a free parameter determined by the
normalization of the expansion rate $H$ during inflation (or
equivalently the scalar field potential $V$). On the other hand, the
relative amplitude of the two spectra is given by
\begin{equation}
\label{rat1}
\frac{A_T^2}{A_S^2} = \epsilon \,.
\end{equation}
Thus, there exists a simple relationship between the relative
amplitude and the tensor spectral index:
\begin{equation}
\label{firstconsistency}
n_T = - 2 \frac{A_T^2}{A_S^2} \,.
\end{equation}
This is the lowest-order consistency equation and represents an
extremely distinctive signature of inflationary models. It is
difficult to conceive of such a relation occurring via any other
mechanism for the generation of the spectra.

Since it is possible for the spectra to have different indices, the
assumption that their ratio is fixed can be true only for a limited
range of scales, but the correction enters at a higher order in the
slow-roll parameters.

This expression is often written in a slightly different form in order
to bring the right hand side closer to observations. Since the spectra
can be defined with arbitrary prefactors, they themselves have no
definite significance. The environment in which each spectrum may have
an effect that allows direct comparison is in large angle microwave
background anisotropies.  In this case the scalar and tensor
fluctuations each contribute independently to the expected value of
the microwave multipoles, $C_l$ (defined and discussed in more detail
in Section \ref{obs}), and in the approximation where only the
Sachs-Wolfe term is included and perfect matter--domination at last 
scattering assumed, this enables one to write the
lowest-order consistency equation as (Liddle and Lyth, 1992, 1993a)
\begin{equation}
\label{firstCl}
\frac{C_l^T}{C_l^S} = -6.2 n_T \,.
\end{equation}
This equation applies for moderate values of $l$ corresponding to
scales that are sufficiently small for the curvature of the last
scattering surface to be negligible and yet are large enough to be
well above the Hubble radius at decoupling\footnote{The exact number
in this relation is sometimes written in different ways. It was first
evaluated exactly as $25(1+48\pi^2/385)/9$ in the scale-invariant
limit by Starobinsky (1985).  This is numerically equal to 6.2. There
is no regime where this strictly holds, as corrections from the `Doppler' 
peak and from the Universe being not perfectly matter dominated at last 
scattering intervene before the asymptote is reached. Other authors evaluate
only part of the expression to approximate it as $2\pi$, or even $6$.
Finally, many authors consider the ratio of contributions to the
quadrupole $l=2$. In this case there is a geometrical correction from
the curvature of the last scattering surface which make the factor
close to $7$.}.

Eqs.~(\ref{scalind1}), (\ref{tensind1}) and (\ref{rat1}) contain all
the information one requires to determine the generic behavior of
inflationary models at this order. Moreover, the current status of
observational data is such that they are sufficient to allow a
reasonable degree of precision to be attained in the study of
large-scale structure and microwave background anisotropies. In the
forthcoming years, however, data quality will inevitably improve and a
higher degree of accuracy in the theoretical calculations will
therefore be required. Indeed, high precision microwave anisotropy
experiments are likely to be the first type of observation demanding
just such an improvement in accuracy.
 
In the next Section we shall show how these improvements may be
implemented.  For the purposes of our present discussion, however,
there are only two input parameters that need to be determined before
one can proceed to investigate inflation-inspired models of structure
formation (Liddle and Lyth, 1993b). The key points are that (a) the
density perturbation spectrum has a power-law form and that (b) some
fraction of the large angle microwave anisotropies might be due to
gravitational waves. These conditions represent two completely
independent parameters, but fortunately, they are the only two new
parameters one requires in the lowest-order approximation. This is
true even though one has complete freedom in choosing the functional
form of the underlying inflationary potential. A large number of
papers have now investigated the implications of these inflationary
parameters for structure formation models such as Cold Dark Matter and
Mixed Dark Matter models. Some only consider the possibility of tilt
(Bond, 1992; Liddle, Lyth, and Sutherland, 1992; Cen et al., 1992, Pogosyan
and Starobinsky, 1995) and some also allow for gravitational waves
(Liddle and Lyth, 1993b; Schaefer and Shafi, 1994; Liddle et al., 1996).

One can classify the generic behavior of all inflationary models
consistent with the lowest-order approximation into six separate
categories, as summarized in Table 1. Each sector is characterized by
the direction of the tilt away from scale invariant density
perturbations and by the relative amplitude of the gravitational
waves. In general, spectra with $n>1$ increase the short-scale power
of the density perturbation spectrum. Such spectra were named {\em
blue} spectra by Mollerach et al. (1994). Conversely, those spectra
with $n<1$ subtract short scale power\footnote{We resist calling them
{\em red} since the usual definition of red spectra is $n < 0$, not $n
< 1$.}. It is a general feature of inflation that $n < 1$ is easier to
produce than $n>1$. The reason for this follows from the definition
Eq.~(\ref{scalind1}) for the scalar spectral index. To lowest-order, a
necessary and sufficient condition for the spectrum to be blue is
simply that $\eta > 2\epsilon$. Since $\epsilon$ is positive by
definition, this condition is not easy to satisfy and this is
particularly so during the final stages of inflation where $\epsilon$
must necessarily begin to approach unity. However, specific inflation
models have been constructed for each possibility, with the exception
of a blue spectrum accompanied by a large gravitational wave
amplitude. This last possibility, while still technically possible, is
particularly hard to realize because it requires a large $\epsilon$
overpowered by a yet larger $\eta$.

\subsection{Reconstructing the potential}

In CKLL1 we developed a framework initiated by Hodges and Blumenthal
(1990) that one might call {\em functional reconstruction}. In this
approach one views the observations as determining the spectra
explicitly as functions of scale. Hodges and Blumenthal (1990)
considered only scalar perturbations, and then Grishchuk and Solokhin
(1991) made an investigation, considering only the tensors, with the
aim of determining the time evolution of the Hubble parameter. In
CKLL1, we provided a unified treatment of both scalars and tensors.
The ultimate aim of such a procedure is to then process the functions
through the differential equations describing the evolution of the
universe during inflation. One thereby determines the potential
driving inflation as a function of the scalar field. If such a
procedure could be carried out exactly, the quantities in the
consistency equation would also be functions of scale.

An important point worth emphasising here is that only by including
the tensors can a full reconstruction be achieved. The scalar
perturbations only determine the potential up to an unknown constant.
As the underlying equations are non-linear, different choices of the
constant lead not just to a rescaling of the potential but to an
entirely new functional form. Thus, there are many potentials which
lead to the same scalar spectrum, and hence no unique reconstruction
of the potential from the scalar spectrum. Any piece of knowledge
concerning the tensors is enough to break this degeneracy.

{}From a practical point of view, one finds that the functional
reconstruction procedure is not very useful, although it does allow
some theoretical insight to be gained. The reason is that {\em exact}
formulae for the amplitudes of the spectra do not exist for an {\em
arbitrary} inflaton potential. Consequently, even though the classical
dynamics of the scalar field can be accounted for exactly, one must
input the information on the spectra using results that depend
directly on the slow-roll expansion. At some level, it is inconsistent
to treat the dynamics exactly and the perturbations approximately, so
formally one should truncate both at the same order of approximation.
Indeed, the next-order calculations we provide in the following
Section show that this joint truncation is indeed preferable. In
general, the next-order correction to the magnitude of the potential
arising from the spectra has an opposite sign and is slightly larger
than the correction to the dynamics. In effect, therefore, an exact
treatment of the dynamics actually leads to a less accurate answer
than that obtained by treating the entire problem to lowest-order in
slow-roll!

We therefore advocate an alternative approach that may be referred to
as {\em perturbative reconstruction}. The fundamental idea behind
perturbative reconstruction follows directly from the fact that the
scalar field must roll sufficiently slowly down its potential if
inflation is to proceed at all.  This is important for the following
reason. Typically, the modes that ultimately lead to observational
effects within our universe first crossed the Hubble radius somewhere
between 50 and 60 $e$-foldings before inflation came to an end. (The
precise number of $e$-foldings depends on the final reheating
temperature, but this does not affect the general features of the
argument). During these 10 $e$-foldings of inflationary expansion, the
change in the value of the inflaton field is typically small. In
effect, therefore, the position of the field in the potential would
have remained essentially fixed at some specific value $\phi_0$. It
follows, that cosmological and astrophysical observations can only
yield information regarding this small segment of the potential.
Hence, it is consistent to expand the underlying inflationary
potential as a Taylor series about the point $\phi_0$. The use of such
a procedure to lowest-order was suggested by Turner (1993a), Copeland
et al. (1993a) and CKLL1. Turner (1993b) then included a next-order
term in the potential. The formalism was then developed fully to
next-order in CKLL2, including a next-order term in the derivatives as
well as the potential and outlining the framework for the general
expansion. This framework was recast into a more observationally-based
language by Liddle and Turner (1994) who further discussed the meaning
of the order-by-order expansion.

Perturbative reconstruction can be performed in a controlled way using
the slow-roll expansion order-by-order. The dynamics can be treated to
arbitrary order in this expansion by employing the formalism developed
by Liddle et al. (1994). In contrast, however, the treatment of
perturbations is presently available only to next-order. In this case
there seems no obvious framework by which one can establish an
order-by-order expansion, and even just obtaining terms to one higher
order is a very difficult task.

Modulo questions of convergence, the perturbative reconstruction
procedure successfully encodes functional reconstruction in the sense
that perturbative reconstruction performed to infinite order is
formally equivalent to functional reconstruction. Perturbative
reconstruction can also be rewritten as an expansion in the observed
spectra. The advantage of considering an expansion of this type is
that it indicates exactly how the features in the observed spectra
yield information on the inflationary potential. Such an explicit
account of the observational expansion has not been given before.
  
Before launching into specific calculation, however, it will be
helpful to identify each observable quantity with some order in the
slow-roll expansion.  This may be achieved by considering which
slow-roll parameters occur in the lowest-order term. Thus, one may
employ the lowest-order expressions for the spectra. One sees by
direct differentiation that the information associated with the
accumulation of observables is as follows: $H$ gives $A_T^2$,
$\epsilon$ gives $A_S^2$ and $n_T$, $\eta$ gives $n$ and $dn_T/d \ln
k$, $\xi$ gives $dn/d \ln k$ and $d^2 n_T/d \ln k^2$, and so on. The
key feature is that the tensor spectrum always remains one step above
the scalar one.  Furthermore, we shall see that an additional
derivative of the Hubble parameter for each order is required to
obtain a higher order expression for each observable.

We shall now proceed to derive expressions for the potential and its
first two derivatives correct to lowest-order in the slow-roll
expansion. We consider the Taylor series
\begin{equation}
\label{taylorexpansion}
V(\phi) = V(\phi_0) + V'(\phi_0) \Delta \phi + \frac{1}{2} V''(\phi_0) 
\Delta\phi^2 + \cdots \,,
\end{equation}
about the point $\phi_0$. At this order, the Hamilton-Jacobi equation
(\ref{H}) reduces to $V(\phi) = 3m_{{\rm Pl}}^2 H^2(\phi)/8\pi$, so
the derivatives in this expansion may be expressed directly in terms
of the slow-roll parameters from Eqs.~(\ref{epsilon}) and (\ref{eta}).
It is only consistent to expand the potential to quadratic order,
because the third derivative will contain terms that are of the same
order as terms that were neglected in the original expressions for the
amplitudes. In other words, the lowest-order expressions do not permit
any higher derivatives to be obtained.

It follows by direct substitution, therefore, that
Eq.~(\ref{taylorexpansion}) may be written as
\begin{equation}
\label{expand}
V(\phi) = \frac{3m_{{\rm Pl}}^2 H_0^2}{8\pi} \left[ 1- (16\pi\epsilon_0 
)^{1/2} \frac{\Delta\phi}{m_{{\rm Pl}}} + 4\pi (\epsilon_0 + \eta_0 ) 
\frac{\left( \Delta\phi \right)^2 }{m_{{\rm Pl}}^2} +{\cal{O}} 
\left( \frac{(\Delta\phi)^3}{m_{{\rm Pl}}^3} \right) \right] \,,
\end{equation}
where a subscript $0$ implies that quantities are to be evaluated at
$\phi =\phi_0$. Hence, $H_0$ represents the expansion rate when the
scale corresponding to this value of the scalar field first crossed
the Hubble radius during inflation.

We write the coefficients that arise in this expansion in terms of the
spectra by employing the expressions Eqs.~(\ref{firstscalar}) and
(\ref{firsttensor}) for the amplitudes, the definition
Eq.~(\ref{scalind1}) for the scalar spectral index and the definitions
of the slow-roll parameters. We find that
\begin{eqnarray}
V(\phi_0) & = & \frac{75 m_{{\rm Pl}}^4}{32} A_T^2(k_0) \,, \\
V'(\phi_0) & = & - \frac{75 \sqrt{\pi}}{8} m_{{\rm Pl}}^3 
	\frac{A_T^3(k_0)}{A_S(k_0)} \,, \\
V''(\phi_0) & = & \frac{25\pi}{4} m_{{\rm Pl}}^2 A_T^2(k_0) \left[ 9
	\frac{A_T^2(k_0)}{A_S^2(k_0)} - \frac{3}{2} (1-n_0) \right] \,,
\end{eqnarray}
where $k_0$ is the scale at which the amplitude and spectral indices
are determined and $n_0$ is the scalar spectral index at $k_0$. As
already implied by Eq.~(\ref{scalind1}), if $n$ exceeds one the
potential must be convex ($V''>0$) at the point being probed. However,
$n$ being less than one says nothing definite about convexity or
concavity.

Perturbative reconstruction can be possible even if it ultimately
transpires that the observations necessary to test the consistency
equation non-trivially cannot acquire sufficient accuracy. Similar
work on reconstruction to this level of approximation has been done by
Adams and Freese (1995), Mielke and Schunck (1995) and Mangano, Miele, and
Stornaiolo (1995).

However, it is clear that a determination of the gravitational wave
amplitude on at least one scale is essential for the reconstruction
program to work.  Presently, such a quantity has not been directly
determined, but we may nevertheless draw some interesting conclusions
from the above calculation. In particular, there are a number of
limiting cases to Eq.~(\ref{expand}) that are of interest. Firstly,
when $\epsilon = \eta$, Eq.~(\ref{expand}) is the expansion for the
exponential potential $V \propto \exp (-\sqrt{16\pi \epsilon} \,
\phi/m_{{\rm Pl}})$. (Without loss of generality we may perform a
linear translation on the value of the scalar field such that
$\phi_0=0$).  Secondly, the potential has the form
\begin{equation}
\label{taylor}
V(\phi) =\Lambda \left[ 1+ 2\pi (n-1) \phi^2/m_{{\rm Pl}}^2 \right] \,,
\end{equation}
in the limiting case where $\epsilon \ll 1$. This class of potentials
produces a negligible amount of gravitational waves, but a tilted
scalar perturbation spectrum. The tilt arises because the curvature of
the potential is significant. The direction of the tilt, as determined
by the sign of $(n-1)$, depends on whether the effective mass of the
inflaton field is real or imaginary.

The dynamics of inflation driven by a potential of the form
Eq.~(\ref{taylor}) for $n>1$ has an interesting property. The kinetic
energy of the inflaton field is determined from $H'(\phi)$ via the
second expression in Eq.~(\ref{field}). As the field rolls down the
potential towards $\phi =0$, $H'$ gradually decreases whilst $H$ tends
towards a positive constant. Hence, the field slows down as it
approaches the minimum, but it loses kinetic energy in such a way that
it can never reach the minimum in a finite time. Hence, the de Sitter
universe is a stable attractor for this model and consequently the
inflationary expansion can never end.

There are two ways of circumventing this difficulty. Firstly, one can
argue that the potential only resembles Eq.~(\ref{taylor}) over the
small region corresponding to cosmological scales. This is rather
unsatisfactory, however, since it requires ad-hoc fine-tuning of the
potential and therefore goes against the overall spirit of inflation.
A much more plausible suggestion is that the first term of
Eq.~(\ref{taylor}) arises because a {\em second} scalar field is being
held captive in a false vacuum state. This is the case, for example,
in Linde's Hybrid Inflation scenario (Linde, 1991, 1994; Copeland et
al., 1994b), and an associated instability can end inflation.

We end this section by quoting formulae appropriate to the situation
where one is given the potential and must calculate the predicted
spectra; in general, one cannot analytically find the $H(\phi)$
corresponding to a given $V(\phi)$. In order to obtain the spectra,
one uses the Friedmann equation Eq.~(\ref{H}) and its derivatives in
combination with the slow-roll approximation. To lowest-order, the
spectral indices were first given by Liddle and Lyth (1992), and are
\begin{eqnarray}
n - 1 & = & - 6 \epsilon_{{\rm V}} + 2 \eta_{{\rm V}} \,, \\
n_T & = & - 2 \epsilon_{{\rm V}} \,,
\end{eqnarray}
where
\begin{equation}
\epsilon_{{\rm V}} = \frac{m_{{\rm Pl}}^2}{16 \pi} \, \left( \frac{V'}{V}
	\right)^2 \quad ; \quad \eta_{{\rm V}} = \frac{m_{{\rm Pl}}^2}{8 
	\pi} \, \frac{V''}{V} \,,
\end{equation}
are slow-roll parameters defined from the potential and differ
slightly from the definitions made in terms of the Hubble parameter
used in the rest of this paper (see Liddle et al. (1994) for more
details).  It is also possible to write down next-order expressions
for the spectral indices in terms of the potential (Stewart and Lyth,
1993; Kolb and Vadas, 1994). Expressions such as these 
written in terms of the potential only make sense because of the 
existence of the inflationary attractor.

\section{~~~Next-Order Reconstruction}
\label{second}

\def\theequation{5.\arabic{equation}}
\setcounter{equation}{0}

The level of accuracy discussed in the previous Section, while
perfectly adequate at present, is unlikely to be sufficient once high
resolution microwave background anisotropy experiments are carried
out. The theoretical benchmark for calculating the radiation power
spectrum from a matter power spectrum has been set at one percent in
order to cope with such observations (Hu et al., 1995). If inflation is
to take advantage of this level of accuracy, it is vital that the
initial power spectrum can be considered to at least a similar level
of accuracy. At the very least, this will require the next-order
expressions for the spectra, which represent the highest level of
accuracy presently achieved.

For many potentials, the next-order corrections may be small, perhaps
smaller than the likely observational errors on the lowest-order
terms. We shall see this in the simulated example later in this paper.
In such a case the next-order calculation is still useful, because it
serves as an estimate of the theoretical error bar on the calculation,
which can be contrasted with the observational error.

We devote this Section to describing the next-order results in detail.

\subsection{The consistency equations}

Let's first consider the next-order version of the lowest-order
consistency equation Eq.~(\ref{firstconsistency}). The best available
calculations of the perturbation spectra are those by Stewart and Lyth
(1993) containing the next-order, which we reviewed extensively in
Section \ref{pert}. To this order, the amplitudes for the scalar and
tensor fluctuations are given by
\begin{eqnarray}
\label{secondscalar}
A_S(k) & = & \frac{4}{5 m_{{\rm Pl}}^2} \left[ 1-(2C+1) \epsilon + 
	C\eta \right] \left. \frac{H^2}{|H'|} \right|_{k=aH} \,, \\
\label{secondtensor}
A_T(k) & = & \frac{2}{5\sqrt{\pi}} \left[ 1-(C+1 ) \epsilon \right] 
	\left. \frac{H}{m_{{\rm Pl}}} \right|_{k=aH} \,,
\end{eqnarray}
respectively, where we choose the same normalizations for $A_S$ and
$A_T$ as in Section \ref{first}. We recall that $C \simeq -0.73$ is a
constant. Once again, the right-hand sides of these expressions are to
be evaluated when the scale in question crosses the Hubble radius
during inflation.

Throughout the remainder of this Section we shall be quoting results
that feature a leading term and a correction term, the next-order
term, which is one order higher in the slow-roll parameters. We shall
utilize the symbol ``$\simeq$'' to indicate this level of accuracy.
The correction terms shall be placed in square brackets, so the
lowest-order equations can always be obtained by setting the square
brackets equal to unity, except in Eqs.~(\ref{scalarindex}) and 
(\ref{V''result}) where it
needs to be set to zero.

To next-order, the scalar and tensor spectral indices may be expressed
in terms of the first three slow-roll parameters by differentiating
Eqs.~(\ref{secondscalar}) and (\ref{secondtensor}) with respect to
wavenumber $k$ and employing Eq.~(\ref{kphi}). Some straightforward
algebra yields (Stewart and Lyth, 1993)
\begin{eqnarray}
\label{scalarindex}
1-n & \simeq & 4 \epsilon -2 \eta +\left[ 8(C+1) \epsilon^2 -(6+10 C) 
	\epsilon \eta +2C \xi^2 \right] \,,\\
\label{tensorindex}
n_T & \simeq & -2 \epsilon \left[ 1+(3+2C) \epsilon -2(1+C) \eta \right] \,.
\end{eqnarray}

A very useful relationship may be derived by considering the ratio of
the tensor and scalar amplitudes and replacing the derivative of the
Hubble expansion rate with $\epsilon$. We find that
\begin{equation}
\label{useful}
\epsilon \simeq \frac{A_T^2}{A_S^2} \left[ 1-2C (\epsilon -
\eta ) \right] \,.
\end{equation} 
This relationship is the next-order generalization of
Eq.~(\ref{rat1}). It plays a central role in deriving the next-order
expressions for the potential and its first two derivatives in terms
of observables. Moreover, substitution of this expression into
Eq.~(\ref{tensorindex}) implies that
\begin{equation}
n_T \simeq -2 \frac{A^2_T}{A^2_S} \left[ 1+3 \epsilon -2 \eta \right] \,.
\end{equation}

Now, since all the quantities in the square brackets of this
expression are accompanied by a lowest-order prefactor, they may be
converted into observables by applying the lowest-order expressions
Eqs.~(\ref{scalind1}) and (\ref{rat1}). We conclude, therefore, that
\begin{equation}
\label{secondconsistency}
n_T \simeq -2 \frac{A^2_T}{A^2_S} \left[ 1-\frac{A^2_T}{A^2_S} +(1-n) 
\right] \,.
\end{equation}

This is the next-order version of the lowest-order consistency
equation $n_T = -2A_T^2/A_S^2$, given first in CKLL2 and translated
into more observational language by Liddle and Turner (1994). It is
interesting to remark that the corrections entering at next-order
depend only on the relative amplitudes of the spectra and on $n$. They
do {\em not} depend on $n_T$ or on any of the derivatives of the
indices, because they can be consistently removed using the
lowest-order version of the same equation.  This has an important
consequence that has only been implicit in the literature thus far. We
anticipate that $n$ will be considerably easier to measure than $n_T$.
It is reasonable to suppose, therefore, that if one has enough
observational information to test the lowest-order consistency
equation, one will also have sufficient data to test the next-order
version as well. In other words, the situation where only the
quantities in the lowest-order consistency equation are known is
unlikely to arise.  Consequently, one should employ the next-order
consistency equation when testing the inflationary scenario, rather
than the more familiar version given by Eqs.~(\ref{firstconsistency})
or (\ref{firstCl}).

Another new feature of extending the observables to allow
reconstruction at this order is that one has an entirely new
consistency equation, being the lowest-order version of the derivative
of the original consistency equation.  One calculates $dn_T/d \ln k$
by differentiating Eq.~(\ref{tensorindex}) with respect to scale $k$
and employing Eqs.~(\ref{epsilon}) and (\ref{kphi}). One finds that
\begin{equation}
\frac{dn_T}{d \ln k} \simeq -4 \epsilon (\epsilon -\eta )  \,.
\end{equation}
Conversion of this expression into observables follows immediately by
substituting in the lowest-order results Eqs.~(\ref{scalind1}) and
(\ref{rat1}), giving
\begin{equation}
\frac{d n_T}{d \ln k} \simeq 2 \frac{A_T^2}{A_S^2} \left( 
2 \frac{A^2_T}{A_S^2} +(n-1) \right) \,.
\end{equation}
This equation was derived by Kosowsky and Turner (1995), though they
did not explicitly recognize it as a new consistency equation.
Unfortunately, the observables appearing in the above expression are far
from promising as regards using it.

\subsection{Reconstruction of the potential to next-order}

Now that we have discussed the formalism necessary for calculating the
dynamics and perturbation spectra up to next-order in the slow-roll
expansion, we shall proceed to consider the reconstruction of the
inflationary potential at this improved level of approximation.

We begin by deriving expressions for the potential and its derivatives
directly from the field equation Eq.~(\ref{H}) and the definitions
Eqs.~(\ref{epsilon}) -- (\ref{xi}) for the slow-roll parameters.
Successive differentiation of Eq.~(\ref{H}) with respect to the scalar
field yields the exact relations
\begin{equation}
\label{V}
V= \frac{m_{{\rm Pl}}^2 H^2}{8\pi} (3-\epsilon )  \,,
\end{equation}
\begin{equation}
\label{V'}
V' = -\frac{m_{{\rm Pl}} H^2}{\sqrt{4\pi}} \epsilon^{1/2} 
(3 - \eta ) \,,
\end{equation}
\begin{equation}
\label{V''}
V'' = H^2 \left( 3\epsilon +3\eta - ( \eta^2 +\xi^2 ) \right) \,,
\end{equation}
Our immediate aim is to consider these expressions at a single point
$\phi_0$ and rewrite them in terms of observable quantities. The
amplitude of the potential is derived by substituting
Eqs.~(\ref{secondtensor}) and (\ref{useful}) into Eq.~(\ref{V}):
\begin{eqnarray}
\label{V_0}
V(\phi_0) & \simeq & \frac{75 m_{{\rm Pl}}^4}{32} A_T^2(k_0) \left[ 1 +
	\left( \frac{5}{3} + 2C \right)
	\frac{A_T^2(k_0)}{A_S^2(k_0)} \right] \,, \\
 & \simeq & \frac{75 m_{{\rm Pl}}^4}{32} A_T^2(k_0) \left[ 1 + 
	0.21 \frac{A_T^2(k_0)}{A_S^2(k_0)}\right] \,.
\end{eqnarray}
At this stage, it is interesting to consider how this result would be
altered if one treated the scalar field dynamics in full generality
rather than truncating at next-order. It follows from the general
expression Eq.~(\ref{V}) for the potential that the numerical factor
on the next-order term in the last expression of Eq.~(\ref{V_0}) would
become $-1/3$. What this means is that the next-order correction to
the potential that is due to the spectra dominates the dynamical
corrections.  This is true for all inflationary models. Since the sign
of the spectral correction is opposite to that of the dynamical ones,
the overall sign of the correction is reversed.

Since the potential's first derivative contains $\eta$, we need
information regarding the value of the scalar spectral index at $k_0$
if we are to obtain $V'(\phi)$. We replace the $H^2$ term in
Eq.~(\ref{V'}) by substituting the tensor amplitude
Eq.~(\ref{secondtensor}) and collecting together the terms containing
$\{ \epsilon , \eta \}$ to linear order. These may then be written in
terms of the spectra via the lowest-order expressions
Eqs.~(\ref{scalind1}) and (\ref{firstconsistency}). The result is
\begin{eqnarray}
V'(\phi_0) & \simeq & - \frac{75 \sqrt{\pi}}{8} m_{{\rm Pl}}^3 \, 
	\frac{A_T^3(k_0)}{A_S(k_0)} \left[ 1 + (C+2) \epsilon +
	(C-1/3) \eta \right] \,, \nonumber \\
 & \simeq & - \frac{75 \sqrt{\pi}}{8} m_{{\rm Pl}}^3 \, 
	\frac{A_T^3(k_0)}{A_S(k_0)} \left[ 1 + 1.27 \epsilon
	-1.06 \eta \right] \,, \nonumber \\
 & \simeq & -  \frac{75 \sqrt{\pi}}{8} m_{{\rm Pl}}^3 \, 
	\frac{A_T^3(k_0)}{A_S(k_0)} \left[ 1 -0.85
	\frac{A_T^2(k_0)}{A_S^2(k_0)} +0.53 (1-n_0) \right] \,.
\end{eqnarray}

The calculation for $V''(\phi_0)$ is much more involved. A new
observable is needed to determine $\xi$; the easiest example being the
rate of change of the scalar spectral index. This will be
substantially harder to measure, though, and it is fortunate that it
only enters at next-order. (However, it would enter at leading order
in $V'''(\phi_0)$, as mentioned in CKLL2 and derived fully in Liddle
and Turner (1994)). We can obtain the next-order correction to
$V''(\phi_0)$ directly in terms of the slow-roll parameters by
employing Eqs.~(\ref{secondtensor}) and (\ref{V''}). We find that
\begin{equation}
\label{VPPSR}
V''(\phi_0) \simeq \frac{75 \pi}{4} m_{{\rm Pl}}^2 A_T^2(k_0) \left(
	\epsilon + \eta \right) \left[ 1 + (2C+2) \epsilon - 
	\frac{1}{3} \left( \frac{\eta^2 +\xi^2}{\eta +\epsilon} 
	\right) \right] \,.
\end{equation}

To proceed, we must convert the prefactor $(\epsilon +\eta )$ into
observables, accurate to next-order. To accomplish this we must employ
the next-order result Eq.~(\ref{scalarindex}) for the scalar spectral
index. A straightforward rearrangement of this latter equation yields
\begin{eqnarray}
\epsilon+\eta & \simeq & 3 \epsilon \left[ 1 + \frac{4}{3} (C+1) 
	\epsilon - \left( \frac{3+5C}{3} \right) \eta + \frac{C}{3}
	\frac{\xi^2}{\epsilon} \right] - \frac{1-n_0}{2} \,, \nonumber \\
 & \simeq & 3 \frac{A_T^2}{A_S^2} \left[ 1 + \frac{1}{3} (4-2C) \epsilon + 
 	\frac{1}{3} (C-3) \eta + \frac{C}{3} \frac{\xi^2}{\epsilon} 
 	\right] - \frac{1-n_0}{2} \,,
\end{eqnarray}
where the second expression follows after substitution of
Eq.~(\ref{useful}).  Substituting this into Eq.~(\ref{VPPSR}) yields
\begin{eqnarray}
\label{***}
V''(\phi_0) & \simeq & \frac{225 \pi}{4} m_{{\rm Pl}}^2 
	\frac{A_T^4(k_0)}{A_S^2(k_0)} \left[ 1 + \frac{4C+10}{3} 
	\epsilon + \frac{C-3}{3} \eta + \frac{C}{3} 
	\frac{\xi^2}{\epsilon} \right] \,, \\
 & & - \frac{75 \pi}{8} m_{{\rm Pl}}^2 A_T^2(k_0) \left(1-n_0 \right) \left[
	1 + (2C+2) \epsilon \right] - \frac{25\pi}{4} m_{{\rm Pl}}^2 A_T^2
	(k_0) 	(\eta^2+\xi^2) \,, \nonumber
\end{eqnarray}
where the last term is entirely next-order. Note that there are two
lowest-order terms. An interesting case is $\eta= -\epsilon$,
corresponding to $H \propto \phi^{1/2}$, for which the lowest-order
term vanishes identically and the final term of Eq.~(\ref{VPPSR}) is
the only one to contribute. The second derivative of the potential is
the lowest derivative at which it is possible for the expected
lowest-order term to vanish.

The final step is to convert the next-order terms into the
observables. As they are already next-order, one only needs the
lowest-order term in their expansion to complete the conversion. From
the lowest-order expression for the scalar spectral index, one finds
to lowest-order that
\begin{equation}
\label{xin'}
\frac{\xi^2}{\epsilon} \simeq - \frac{1}{2\epsilon} \left. 
	\frac{dn}{d\ln k} \right|_{k_0} + 5 \eta - 4 \epsilon \,.
\end{equation}
Note that the derivative of the spectral index is of order
$\epsilon^2$.  Finally, substitution of Eqs.~(\ref{scalind1}),
(\ref{firstconsistency}) and (\ref{xin'}) into Eq.~(\ref{***}) yields
\begin{eqnarray}
\label{V''result}
V''(\phi_0) & \simeq & \frac{25\pi}{4} m_{{\rm Pl}}^2 A_T^2(k_0) \left\{ 9 
\frac{A_T^2(k_0)}{A_S^2(k_0)} - \frac{3}{2} (1-n_0) + 
 \left[ (36C+2) \frac{A_T^4(k_0)}{A_S^4(k_0)} \right. \right. \nonumber \\
& - & \left. \left. \frac{1}{4} (1-n_0)^2 - (12C-6) 
\frac{A_T^2(k_0)}{A_S^2(k_0)} (1-n_0) -\frac{1}{2} (3C-1)
\left. \frac{dn}{d \ln k} \right|_{k_0} \right] \right\} \,,
\end{eqnarray}
where the first two terms in the curly brackets represent the
lowest-order contribution.

Before we conclude this section, it is worth remarking on a point that
has perhaps been implicit in the existing literature but has not been
stated explicitly before. A determination of each successive
derivative of the potential requires an extra piece of observational
information. In particular, for the case of lowest-order perturbative
reconstruction, we conclude that the first term in the Taylor
expansion requires only $A_T$, but the second requires both $A_S$ and
$A_T$. The third term, on the other hand, needs both of these together
with $n_0$. The ability to make the observations therefore dictates
how many derivatives we can determine. On the other hand, a comparison
of the lowest-order and next-order expressions for the derivatives
implies the following: the new piece of information necessary for the
derivation of the lowest-order term in $V'$ is also sufficient to
yield the next-order term in $V$. Likewise, the next observation will
give the lowest-order term in $V''$ and this is enough to give the
next-order term in $V'$. Furthermore, it is also sufficient, {\em in
principle}, to give the third-order term in $V$. We stress {\em in
principle} because the theoretical machinery has not been developed to
allow the calculation of a third-order term in the potential or its
derivatives to be performed. Hence, while observational limitations
constrain how high a derivative we can reach, it may be theoretical
rather than observational limitations which prevent higher accuracy in
the lower derivatives. This will be the case even though the necessary
observational information may become available.

Table 2 lists the inflation parameters required for reconstruction of a 
given derivative of the potential. Reconstruction requires the inflation 
parameters in terms of observables. Relations between inflationary 
parameters and observables are given in Tables 3 and 4. A combination of 
information from Table 2 and Table 4 results in Table 5, the observables 
needed to reconstruct a given derivative of the potential to a certain
order.  Although we know the information required for the next-to-next order 
given in Table 5, we don't know the coefficients of the expansion.

\section{~~~The Perturbative Reconstruction Framework}
\label{genfram}

\def\theequation{6.\arabic{equation}}
\setcounter{equation}{0}

Although the next-order results of the previous Section represents the
theoretical state-of-the-art, it is possible to see how the general
pattern goes. We discuss this in this Section and also introduce an
expansion of the observations corresponding to perturbative
reconstruction.

\subsection{A variety of expansions}

During reconstruction, there are three types of expansion being
carried out.  There is an expansion in terms of observables, an
expansion in terms of slow-roll parameters and an expansion of the
potential itself.

Since the underlying theme behind the reconstruction program is that
one is driven by observations, let us first consider what information
might be available. The reconstruction program assumes some
measurements of $A_S(k)$ and $A_T(k)$ are available over some range of
scales. In practice, the likely range of observations for the scalars
will probably be no greater than $-5 < \ln(k/k_0) < 5$, with a much
shorter range for the tensors. In accordance with the perturbative
reconstruction strategy, the spectra should be expanded about some
scale $k_0$ which corresponds to the scale at horizon crossing when
$\phi = \phi_0$. The appropriate expansion is in terms of
$\ln(k/k_0)$, and of course it makes best sense to carry out the
expansion about a wavenumber close to the middle of the available
data.

In general, the expansions can be written as
\begin{eqnarray}
\label{obse}
\ln A_S^2(k) & = & \ln A_S^2 (k_0) + (n(k_0)-1) \ln \frac{k}{k_0} +
	\frac{1}{2} \left. \frac{dn}{d\ln k} \right|_{k_0} \ln^2
	\frac{k}{k_0} + \cdots \,,\\
\ln A_T^2(k) & = & \ln A_T^2 (k_0) + n_T(k_0) \ln \frac{k}{k_0} + 
\frac{1}{2}
	\left. \frac{dn_T}{d\ln k} \right|_{k_0} \ln^2 \frac{k}{k_0} +
	\cdots \,,
\end{eqnarray}
where the coefficients continue as far as the accuracy of observations
permit. There is no obligation for the two series to be the same
length.  Indeed, we anticipate that information associated with the
scalars will be considerably easier to obtain in practice.

The range of $\ln k$ over which data are available leads to the range
of $\phi$ over which the reconstruction converges well. Notice that
since we believe $\ln(k/k_0)$ can be somewhat greater than unity,
convergence of this type of series will only occur if the successive
coefficients become smaller.  Fortunately, we have already seen in
Section \ref{first} that the lowest-order inflationary predictions
attach an extra slow-roll parameter to each higher derivative of the
spectra taken, so convergence can still occur as long as the slow-roll
parameters are smaller than $1/\max{|\ln(k/k_0)|}$. This forms a good
guide as to how wide a range of scales can be addressed via
perturbative reconstruction. The observation that the spectral index
(at least of the scalars) is not too far from unity suggests that the
slow-roll parameters are small. Hence, the observational expansion
might continue to converge well outside the range of $\ln k$ actually
observed. The equivalent statement regarding the potential would be to
say that if it is reconstructed very smoothly for the range $\Delta
\phi$ corresponding to observations, one should feel fairly confident
in continuing the extrapolation of the potential beyond the region
where direct observations were available (though in a practical sense
this does not correspond to any extra information).

The observational expansion discussed above is closely related to the
slow-roll expansion. In particular, we may consider the expansion of
the spectra at a given $k$ in terms of slow-roll parameters, as
discussed in Section \ref{pert}. A qualitative comparison of the two
expansions then yields a general pattern. Each term from the scalars
allows the determination of one extra slow-roll parameter. With regard
to the tensors, a single piece of information (presumably the
amplitude) is necessary before one can proceed at all, as we have
discussed previously. Beyond that, however, extra terms for the
tensors do not provide new slow-roll parameters. Instead, they lead to
degenerate information and hence consistency relationships.  If one
has the first two terms for the tensors and the first scalar term, one
can test the single familiar consistency equation $n_T = - 2
A_T^2/A_S^2$.  Further tensor terms result in a whole hierarchy of
consistency equations, as we shall discuss further in the next
Subsection.

By including terms consisting of products of more and more slow-roll
parameters, one builds up a more accurate answer. However, there are
two separate factors that prevent arbitrary accuracy from being
obtained. The first is observational limitations. For a practical
observational data set with error bars, the observational expansion
discussed above can only be carried out to some term, beyond which the
coefficients are determined as being consistent with zero within the
errors. (If the error bars are still small when this happens, it may
still correspond to useful information). This reflects directly on the
number of slow-roll parameters $\epsilon$, $\eta$, $\xi$, etc, that
one can measure. In general, however, there are an infinite number of
slow-roll parameters, and formally they are all of the same order
(meaning that for a `generic' potential, one expects them all to be of
similar size). This appears to be rather problematic, since a finite
number of terms in the observational expansion cannot constrain an
infinite number of slow-roll parameters. Fortunately, however, only a
finite (and usually small) number of such terms ever appear when a
specific expression is considered.

The second restriction is that current technical knowledge concerning
the generation of the spectra, as reviewed in Section \ref{pert}, only
allows the calculation of a lowest-order term plus a correction
involving single slow-roll parameters. In general, one anticipates
further corrections including products of two or more slow-roll
parameters, but that has not been achieved. It follows, therefore,
that the number of derivatives in the potential that may be calculated
is determined by observational restrictions, whilst the accuracy of
each derivative is also constrained by theoretical considerations.

It should be emphasized that once an expression written as an
expansion in slow-roll parameters has been found, it can be
differentiated an arbitrary number of times. It is interesting that
the derivatives are accurate to the same number of orders in the
slow-roll parameters. This follows because differentiation respects
the order-by-order expansion. However, differentiation introduces
higher and higher slow-roll parameters from the infinite hierarchy. An
important point here is that the `lowest-order' can be a product of
any number of slow-roll parameters; the phrase is not synonymous with
setting the slow-roll parameters all to zero.

Having started with the observations, we now come round to the crux of
the reconstruction process: the inflaton potential. In perturbative
reconstruction, one aims to calculate the potential and as many of its
derivatives as possible {\em at a single point} to some level of
accuracy in slow-roll parameters. The ultimate goal is to use this
information to reconstruct some portion of the potential about this
point, by carrying out some expansion of $V(\phi)$ about the point
$\phi_0$. The simplest strategy is to use a Taylor series
\begin{equation}
V(\phi) = V(\phi_0) + V'(\phi_0) \Delta \phi + \frac{1}{2} V''(\phi_0) 
\Delta\phi^2 + \cdots \,,
\end{equation}
and we shall only consider that case here. The literature does include
more ambitious strategies such as Pad\'{e} approximants and these may
become useful when specific data are available (Liddle and Turner,
1994). The success of this expansion is governed by how far away from
$\phi_0$ one hopes to go, which ultimately arises from the range of
observations one has available, as well as on how accurately the
individual derivatives are determined.

This expression shows us that perturbative reconstruction of the
potential actually involves {\em two} expansions. We have already seen
that the potential is obtained up to some accuracy in the slow-roll
expansion.  However, for reconstruction to be successful, it is also
imperative to consider how accurate the expansion in $\Delta \phi$
might be. Determining the coefficients of only the first one or two
terms may be completely useless if $\Delta \phi$ turns out to be
large.

The key to investigating this is to rewrite $\Delta \phi$ in terms of 
$\Delta 
\ln k$, the range of scales over which observations can realistically be 
expected to cover\footnote{Turner (1993b) and Liddle and Turner (1994)
carried out a similar analysis using $\Delta N$, the number of
$e$-foldings.  This is perfectly valid but somewhat harder to
interpret in terms of observable scales since it is only formally
equivalent in a lowest-order approximation. In this work, however, we
desire a simple interpretation of the next-order results.}. Broadly
speaking this corresponds to the interval from $1$ Mpc to about $10^4$
Mpc, so assuming a center point in the middle of this region implies a
range for $\Delta \ln k$ between $\pm 5$. This may be biased through
tensor data only being available on large scales, though it will also
be of considerably lower quality than the scalar data. The
relationship that allows one to achieve the comparison between $\Delta
\phi$ and $\Delta \ln k$ is the exact formula Eq.~(\ref{kphi})
presented earlier
\begin{equation}
\frac{d\phi}{d\ln k} = \frac{m_{{\rm Pl}}^2}{4 \pi} \frac{H'}{H} \,
	\frac{1}{\epsilon -1} = \frac{m_{{\rm Pl}}}{\sqrt{4\pi}} \,
	\frac{\sqrt{\epsilon}}{\epsilon -1} \,,
\end{equation}
together with its derivatives. One can then expand $\Delta \phi$ in
terms of $\Delta \ln k$, expanding each coefficient up to some order
in the slow-roll expansion. Such an expansion begins
\begin{eqnarray}
\label{phitok}
\Delta \phi & = & - \frac{m_{{\rm Pl}}}{\sqrt{4\pi}} \sqrt{\epsilon} 
	\left[1+\epsilon + \cdots \right] \Delta \ln k \\ \nonumber
	&& + \frac{m_{{\rm Pl}}}{\sqrt{16\pi}} \sqrt{\epsilon} 
	\left[\epsilon - \eta + \cdots \right] (\Delta \ln k)^2 + \cdots \,,
\end{eqnarray}
where, for illustrative purposes, the first coefficient has been given
to next-order in slow-roll and the second one to lowest-order. The
signs are chosen in accordance with our convention that $V'<0$.

For clarity we shall employ $\beta$ to represent a generic slow-roll
parameter. One can then schematically represent the double expansion
(one in $\Delta \phi$ and one in the slow-roll parameters), as
\begin{eqnarray}
\label{cansee}
\frac{V(\phi)}{A_T^2(k_0)} & \sim & \left[ 1 + \beta + \cdots \right] \,,
	\nonumber \\
& + & \beta \Delta \ln k \left[ 1+ \beta + \cdots \right] \left\{ 1 + \beta 
	+ \beta \Delta \ln k + \cdots \right\} \,, \nonumber \\
& + & \beta^2 (\Delta \ln k)^2 \left[ 1 + \beta + \cdots \right] \left\{ 
	1 + \beta + \beta \Delta \ln k + \cdots \right\} \,,
\end{eqnarray}
where numerical constants have not been displayed. The square brackets
represent the expansion of the potential and its derivatives at
$\phi_0$, while the curly brackets represent the $\Delta \phi$, which
itself is written as an expansion in $\Delta \ln k$ with coefficients
expanded in slow-roll.

For the slow-roll expansion to make sense, we need $\beta \ll 1$. One
can see from the schematic layout of Eq.~(\ref{cansee}) that
convergence of the expansion will fail unless $\beta \Delta \ln k \ll
1$, as successively higher-order terms will otherwise become more and
more important. However, we have agreed that $\Delta \ln k$ itself
need not be small. In regions where it is, it is clear that the best
results are obtained by calculating the low derivatives of the
potential as accurately as possible. In regions where $\Delta \ln k$
is not small, however, it is more fruitful to calculate higher
derivatives.

\subsection{The consistency equation hierarchy}

In the previous Subsection, we stated that there exists an infinite
hierarchy of consistency equations. It is not difficult to see why
such a hierarchy should exist. Even though exact expressions for the
spectra as a function of scale are not presently available, one can
imagine having such expressions, at least in principle. In this case,
one could then write down a consistency equation in the full
functional reconstruction framework that applied over all available
scales. This equation could then be represented in the perturbative
reconstruction framework by performing a Taylor (or similar) expansion
on both sides of it. The perturbative consistency equations could then
be derived by equating the coefficients of the expansions. The key
idea here is that the full functional consistency equation and all its
derivatives must be satisfied at the point about which perturbative
reconstruction is being attempted. The equality of each derivative at
this point, however, represents a separate piece of information.

In Section \ref{first} we presented the consistency equation
Eq.~(\ref{firstconsistency}) for lowest-order perturbative
reconstruction.  The connection between the tensor--scalar ratio and
the tensor spectral index was first presented by Liddle and Lyth (1992)
and has been much discussed in the literature. This consistency
equation is simply the (unknown) full functional consistency equation
applied at a single point, and moreover, it is the version of that
equation truncated to lowest-order in slow-roll. Indeed, it does not
require a determination of $n$ and it corresponds to the lowest,
non-trivial truncation of the expansion of the observed spectra.

The next order in slow-roll introduces $n$ and $dn_T/d \ln k$. This
not only supplies enough information to impose a next-order version of
the original consistency equation, but is also enough to impose a
lowest-order version of the {\em derivative} of the consistency
equation. The next-order versions of the original consistency equation
were supplied by CKLL2 and Liddle and Turner (1994) and we discussed
these in Section \ref{second}. We also discussed the lowest-order
version of the derivative of the consistency equation in that Section.
This equation was first given by Kosowsky and Turner (1995).

This pattern continues at all orders in the expansion. One can ask why
this has not been emphasised before. One reason is that until now a
clear understanding has not been established regarding the type of
{\em observational} information that appears at each order in the
expansion. At the same stage that one introduces $n$ in the slow-roll
expansion, one should also introduce the rate of change of the tensor
spectral index. The latter does not provide any new information
regarding the reconstruction, in the same way that $n_T$ did not
provide new information at lowest-order in slow-roll. However, it is
subject to the new consistency equation.  Researchers have not paid
attention to the new consistency equation because it requires
$dn_T/d\ln k$ and it seems very unlikely that this could ever be
measured.

This concludes our discussion of the theoretical framework for
perturbative reconstruction. In the following Section, therefore, we
shall discuss whether the observations are likely to reach an adequate
level of sophistication in the foreseeable future and then consider a
worked example that illustrates how the reconstruction programme might
be applied in practice.

\section{~~~Worked Examples of Reconstruction}
\label{obs}

\def\theequation{7.\arabic{equation}}
\setcounter{equation}{0}

\subsection{Prospects for reconstruction}

In this Subsection, we shall consider the long-term prospects for
reconstructing the inflaton potential. It is clear that one must
determine the amplitudes of the {\em primordial} power spectra of
scalar and tensor fluctuations on at least one scale, together with
the slope of the scalar spectrum at that scale. Such information would
provide enough information to reconstruct the potential and its first
two derivatives to lowest--order.  However, a measurement of $n_T$ is
also required if one is to test the inflationary hypothesis via the
consistency equation. If such information becomes available at all, it
will probably be {\em after} $A_S$, $A_T$ and $n$ have themselves been
determined, so reconstructing to lowest-order should prove easier to
accomplish than testing the scenario via the consistency equation.

It is convenient to separate the full cosmological parameter space
into two sectors. The first contains the inflationary parameters
essential for reconstructing the potential and testing the consistency
equations. They are
\begin{equation}
\label{inflationparameter}
(A_S, r, n, n_T, \cdots) \,,
\end{equation}
where all are evaluated at $k_0$, and the list extends to as many
derivatives of the spectra as one wishes to consider. The
tensor-scalar ratio $r \equiv 12.4A_T^2 /A_S^2$ is defined so that
$r=1$ corresponds to an equal contribution to large angle microwave
anisotropies from the scalar and tensor fluctuations, as follows from
Eqs.~(\ref{firstconsistency}) and (\ref{firstCl})).

The second set consists of the other cosmological parameters:
\begin{equation}
\label{cosmologicalparameter}
(\Omega_0, \Omega_{\Lambda}, \Omega_{{\rm CDM}}, \Omega_{{\rm HDM}} ,
\Omega_{\rm B} h^2, h , z_R , \ldots ) \,,
\end{equation}
where the $\Omega$ represent the densities in matter of various sorts,
respectively the total matter density, cosmological constant, cold
dark matter, hot dark matter and baryonic matter. Here $z_R$
represents the redshift of recombination; it may be that this single
parameter is adequate or the full ionization history may have to be
taken into account. In the standard cold dark matter (CDM) model these
parameters take the values $(A_S(k_0),0,1,0)$ and $(1, 0, 0.95, 0,
0.0125, 0.5)$ respectively (further parameters concerning derivatives
of the spectral indices in the first set being zero); that is, the
scalar amplitude is the only free parameter available to fit to
observations. The standard ionization history of the universe is also
assumed.

Experiments measuring microwave background anisotropies offer the most
promising route towards acquiring such information to within the
desired level of accuracy. Although redshift surveys provide valuable
insight into the nature of the scalar spectrum at the present epoch,
uncertainties in the mass--to--light ratio of galaxy distributions
imply that it is very difficult to determine the primordial spectrum
from these observations alone. There are further complications
associated with uncertainties in the type of non-baryonic dark matter
in the universe. These can lead to significant modifications in the
form of the transfer function. One crucial advantage that microwave
background experiments have, however, is that the level of anisotropy
above 10 arcmin is almost independent of whether the dark matter is
hot or cold (Seljak and Bertschinger, 1994; Stompor, 1994; Ma and
Bertschinger, 1995; Dodelson, Gates, and Stebbins, 1996). Moreover, as we
shall see in Section \ref{otherways}, a direct detection of the
stochastic background of gravitational waves by laser interferometers
seems highly improbable. Thus, the microwave background anisotropies
appear to be the only practical route at present towards determining
the gravitational wave amplitude.

It is conventional to expand the temperature distribution on the sky
in terms of spherical harmonics
\begin{equation}
\frac{\Delta T}{T_0} =\sum_{l=2}^{\infty} 
	\sum_{m=-l}^l a_{lm}(r)Y_{lm} ({\bf x}) \,,
\end{equation}
where the monopole and dipole terms have been subtracted out and $T_0
= 2.726$K is the present mean background temperature. The $l$-th
multipole corresponds loosely to an angular scale of $\pi/l$, and a
comoving length scale of $100 h^{-1}$ Mpc at the last scattering
surface subtends an angle of about one degree (for $\Omega_0 = 1$).

Inflation predicts that the $a_{lm}$ are gaussian random variables,
with a rotational invariant expectation value for their variance $C_l
\equiv \langle |a_{lm}|^2 \rangle$. The radiation power spectrum is
defined to be $l(l+1) C_l$; this is exactly constant in the case of a
scale--invariant density perturbation spectrum $(n=1, r=0)$ when the
Sachs--Wolfe effect is the sole source of anisotropy (Sachs and Wolfe,
1967; Bond and Efstathiou, 1987).  In general, both tensor and scalar
perturbations contribute to the observed radiation power spectrum, and
for inflation these contributions are independent, so
$C_l=C^S_l+C^T_l$.

Accurate calculations of the $C_l$ from both scalar and tensor modes
require numerical solutions using a Boltzmann code (Bond and Efstathiou,
1987), and this can now be done to an extremely high accuracy, of around one 
percent or so (Hu et al., 1995). A recent innovation is a new algorithm 
based on an integral solution of the Boltzmann equation (Seljak and 
Zaldarriaga, 1996a), which obtains this level of accuracy at much less 
computational expense. In principle high quality observations can approach 
this accuracy though the question of foreground remains a delicate one (Hu 
et al., 1995; Tegmark and Efstathiou, 1996) and so the true observational
accuracy will be less. These types of numerical study seem essential
for high accuracy work, although they are complemented by analytical
approaches, which can be made both for scalars (Hu and Sugiyama, 1995)
and for tensors. The latter case is the easier for two reasons;
firstly, only gravitational effects need to be considered and
secondly, gravitational waves redshift away once they are inside the
Hubble radius, so their main influence is only on the lower
multipoles, up to $l \simeq 100$.  Analytic studies, of increasing
sophistication, have been made by Abbott and Wise (1984a, 1984b),
Starobinsky (1985), Turner, White, and Lidsey (1993), Atrio-Barandela and
Silk (1994), Allen and Koranda (1994), Koranda and Allen (1994) and Wang
(1996). These results show good agreement with the numerical
calculations of Crittenden et al. (1993a) and Dodelson, Knox,
and Kolb (1994), who evolve the photon distribution function by applying 
first-order perturbation theory to the general relativistic Boltzmann 
equation for radiative transfer. 

With this calculational power in place, there are two main obstacles
to determining the primordial spectra. These are known as `cosmic
variance' and `cosmic confusion', respectively.

\vspace{.1in}

{\em Cosmic Variance}: A given inflationary model predicts the
quantities $C_l=\langle |a_{lm}|^2 \rangle$, but the observed
multipoles measured from a single point in space are $a_l^2=
\sum_{m=-l}^{+l} |a_{lm}|^2/4\pi$. These only represent a single
realization of the $C_l$. It is well known that a finite sampling of
events generated from a random process leads to an intrinsic
uncertainty in the variance even if the experiment is perfectly
accurate; this is sometimes called sample variance.  In the limit of
full sky coverage this uncertainty is known as cosmic variance.

More precisely, the $a_l^2$ are a sum of $2l+1$ Gaussian random
variables and therefore have a probability distribution that is a
$\chi^2$ distribution with $2l+1$ degrees of freedom. Thus, for each
multipole there are $2l+1$ samples, so the uncertainty in the $C_l$ is
given by
\begin{equation}
\label{cosvar}
\frac{\Delta C_l}{C_l} = \sqrt{\frac{2}{2l+1}} \,. 
\end{equation}
This implies that cosmic variance is proportional to $l^{-1/2}$ and is
therefore less significant on smaller angular scales. However, for any
given experiment, the beam width limits how high an $l$ can be
obtained before experimental noise intervenes, and anyway in standard
cosmological models the predicted signal cuts off rapidly beyond $l
\sim 1000$ due to the finite thickness of the last scattering surface. 
Thus, the information on the tensor components is limited because there is 
very little signal in near--scale invariant models for $l \ge 200$ where the 
effects of cosmic variance are less significant. 

\vspace{.1in}

{\em Cosmic Confusion}: The anisotropy below $l \le 60$ is essentially
determined by the inflationary parameters in
Eq.~(\ref{inflationparameter}), and by $\Omega_0$ and
$\Omega_{\Lambda}$, since it is dominated by the purely gravitational
terms rather than the details of the matter content of the universe.
On the other hand, the anisotropies are highly model dependent for
$l>60$ due to the complexity of the operating physical processes. In
particular, the precise level of anisotropy in this range depends
sensitively on the values of the cosmological parameters listed in
Eq.~(\ref{cosmologicalparameter}). Bond et al. (1994) have suggested
that different sets of values for these parameters sometimes lead to
power spectra which are extremely similar (for a review see Steinhardt,
1994). This leads to degeneracies in determined parameters, which Bond
et al. refer to as `cosmic confusion'. Cosmic confusion is problematic
for the reconstruction program and the degeneracy must be lifted
before it can proceed. Fortunately, things have moved on since the
Bond et al. discussion, and it is now acknowledged that observations
can be carried out at such a high accuracy that the degeneracy is
lifted (Hu et al., 1995, Jungman et al., 1996). Tegmark and Efstathiou 
(1996) have found that the microwave background anisotropies can be 
determined to very high precision even in the presence of multi-component 
foreground noise by the COBRAS/SAMBA satellite.

It should also be noted that other methods are available for
determining cosmological parameters. For example, the primordial light
element abundances imply that $0.009 \le \Omega_{\rm B}h^2 \le 0.022$
and these limits may become stronger as observations of deuterium in
quasar absorption lines improve (Olive et al., 1990; Copi, Schramm, and
Turner, 1995). Furthermore, an accurate measurement of $h$, certainly
to within 10\% , seems achievable with the Hubble Space Telescope
(Freedman et al., 1994), whilst polarization of the microwave
background may provide insight into the ionization history of the
universe (Crittenden, Davis, and Steinhardt, 1993b; Frewin, Polnarev, and
Coles, 1994; Crittenden, Coulson, and Turok, 1994; Kosowsky, 1996). There 
has also recently been improved understanding of the possibility of using 
polarization to probe gravitational waves (Kamionkowski, Kosowsky, and 
Stebbins, 1996; Seljak and Zaldarriaga, 1996b; Zaldarriaga and Seljak, 
1996). Because gravitational waves typically contribute more (relative to 
density perturbations) to the polarization than to the total anisotropy, and 
indeed because one can identify a combination of the polarization 
parameters which cannot be induced by density perturbations at all, it may 
ultimately be possible to use polarization to do better than the 
cosmic-variance limited studies of the temperature alone which we discuss 
below.

\vspace{.1in}

In view of this, it is important to consider to what degree the next
generation of satellites will be able to determine the inflationary
parameters in Eq.~(\ref{inflationparameter}). Knox and Turner (1994)
have considered what might be deduced from two experiments $A$ and $B$
whose window functions are centered around $l_A \approx 55$ and $l_B
\approx 200$, respectively.  Experiment $B$ only measures anisotropy
due to the scalar fluctuations, whereas $A$ will be sensitive to both
scalar and tensor fluctuations. They considered `standard'
cosmological parameters $h=0.5$, $\Omega_{\rm B} \approx 0.05$,
$\Omega_{\Lambda} =0$ and a scale--invariant spectrum. They concluded
that if the tensor-scalar ratio $r \ge 0.14$, one should be able to
rule out $r=0$ with 95\% confidence 95\% of the time. Thus, the
gravitational wave amplitude should be quantitatively measurable for
$r \ge 0.14$. If $n$ is reduced, the limit is improved slightly to $r
\ge 0.1$. Knox and Turner (1994) further conclude that full--sky
measurements on angular scales $0.5^o$ and $3^o$ should acquire the
sensitivity required for making such a detection.

For reconstruction to proceed at lowest--order, however, one also
requires $C_l^S$ for some $l$ and also the spectral index $n$. Knox
(1995) has simulated a set of microwave background experiments within
the context of chaotic inflation driven by a $\phi^4$ potential. This
model predicts $n=0.94$, $n_T=-0.04$ and $r=0.28$. He considers a
third measurement made on a smaller angular scale than those of $A$
and $B$. It is this measurement that determines $C_l^S$ and this may
be combined with the measurement at the intermediate scale $l_B$ to
determine the slope $n$. Finally, $r$ is inferred by identifying the
`excess power' arising in measurement $A$ with the gravitational
waves. He concludes that the quantity $C_2^S \,130^{1-n}$ could be
measured to an accuracy of $\pm 0.3$\% and the error in the slope of
the scalar spectrum could be as small as $\pm 0.02$. If $n \approx 1$,
the error on $r$ is $\pm 0.1$ and improves slightly for smaller $n$. A
full--sky experiment designed with current technology and with a $20'$
beam should be able to achieve such precision.

However, these results are derived on the assumption that the
cosmological parameters have been accurately determined by other
means. Indeed, to achieve the above precision on $r$ and $n$, one
requires the errors in $\Omega_{\rm B}h^2$ to be no more than $10$ \%
and $6 \% $, respectively (Knox, 1995).  Furthermore, the Hubble
parameter will have to be determined to within $6\% $ or $14$\%
respectively if $\Omega_{\Lambda} =0.8$ and the uncertainty in
$\Omega_{\Lambda}$ must be below $7\%$.

More recently, Jungman et al. (1996) have carried out an analysis
where all inflationary and cosmological parameters are allowed to
vary. They confirm the expectation that the estimates provided by Knox
(1995) are very optimistic. If all the other cosmological parameters
are left completely free, it is impossible to get any useful
information on the gravitational waves at all --- the required value of $r$ 
is somewhat larger than mentioned above, and $n_{{\rm T}}$ would have to be 
extremely large. However, that represents a somewhat pessimistic assessment, 
because certainly many of the cosmological parameters will be constrained by 
other types of observations, and more importantly one may also feel content 
to live within a subset of cosmological parameter space (for example, 
critical density universes with only cold dark matter).

The accuracy to which the above parameters can be observationally
determined will decide whether the information is good enough to push
any of the expressions beyond lowest-order. Another possibility is
that a more sophisticated observable may become available; Kosowsky and
Turner (1995) have considered the possibility that $dn/d \ln k$ might
be observable in the microwave background. For most models this seems
unlikely as the effect will be small, but there do exist inflationary
models leading to an effect that is large enough to be observable.
Whether this parameter generates any degeneracies with other
inflationary or cosmological parameters in the shape of the $C_l$
remains to be addressed.

\subsection{Toy model reconstructions with simulated data}

We devote this subsection to carrying out a worked example of
reconstruction on a faked data set, to indicate the kind of accuracy
that might be possible. We have tried to make the outcome of analyzing
the simulated data at least reasonably indicative of the sort that
high resolution microwave background experiments might achieve, based
on the analysis by Knox (1995) [see also Jungman et al., 1996].
However, our approach is strictly a toy model; it is not intended to
bear any resemblance to what one might actually do with high accuracy
observations. It seems very unlikely that observations such as CMB
anisotropies might be used to directly estimate the $k$-space spectra
(though such an approach is common with galaxy redshift surveys); the
expectation is that if suitable quality data are obtained then the
appropriate procedure will be to push the theory forward from the
spectra rather than try to calculate the primordial spectra directly
from the observations. That is, some analysis such as a likelihood
analysis would be used to find best fitting parameters such as the
amplitude and spectral indices of the scalars and tensors directly.
Knox (1995) has taken some first steps in this direction.

Perturbative reconstruction requires an expansion of the observations
about a single scale, which will end up corresponding to the location
$\phi_0$ on the potential about which it is to be reconstructed. As
discussed earlier, an expansion of the logarithm of the spectra in
terms of the logarithm of the wavenumber is the best way to proceed.
It will always make the most sense to choose the scale $k_0$ about
which the expansion is done to be near the `central' point of the
logarithmic $k$-interval\footnote{The word `central' is in quotes to
indicate that the effective center point of the data may be biased
through tensors only being available on large scales, plus
scale-dependent error bars on both scalars and tensors. The word is
intended to refer to the point best determined by the data assuming
the type of fit attempted.}. Thus we write
\begin{eqnarray}
\ln A_S^2(k) & = & \ln A_S^2 (k_0) + (n(k_0)-1) \ln \frac{k}{k_0} +
	\frac{1}{2} \left. \frac{dn}{d\ln k} \right|_{k_0} \ln^2
	\frac{k}{k_0} + \cdots \,,\\
\ln A_T^2(k) & = & \ln A_T^2 (k_0) + n_T(k_0) \ln \frac{k}{k_0} +
	\frac{1}{2} \left. \frac{dn_T}{d\ln k} \right|_{k_0} \ln^2
	\frac{k}{k_0} + \cdots \,,
\end{eqnarray}
where we have written in explicitly the observational quantities to
which the coefficients of the expansion correspond.

A given observational program produces some finite set of data with
error bars, such as a list of galaxy redshifts and sky positions, or a
pixel map of the microwave sky. As we said above, it is unlikely to be
a useful strategy to try and obtain the power spectra from these, and
then use these to reconstruct. Rather, one should push the theory
towards the data by parametrizing the spectra and fitting for those
parameters, as has been done so successfully with COBE. Other
parameters which affect the data interpretation, such as the
cosmological parameters, can be fixed or simultaneously fitted as
required. The general reconstruction framework we have described
indicates an efficient parametrization of the spectra that could be
used. 

Despite the above, for our illustrative examples we have chosen
to simulate data for the spectra themselves, as it is the simplest
thing to do. Enough is known (Knox, 1995; Jungman et al., 1996) about
the capabilities of CMB satellites in particular to enable a fairly
realistic example (in terms of the observational uncertainties) to be
constructed. To do anything else would obscure the principal issues. Our aim 
therefore is to simulate a set of data, with errors, for the spectra, which 
when fitted give similar errors on parameters to those expected had we 
carried out the full task of simulating say a microwave sky and fitting 
directly for the spectral parameters. It is well outside the scope of this 
paper to attempt a realistic simulation of what future data might actually 
look like.

As a simple test, we have simulated fake data sets for two different
models, as follows:
\begin{enumerate}
\item A power-law inflation model with power-law index $p = 21$, chosen
to yield $n-1 = n_T = -0.1$. Since power-law inflation can be solved
exactly we know the precise amplitude of the spectra corresponding to
a given normalization of the spectra, Eqs.~(\ref{scalaramp}) and
(\ref{tensoramp}). This particular model has been advocated by White
et al. (1995) as providing a good fit to the current observational
data. 
\item An intermediate inflation model (Barrow and Liddle, 1993), which
gives a scale-invariant spectrum of density perturbations but still
possesses significant gravitational waves. We choose a version where
scalars and tensors contribute equally to COBE (to be precise, their
contributions to the tenth multipole are chosen to be the same). In
this case, a precise calculation of the spectra cannot be made, so we
compromise by using the next-order approximation to generate the
spectra from the underlying model.
\end{enumerate}
These models both have quite substantial gravitational waves. They
have been chosen to be compatible with present observational data,
though they can be regarded as rather extreme cases which maximize the
chance of an accurate reconstruction.

The simulated data are constructed by the following procedure.
\begin{itemize}
\item The overall normalization reproduces the COBE result.
\item The scalar error bars are consistent with cosmic variance limited 
microwave anisotropy observations up to $l = 200$ (except that for
simplicity we have modeled the errors by a gaussian rather than the
formally correct $\chi^2_{2l+1}$ distribution). Other cosmological
parameters, which affect the microwave anisotropy spectrum, are
assumed fixed. The COBRAS/SAMBA satellite can go to much higher $l$,
but of course the other cosmological parameters will be uncertain
which limits the estimation of the inflationary parameters. By
stopping at $l = 200$, we find that the accuracy we obtain is similar
to that suggested by Jungman et al. (1996) for the full problem, so it
serves as a reasonable compromise.
\item For the tensors, reasonable {\it a priori} estimates for the
error bars are harder to establish. We have assumed data corresponding
to $l$ up to 40, which is where the tensor contribution to $C_l$
begins to cut off, and we have chosen error bars so as to reproduce
the observational uncertainty in the tensor amplitude suggested by
Knox (1995). We then accept whatever uncertainty in the tensor
spectral index this gives us, and it happens to be in reasonable
agreement with that suggested by Knox.
\end{itemize}

The simulated data for Model 1 are shown in Figure 2, along with the
best fit reconstructions. Since scalar data runs from $l = 2$ to
$200$, it covers two orders of magnitude in wavenumber, corresponding
to $\Delta \ln k \simeq 4.6$. The input and output parameters are
shown in Table 4. We performed two fits, the first being a power-law
fit and the second also allowing for a variation in the scalar
spectral index (though in fact the underlying spectrum has none). The
Figures and subsequent discussion use the former.

The results for Models 1 and 2 contain no particular surprises. 
Although this is intended only to be indicative and certainly falls way 
short of the sophistication that can be brought into play on realistic data, 
the error bars are probably fairly reasonable. As expected, the tensor
spectral index is the real stumbling block, but at least with these
models one obtains a strong handle on $A_T^2$, thus allowing a unique
reconstruction. For these reconstructions, we find that the lowest-order
consistency equation Eq.~(\ref{firstconsistency}) is indeed satisfied
\begin{equation}
0.108 \pm 0.013 = 2 \frac{A_T^2}{A_S^2} = - n_T = 0.25 \pm 0.10 \,,
\end{equation}
for Model 1 and 
\begin{equation}
0.14 \pm 0.02= 2 \frac{A_T^2}{A_S^2} = - n_T = 0.12 \pm 0.11 \,,
\end{equation}
for Model 2. The same is true for 
the next-order version Eq.~(\ref{secondconsistency}). For Model 1 we obtain
\begin{equation}
0.114 \pm 0.014 = 2 \frac{A_T^2}{A_S^2}\left[1-\frac{A_T^2}{A_S^2} + (1-n) 
	\right] = - n_T = 0.25 \pm 0.10 \,,
\end{equation}
whereas for Model 2 we find
\begin{equation}
0.13 \pm 0.02 = 2 \frac{A_T^2}{A_S^2}\left[1-\frac{A_T^2}{A_S^2} + (1-n) 
	\right] = - n_T = 0.12 \pm 0.11 \,.
\end{equation}

While encouraging, we see that the test is not particularly strong due
to the poorly determined $n_T$. In models where the tensors are even
weaker than considered here, the task of testing the consistency
equation will be yet harder.

Proceeding on to the reconstruction, Table 5 shows lowest-order and
next-order reconstructions, in comparison to the exact underlying
potential for both Models. The consistency equation has been used to 
eliminate $n_T$ as it is the most poorly determined quantity. A next-order 
version of $V''(\phi_0)$ cannot be obtained without a value for $dn/d\ln
k|_{k_0}$, though the size of the correction could be bounded from the
error bars on the null result. The reconstructed potentials, both
lowest-order and next-order, for Model 1 are shown in Figure 3 in
comparison to the underlying potential. A Taylor series has been used
to generate them, and the range of $\phi$ shown corresponds to the
range of observational data (a range of two orders of magnitude in
$k$) determined using Eq.~(\ref{phitok}).

We see that in both models the lowest-order reconstruction has been very 
successful. The errors are dominated by those in measuring the tensor 
amplitude. However, in neither case does the next-order result offer a 
significant improvement, given the observational error bars. The main 
importance of the next-order result appears therefore to be in bounding the 
theoretical error, rather than in providing improved accuracy in the overall 
reconstruction.

Figure 3 can be compared to a similar figure in Liddle and Turner
(1994), who investigated reconstruction of a similar exponential
potential. However, they did not include any observational errors,
concentrating instead on the theoretical errors and on the efficacy of
different expansion techniques for the potential. They also assumed
reconstruction over a wider range of scales, and had somewhat poorer
convergence of the reconstructed potential through expanding about one
end of the data (the quadrupole) rather than the center.

\section{~~~Other Ways to Constrain the Potential}
\label{otherways}

\def\theequation{8.\arabic{equation}}
\setcounter{equation}{0}

Up until now we have concentrated, at least implicitly, on
observations connected to large-scale structure in the universe,
including microwave background anisotropies. These certainly provide
the best source of constraints on the inflationary potential, and one
should be very pleased at the prospect of obtaining such constraints.
However, they do cover only a small portion of the full inflationary
potential. There is of course no way of uncovering information about
the potential relevant to larger scales (beyond waiting the relevant
number of Hubble times!), but in principle there are a variety of ways
of constraining the potential appropriate to smaller scales. We shall
discuss such possibilities in this Section. In particular, one may
constrain the potential from the fact that inflation must come to an
end some 50 $e$-foldings after the large-scale structure scales pass
outside the Hubble radius. Further constraints are associated with the
scalar and tensor perturbations on small scales. In principle, laser
interferometers could observe the tensor spectrum as a stochastic
background, though we shall see that this is not promising. The
possible overproduction of primordial black holes (PBHs) immediately
after inflation places upper limits on the amplitude of the last
scalar fluctuation to cross the Hubble radius just before inflation
ends, while distortions to the microwave background spectrum limit
scalar fluctuations on mass scales well below large-scale structure
scales.

\subsection{To the end of inflation and the area law}

In traditional inflation models, inflation can come to an end in one
of two ways. The first is via some drastic event, such as a quantum
tunneling (for example in extended inflation) or a sudden instability
(probably connected to a second field, as in hybrid inflation). If
this happened, probably little information can be drawn from the
behavior approaching the end of inflation.  The second way inflation
may come to an end is simply by the potential becoming too
(logarithmically) steep to sustain inflation any longer, as in generic
chaotic inflation models, so that $\epsilon$ reaches unity.

Let us see what one can conclude in the latter case. For definiteness,
let us assume that 50 $e$-foldings are supposed to occur after the
scale $k_0$, about which reconstruction is attempted, leaves the
horizon. The modest dependence of this number on the details of
reheating will not be important.  By assumption, inflation will end
precisely when $\epsilon = 1$. The number of $e$-foldings which occur
between two scalar field values is given exactly by
\begin{equation}
N = \sqrt{\frac{4\pi}{m^2_{{\rm Pl}}}} \int_{\phi_1}^{\phi_2} 
\frac{1}{\sqrt{\epsilon(\phi)}} \, d\phi \,.
\end{equation}
For our purposes, this can be neatly written as an integral constraint 
(Liddle, 1994a)
\begin{equation}
\int_{\phi_0}^{\phi_{{\rm end}}} \frac{1}{\sqrt{\epsilon(\phi)}} \, 
	\frac{d\phi}{m_{{\rm Pl}}} = \frac{50}{\sqrt{4\pi}} \,.
\end{equation}

This can most easily be thought of graphically. We have reconstructed
the value of $\epsilon$ and its derivative at $\phi_0$, and know
$\epsilon(\phi_{{\rm end}}) = 1$. As shown in Figure 4, if we plot the
curve of $1/\sqrt{\epsilon}$ against $\phi/m_{{\rm Pl}}$, it must be
such that it reaches unity just as the area under it reaches
$50/\sqrt{4\pi}$. While there remain many ways in which the curve may
do this, it does exclude some possibilities such as a sudden
flattening of the potential after observable scales leave the
horizon.\footnote{It appears that this can be used to derive an upper
limit, albeit a weak one, on $(\phi_{{\rm end}} - \phi_0)$, from the
knowledge that $\epsilon \leq 1$. In fact this is not the case, since
$H$ starts to exhibit strong variation when $\epsilon$ approaches one.
The number of $e$-foldings should then strictly be characterized by
the increase in $aH$ rather than $a$ alone (see Liddle et al. (1994)
for details).  In principle, a yet weaker constraint may be derived by
using energy scale arguments to limit how much $H$ can decrease in the
late stages of inflation, but such a constraint seems too weak to be
worth pursuing.}

\subsection{Local detection of primordial gravitational waves}

A number of authors have examined the possibility that the stochastic
background of primordial gravitational waves produced during inflation
could be detected locally (Allen, 1988; Grishchuk, 1989; Sahni, 1990;
Souradeep and Sahni, 1992; White, 1992; Turner et al., 1993; Liddle, 1994b;
Bar-Kana, 1994). In general, the wavenumber of the gravitational waves
is related to the value of the inflaton field during inflation via the
relation $\ln (k /k_0 ) = 60 - N$, where $N$ is the number of
$e$-foldings before the end of inflation and $k_0 = a_0 H_0 \approx 3
\times 10^{-18} h$ Hz is the wavenumber of the mode that is just
reentering the Hubble radius at the present epoch. Thus, the modes
with wavenumbers associated with the maximum sensitivity of typical
beam-in-space experiments $(\sim 10^{-3} {\rm Hz})$ first crossed the
Hubble radius approximately 25 $e$-foldings before the end of
inflation. A direct detection of such waves would therefore provide
unique insight into a region of the inflationary potential that cannot
be probed by large-scale structure observations. However, we shall see
that this is unlikely to be possible.

There are a number of gravitational wave detectors currently under
construction or proposal (see e.~g.~Thorne, 1987, 1995). The
ground-based Laser Interferometer Gravitational Wave Observatory
(LIGO) should have a peak sensitivity of $\Omega_g \approx 10^{-11}
h^{-2}$ at 10 Hz (Christensen, 1992), where $\Omega_g$ is the energy
density per logarithmic frequency interval. The proposed space-based
interferometers, the Laser Gravitational Wave Observatory in Space
(Faller et al., 1985; Stebbins et al., 1989) and the Laser
Interferometer Space Antenna (Danzmann, 1995) probe lower
frequencies, but with a sensitivity to flat spectrum stochastic
sources which is less than that of LIGO.

After inflation, the evolution of the gravitational wave perturbation
is determined by Eq.~(\ref{fieldeqn}). We have already studied the
effect of modes which have wavelengths greater than the Hubble radius
by the time of last scattering, which contribute to microwave
background anisotropies.  However, the scales which can be detected
locally will have re-entered the Hubble radius before the onset of
matter domination. In this regime they behave as radiation, so their
energy density stays fixed during the radiation era but falls during
the matter era. This suppression factor is directly measured by the
radiation density today, $\Omega_{\rm rad} = 4 \times 10^{-5} h^{-2}$.
Thus the predicted amplitude on scales re-entering before
matter-radiation equality is (Allen, 1988; Sahni, 1990; Liddle, 1994b)
\begin{equation}
\Omega_g h^2 = \frac{2}{3\pi} \left( \frac{H}{m_{{\rm Pl}}} \right)^2 
	\times 4 \times 10^{-5} \,.
\end{equation}

For the inflation models we have been discussing, $H$ always decreases
with time, and hence the primordial amplitude on short scales is
always less than that on large scales\footnote{`Superinflation' models
have been considered within the context of superstring motivated
cosmologies, and it appears that in that case the gravitational wave
amplitude could rise sufficiently on short scales to be detectable
(Brustein et al., 1995). However, no complete model, demonstrating how
superinflation might successfully end, has been constructed thus far
(Brustein and Veneziano, 1994; Levin, 1995a).}. The quadrupole anisotropy
already places an extremely stringent limit on the amplitude of the
spectrum at large scales, and this immediately translates into a
conservative, but robust, constraint across all short scales of
(Liddle, 1994b)
\begin{equation}
\Omega_g h^2 \leq 4 \times 10^{-15} \,.
\end{equation}
This puts the inflationary signal well out of reach of any of the
proposed experiments.

\subsection{Primordial black holes}

It has been conjectured that primordial black holes (PBHs) may form
during the reheating phase immediately after inflation (Khlopov,
Malomed, and Zel'dovich, 1985; Carr and Lidsey, 1993; Carr, Gilbert, and
Lidsey, 1994; Randall, Solja\u{c}i\'{c}, and Guth, 1996; Garc\'{\i}a-Bellido 
et 
al., 1996).  While there are considerable theoretical uncertainties
attached to this possibility, if such formation does occur, it can
constrain the scalar spectrum at very short scales. During inflation
the first scales to leave the Hubble radius are the last to come back
in and this implies that the very last fluctuation to leave will be
the first to return. In some regions of the post-inflationary
universe, the fluctuation will be so large that one expects that the
collapse of a local region into a black hole will become inevitable.
The higher the rms amplitude the larger the fraction of the universe
forming PBHs. The observational consequences of the evaporation of
these black holes then leads to upper limits on the number that may
form and hence on the magnitude of the spectrum on the relevant
scales. Thus, one may constrain the amplitude of the density spectrum
on scales many orders of magnitude smaller than those probed by
large-scale structure observations and microwave background
experiments. These constraints lead to an upper limit on the spectral
index and may therefore provide insight into features of the
inflationary potential towards the end of inflation.

We parametrize the density spectrum in terms of the mass scale $M$
associated with the Hubble radius when a given mode reenters. Hence,
$\delta(M) \propto M^{(1-n)/6}$ defines the scalar spectral index.
PBHs are never produced in sufficient numbers to be interesting if
$n<1$, but they could be if the spectrum is `blue' with $n>1$.

When an overdense region with equation of state $p=\gamma \rho$ stops
expanding, it must have a size greater than $\sqrt{\gamma}$ times the
horizon size in order to collapse against the pressure. The
probability of a region of mass $M$ forming a PBH is (Carr, 1975)
\begin{equation}
\label{beta0}
\beta (M) \approx \delta (M) \exp \left( -\frac{\gamma^2}{2\delta^2 (M)}
\right) \,.
\end{equation}

The constraints on $\beta (M)$ in the range $10^{10}{\rm g} \le M\le
10^{17}{\rm g}$ have been summarized by Carr and Lidsey (1993).  In
particular, PBHs with an initial mass $\sim 10^{15}$g will be
evaporating at the present epoch and may therefore contribute
appreciably to the observed gamma-ray and cosmic-ray spectra at 100
MeV (MacGibbon and Carr, 1991).  On the other hand, $10^{10} $g PBHs
have a lifetime $\sim$ 1 sec and, if produced in sufficient numbers,
would lead to the photodissociation of deuterium immediately after the
nucleosynthesis era (Lindley, 1980).  PBHs of mass slightly below
$10^{10}$g could alter the photon--to--baryon ratio just prior to
nucleosynthesis. An upper limit therefore arises by requiring that
evaporating PBHs do not generate a photon--to--baryon ratio exceeding
the current value $S_0 = 10^9$ (Zel'dovich and Starobinsky, 1976).

Carr et al. (1994) have considered the constraints on $\beta (M)$
below $10^{10}$g. In this region the strongest constraint arises if
evaporating PBHs leave behind stable Planck mass relics (MacGibbon,
1987; Barrow, Copeland, and Liddle, 1992). The observational constraint
from the relics derives from the fact that they cannot have more than
the critical density at the present epoch, $\Omega_{\rm rel} <1$.

The upshot of this analysis is that the spectral index is typically
constrained to be less than about 1.5, depending weakly on assumptions
as to the reheat temperature after inflation and whether one takes
into account the black hole relic constraint. Because the constraint
applies at the end of inflation, on scales greatly separated from the
microwave anisotropies, it is independent of the COBE normalization
and also of the choice of dark matter.  However, in this form it
relies on the spectral index being constant right across those scales
(which it would be in the hybrid inflation model (Copeland et al.,
1994b)).  For general inflation models it should be reinterpreted as a
specific constraint on the amplitude at the short scales being
sampled.

Finally, a constraint on the amplitude of the spectrum at a scale 
corresponding to an horizon mass $\approx 0.1M_{\odot}$ can in principle 
be derived from the recent observations of massive compact halo objects 
(MACHOs) (Alcock et al., 1993; Aubourg et al., 1993). The estimated mass 
range 
of these objects suggests that they constitute about 0.1 per cent of the 
critical density. Although the favored explanation for these microlensing 
events is that they are due to substellar baryonic brown dwarfs, it 
is quite possible that MACHOs may be primordial black holes and therefore 
non--baryonic in nature (Nasel'skii and Polnarev; Ivanov, Nasel'skii,
and Novikov, 1994; Yokoyama, 1995). Such PBHs could form from vacuum 
fluctuations in the manner discussed above if the amplitude of 
spectrum is sufficiently high on the appropriate scale. This may be 
possible, for example, if the potential has a suitable form (Ivanov et al., 
1994). Alternatively, a spike may be imposed on the underlying spectrum by 
the quantum fluctuations of a second scalar field (Yokoyama, 1995; Randall 
et al., 1996; Garc\'{\i}a-Bellido et al., 1996). If the amplitude is too 
high on this particular scale, however, it would lead to the overproduction 
of MACHO-PBHs. Consistency with the observations therefore constrains both 
the spectrum and the inflationary potential.

\subsection{Spectral distortions}

A further constraint on $\delta (M)$ over mass scales considerably
smaller than those corresponding to large-scale structure may be
derived by considering departures of the microwave spectrum away from
a pure blackbody. (For detailed reviews see e.~g.~Danese and de Zotti
(1977) and Sunyaev and Zel'dovich (1980)). Above a redshift of $z_y
\approx 2.2\times 10^4 \left( \Omega_{\rm B} h^2 \right)^{-1/2}$,
Compton scattering is able to establish local thermodynamic
equilibrium whenever there is a sudden redistribution or release of
energy into the universe (Burigana, Danese, and de Zotti, 1991). This
produces a Bose--Einstein spectrum $n \propto \left[ \exp (x + \mu )
-1 \right]^{-1}$ that is characterized by a chemical potential $\mu$,
where $x = h \nu /k T$. (A Planck spectrum corresponds to $\mu =0$).
On the other hand, equilibrium cannot be established for redshifts
just below $z_y$. The distribution of energy at this time could
therefore lead to observable spectral distortions $(\mu \ne 0)$ in the
microwave background at the present epoch. The Far Infrared Absolute
Spectrophotometer (FIRAS) aboard COBE has constrained the spectral
distortion to be $|\mu | < 3.3 \times 10^{-4}$ (Mather et al., 1994),
whilst Hu, Scott, and Silk (1994) have strengthened this limit by
considering the COBE measurement of temperature fluctuations on
$10^{\circ}$ (Bennett et al., 1994). They find that $\mu < 5.0
\times 10^5 (\Delta T/T)^2_{10^{\circ}} \approx 6.3 \times 10^{-5}$. 

These limits imply that photon diffusion would have been the dominant
mechanism for producing spectral distortions (Daly, 1991). Silk (1967)
first showed that the damping of adiabatic fluctuations can proceed if
their mass scales are below a characteristic mass known as the Silk
mass. At sufficiently early times, the photons and baryons in the
universe are strongly coupled through Thomson scattering and they
therefore behave as a single viscous fluid. When adiabatic
fluctuations reenter the Hubble radius, they set up pressure gradients
and these result in pressure waves that oscillate as sound waves. As
the epoch of recombination approaches, however, the mean--free--path
of the photons increases and the photons are able to diffuse out of
the overdense regions into underdense regions. Thus, the
inhomogeneities in the photon--baryon fluid are damped. The energy
stored in the fluctuations is redistributed by the diffusion of
photons and it is this transfer of energy during the epoch near to
$z_y$ that produces the spectral distortions. The fluctuations that
lead to these potentially observable effects have mass scales in the
range $10^{-3} < M/ {\rm M}_{\odot} < 10^3$ (Sunyaev and Zel'dovich,
1970; Barrow and Coles, 1991).

The observational upper limit on $\mu$ implies an upper limit on the
amplitude of the pressure wave and therefore a limit on $\delta (M)$.
The energy density in a linear sound wave is $\rho u^2$, where $u
\approx c/\sqrt{3}$ is the sound speed. Thus, the dimensionless energy
release caused by the damping is $q \approx \delta^2/3$. It can be
shown that the spectral distortion is given by $\mu \approx 1.4 q$ and
it follows, therefore, that $\delta < 1.46 \sqrt{\mu} \approx 0.01$.

By normalizing the spectrum at COBE scales $(\sim 10^{22}M_{\odot})$,
an upper limit on the spectral index may be derived. Barrow and Coles
(1991) and Daly (1991) assume that the distortion is entirely due to
the largest amplitude wave and deduce a limit of $n< 1.8$ for $M \sim
10^{-3} {\rm M}_{\odot}$. (The limit becomes weaker for larger
scales). Hu et al.  (1994) have derived a stronger constraint of
$n<1.5$ by refining these calculations. This is comparable to the PBH
constraints we have just discussed (though somewhat weaker if one
believes the PBH relic constraint).  However, it is probably more
reliable because it is based on physics that is relatively well
understood and requires a less severe extrapolation to smaller scales.

\section{~~~Conclusions}
\label{conc}

\def\theequation{9.\arabic{equation}}
\setcounter{equation}{0}

In this paper, we have reviewed the relationship between observations
of microwave aniso\-tropies and of large-scale structure and the
possibility of connecting them to the potential energy of a scalar
field driving inflation.  We have argued that, given suitable quality
observations, the inflationary idea can be tested and then features of
the inflationary potential can be directly measured. In many ways this
is remarkable, given that it is impossible, by many orders of
magnitude, for an Earth-based accelerator to pursue this task.

It is predicted that inflation produces both gravitational waves and
density perturbations. Consequently, the employment of observations
may be divided into two main parts. The most challenging is the test
of the inflationary consistency relations; if these prove testable and
are confirmed, it will provide a powerful vindication of the 
chaotic inflation paradigm. One could then feel confident in following the 
less observationally challenging task of employing observations to
discern information regarding the inflationary potential, in the form
of its value and that of its first few derivatives at a single point.

We have indicated the different approximation schemes that must be
invoked.  Of paramount importance is the slow-roll expansion, but this
must also be coupled to an expansion of the observables. In the
simplest instance this latter expansion corresponds to the
approximation of power-law spectra. The lowest levels of approximation
are certainly able to cope with present-day observations of both
microwave anisotropies and large-scale structure.  However, in this
work we have been forward looking, since the demands that will be
imposed on theoretical accuracy by future observations, especially
satellite-based microwave background anisotropy measurements, will be
high.  Indeed, they could in principle threaten the limits of
present-day theoretical knowledge regarding the calculation of the
spectra.

We must emphasize that our calculations have all been implemented
within the standard paradigm for chaotic inflation. The vast majority
of known viable models can be expressed within this class, either
trivially or by cunning manipulation, but one should bear in mind that
there exist some models of inflation for which this is not the case.
In some examples, such as old versions of the open inflationary
scenario or some multi-field theories, this is because the predictions
turn out to be dependent on initial conditions.  Although such a
situation would be unfortunate it is not logically excluded.  Other
theories, such as the recently investigated single-bubble open
inflationary models, rely on dynamics that are much more complicated
than that of the standard scenario (Sasaki et al., 1993; Bucher et al.,
1994; Linde, 1995). They therefore lead to a more complicated
relationship between theory and observations. Furthermore, even if the
inflationary hypothesis is indeed correct, it may be the case that the
actual model produces a very low amplitude of gravitational waves (Lyth 
1996). This would make them impossible to measure and such a situation would
remove the ability to make a consistency check and thus eliminate most
of the potential for reconstruction. 

Finally, there remains every possibility that the entire inflationary idea 
is incorrect; if so, one can at least hope that this is manifested in a 
failure of the consistency relations. However, it may not prove 
possible to test the consistency relations; might one then blunder into 
reconstructing a non-existent object? With sufficiently good observations, 
such as a CMB satellite will provide, the answer should be no. The $C_l$ 
spectrum, when it is observed, will contain huge amounts of degenerate 
information. If the correct underlying theory is topological defects, (see 
for example Vilenkin and Shellard, 1994), the spectral shape should be very 
different to any simple inflation model for any values of the cosmological 
parameters. One can certainly reconstruct a `potential' which would give the 
observed $C_l$, but it would probably be of such a complex form as to have 
little particle physics motivation for it, leaving people to search for 
other explanations. 

In a standard inflation scenario, the $C_l$ give a complete description of 
the gaussian perturbations generated. This prediction can also be tested 
against the observations; present observations are compatible with 
gaussianity though they are not strong enough to give a convincing test. In 
the future we can expect such tests to be widely applied. While in principle 
it is possible to construct inflation models giving non-gaussian 
perturbations, in practice such models are so contrived that again, were 
such features detected, one would quickly be looking for a more plausible 
theory for the origin of perturbations. It might well also be that the shape 
of the power spectrum might be incompatible with the non-gaussian nature, 
within the general context of inflation.

The bulk of this review has covered work already discussed in the
literature. We have given an extensive account of the Stewart and Lyth
(1993) calculation of the perturbation spectra, which provides the
accuracy needed to discuss anticipated observations. The
reconstruction framework has then been described to an accuracy which
ought to be sufficient for years to come.  However, as well as the
review material, we have brought to light a few new results and
viewpoints and we summarize these here.
\begin{itemize}
\item The consistency equation discussed in the present literature is just 
one of an infinite hierarchy of consistency equations, each of which
can be taken (in principle) to arbitrary accuracy in the slow-roll
expansion.  Kosowsky and Turner (1995) have written down the form for
the second member and we have reproduced it here. However, it is
probable that only the first consistency equation will ever be tested.
\item We have indicated that since scalar perturbations are much easier to 
measure than tensor ones, the appropriate form of the first
consistency equation to consider is not the lowest-order version, but
rather the next-order version. One requires $n_T$ to test the
lowest-order version and it is very unlikely that such observations
would be available without there also being the appropriate ones to
include the next-order version as well. (The only new ingredient in
the next-order version over and above those quantities in the
lowest-order version is $n$).
\item We have been more explicit than previous work as to how observations 
of the primordial spectra should be handled in terms of an expansion in
$\ln k$. We discussed how this expansion relates to the slow-roll
expansion. A worked example on simulated data has illustrated these
ideas in action.
\end{itemize}

In conclusion, therefore, the relationship between inflationary
cosmology and large-scale structure observations is well understood
and the theoretical machinery necessary for taking advantage of high
accuracy observations is now in place. These promise the possibility
of constraining physics at energies inaccessible to any other form of
experiment. Such observations are eagerly awaited.

\section*{Acknowledgments}

JEL is supported by the Particle Physics and Astronomy Research
Council (UK). ARL is supported by the Royal Society. EWK and MA are
supported at Fermilab by the DOE and NASA under Grant NAG 5--2788. TB
is supported by JNICT (Portugal). We are grateful for many helpful
discussions with John Barrow, Robert Caldwell, Bernard Carr, Scott
Dodelson, John Gilbert, Martin Hendry, Lloyd Knox, David Lyth, Douglas
Scott, Paul Steinhardt, Reza Tavakol, Michael Turner and Martin White.

\frenchspacing
\section*{References}

\begin{description}

\item Abbott, L. F. and M. B. Wise, 1984a, Nucl. Phys. B {\bf 244}, 541.

\item Abbott, L. F. and M. B. Wise, 1984b, Phys. Lett. B {\bf 135}, 279.

\item Adams, F. C., J. R. Bond, K. Freese, J. A. Frieman and 
	A. Olinto, 1993, Phys. Rev. D {\bf 47}, 426.

\item Adams, F. C. and K. Freese, 1995, Phys. Rev. D {\bf 51}, 6722. 

\item Albrecht A. and P. J. Steinhardt, 1982, Phys. Rev. Lett. { \bf 48},
	1220.

\item Alcock, C. et al., 1993, Nat. {\bf 365}, 621.

\item Allen, B., 1988, Phys. Rev. D {\bf 37}, 2078.

\item Allen, B. and S. Koranda, 1994, Phys. Rev. D {\bf 50}, 3713.

\item Allen, T. J., B. Grinstein, and M. B. Wise, 1987, Phys. Lett. B
	{\bf 197}, 66.

\item Arnowitt, R., S. Deser and C. W. Misner, 1962, in Gravitation: 
	An Introduction to Current Research, ed L. Witten 
	(John Wiley, New York).

\item Atrio--Barandela, F. and J. Silk, 1994, Phys. Rev. D {\bf 49}, 1126.

\item Aubourg, E. et al., 1993, Nat. {\bf 365}, 623. 

\item Bar--Kana, R., 1994, Phys. Rev. D {\bf 50}, 1157.

\item Bardeen, J. M., 1980, Phys. Rev. D {\bf 22}, 1882.

\item Bardeen, J. M., P. J. Steinhardt and M. S. Turner, 1983, 
	Phys. Rev. D {\bf 28}, 679.

\item Barrow, J. D. and R. A. Matzner, 1977, Mon. Not. R. astr. Soc. 
	{\bf 181}, 719.

\item Barrow, J. D. and S. Cotsakis, 1988, Phys. Lett. B {\bf 214}, 515.

\item Barrow, J. D. and P. Coles, 1991, Mon. Not. R. astr. Soc. 
	{\bf 248}, 52.

\item Barrow, J. D., E. J. Copeland and A. R. Liddle, 1992, Phys. Rev. 
	D {\bf 46}, 645.

\item Barrow, J. D. and A. R. Liddle, 1993, Phys. Rev. D {\bf 47}, R5219 

\item Bennett, C. L. et al., 1994, Astrophys. J. {\bf 436}, 423.

\item Bennett, C. L. et al., 1996, Astrophys. J. {\bf 464}, L1.

\item Bergmann, P. G., 1968, Int. J. Theor. Phys. {\bf 1}, 25.

\item Birrell, N. and P. C. W. Davies, 1982, {\em Quantum Fields in 
	Curved Space} (Cambridge Univ. Press, Cambridge).

\item Bond, J. R., 1992, in Highlights in Astronomy Vol 9, Proc of the 
	IAU Joint Discussion, ed J. Bergeron.

\item Bond, J. R. and G. Efstathiou, 1987, Mon. Not. R. astr. Soc. 
	{\bf 226}, 665.

\item Bond, J. R., R. Crittenden, R. L. Davies, G. Efstathiou and 
	P. J. Steinhardt, 1994, Phys. Rev. Lett. {\bf 72}, 13.

\item Brustein, R. and G. Veneziano, 1994, Phys. Lett. B {\bf 329}, 429.

\item Brustein, R., M. Gasperini, M. Giovannini and G. Veneziano, 1995,
	Phys. Lett. B {\bf 361}, 45.

\item Bucher M., A. S. Goldhaber and N. Turok, 1995, Phys. Rev.
	D {\bf 52}, 3314.

\item Bucher M., A. S. Goldhaber and N. Turok, 1995, Nucl. Phys. S {\bf 43}
	173.

\item Burigana, C., L. Danese and G. de Zotti, 1991, Astr. Astrophys. 
	{\bf 246} 49.

\item Carr, B. J., 1975, Astrophys. J. {\bf 205}, 1.

\item Carr, B. J. and J. E. Lidsey, 1993, Phys. Rev. D {\bf 48}, 543.

\item Carr, B. J., J. H. Gilbert and J. E. Lidsey, 1994, Phys. Rev. 
	D {\bf 50}, 4853.

\item Cen, R., N. Y. Gnedin, L. A. Kofman and J. P. Ostriker, 1992,
	Astrophys. J. Lett. {\bf 399}, L11.

\item Christensen, N., 1992, Phys. Rev. D {\bf 46}, 5250.

\item Copeland, E. J., E. W. Kolb, A. R. Liddle and J. E. Lidsey, 1993a, 
	Phys. Rev. Lett. {\bf 71}, 219.

\item Copeland, E. J., E. W. Kolb, A. R. Liddle and J. E. Lidsey, 1993b, 
	Phys. Rev. D {\bf 48}, 2529 [CKLL1].

\item Copeland, E. J., E. W. Kolb, A. R. Liddle and J. E. Lidsey, 1994a, 
	Phys. Rev. D {\bf 49}, 1840 [CKLL2].

\item Copeland, E. J., A. R. Liddle, D. H. Lyth, E. D. Stewart and 
	D. Wands, 1994b, Phys. Rev. D {\bf 49}, 6410.

\item Copi, C. J., D. N. Schramm and M. S. Turner, 1995, Science {\bf 267}, 
	192.

\item Crittenden, R., J. R. Bond, R. L. Davis, G. Efstathiou and P. J.
	Steinhardt, 1993a, Phys. Rev. Lett. {\bf 71}, 324.

\item Crittenden, R., R. L. Davis and P. J. Steinhardt, 1993b, 
	Astrophys. J. Lett. {\bf 417}, L13.

\item Crittenden, R., D. Coulson and N. Turok, 1994, Phys. Rev. Lett. 
	{\bf 73}, 2390.

\item Daly, R. A., 1991, Astrophys. J. {\bf 371}, 14.

\item Danese, L. and G. de Zotti, 1977, Rivista del Nuovo Cimento 
	{\bf 7}, 277.

\item Danzmann, K., 1995, Ann. N. Y. Acad. Sci. {\bf 759}, 481.
 
\item Davis, R. L., H. M. Hodges, G. F. Smoot, P. J. Steinhardt and M. S.
	Turner, 1992, Phys. Rev. Lett. {\bf 69}, 1856.

\item Dodelson, S., L. Knox and E. W. Kolb, 1994, Phys. Rev. Lett. 
	{\bf 72}, 3444.

\item Dodelson, S., E. Gates and A. Stebbins, 1996, Astrophys. J. {\bf 
	467}, 10.

\item Dolgov, A. and J. Silk, 1993, Phys. Rev. D {\bf 47}, 2619.

\item Durrer, R. and M. Sakellariadou, 1994, Phys. Rev. D {\bf 50}, 6115.

\item Easther, R., 1996, Class. Quant. Grav. {\bf 13}, 1775.
	
\item Easther, R., K. Maeda and D. Wands, 1996, Phys. Rev. D {\bf 53},
	4247. 

\item Efstathiou, G. P., 1990, in Physics of the Early Universe, eds 
	A. T. Davies, A. Heavens and J. Peacock, SUSSP publications
	(Edinburgh).

\item Faller, J. E., P. L. Bender, J. L. Hall, D. Hils and M. A. Vincent.,
	1985, in Proceedings of the Colloquium Kilometric Optical 
	Arrays in Space (European Space Agency, Noordwijk).

\item Freedman, W. L. et al., 1994, Nat. {\bf 371}, 757.

\item Frewin, R. A., A. G. Polnarev and P. Coles, 1994, Mon. Not. R. 
	astr. Soc. {\bf 266}, L21.

\item Garc\'{\i}-Bellido, J. and D. Wands, 1995, Phys. Rev. D {\bf 52}, 
	6739.

\item Garc\'{\i}-Bellido, J. and D. Wands, 1996, Phys. Rev. D {\bf 53},
	5437.

\item Garc\'{\i}-Bellido, J., Linde A. D. and D. Wands, 1996,
	``Density Perturbations and Black Hole Formation in Hybrid 
        Inflation'', Phys. Rev. D, to appear, astro-ph/9605094.

\item Gasperini, M. and G. Veneziano, 1993a, Astropart. Phys. {\bf 1}, 
	317. 

\item Gasperini, M. and G. Veneziano, 1993b, Mod. Phys. Lett. A {\bf 8},
	3701. 

\item Gasperini, M. and G. Veneziano, 1994, Phys. Rev. D {\bf 50}, 2519. 

\item Gasperini, M., M. Giovannini  and G. Veneziano, 1995, Phys. Rev.
	D {\bf 52}, R6651.

\item Gott, J. R., 1982, Nature {\bf 295}, 304.

\item Gott, J. R. and T. S. Statler, 1984, Phys. Lett. B {\bf 136}, 157.

\item Grib, A. A., S. Mamaev and V. Mostepanenko, 1980, {\em Quantum 
	Effects in Strong External Fields} (Atomizdat, Moscow).

\item Grishchuk, L. P., 1974, Zh. Eksp. Teor. Fiz. {\bf 67}, 825 (Sov. 
	Phys. JETP {\bf 40}, 409).

\item Grishchuk, L. P., 1977, Ann. N. Y. Acad. Sci. {\bf 302}, 439.

\item Grishchuk, L. P., 1989, Sov. Phys. Usp. {\bf 31}, 940.

\item Grishchuk, L. P. and Yu. V. Sidorav, 1988, in Fourth Seminar on 
	Quantum Gravity, eds M. A. Markov, V. A. Berezin and V. P. 
	Frolov (World Scientific, Singapore).

\item Grishchuk, L. P. and M. Solokhin, 1991, Phys. Rev. D {\bf 43}, 2566.

\item Guth, A. H., 1981, Phys. Rev. D {\bf 23}, 347.

\item Guth, A. H. and S. -Y. Pi, 1982, Phys. Rev. Lett. {\bf 49}, 1110.

\item Harrison, E. R., 1970, Phys. Rev. D {\bf 1}, 2726.

\item Hawking, S. W. and G. F. R. Ellis, 1973, {\em The Large 
	Scale Structure of Space--time} (Cambridge University Press,
	Cambridge). 

\item Hawking, S. W., 1982, Phys. Lett. B {\bf 115}, 295.

\item Hawking, S. W. and I. G. Moss, 1982, Phys. Lett. B {\bf 110}, 35.

\item Higgs, P. W., 1959, Nuovo Cimento {\bf 11}, 816.

\item Hodges, H. M. and G. R. Blumenthal, 1990, Phys. Rev. D {\bf 42}, 
	3329.
 
\item Hu, W., D. Scott and J. Silk, 1994, Astrophys. J. Lett. {\bf 430}, L5.

\item Hu, W. and N. Sugiyama, 1995, Phys. Rev. D {\bf 51}, 2599.

\item Hu, W., D. Scott, N. Sugiyama and M. White, 1995, Phys. Rev.
	D {\bf 52}, 5498.

\item Ivanov, P., P. Nasel'skii and I. Novikov, 1994, Phys. Rev. 
	D {\bf 50}, 7173. 

\item Jungman, G., M. Kamionkowski, A. Kosowsky and D. N. Spergel,
	1996, Phys. Rev. D {\bf 54}, 1332.  

\item Kalara, S., N. Kaloper and K. A. Olive, 1990, Nucl. Phys. B {\bf 
	341}, 252.

\item Kaloper, N., R. Madden and K. A. Olive, 1995, Nucl. Phys. B
	{\bf 452}, 677. 

\item Kaloper, N., R. Madden and K. A. Olive, 1996, Phys. Lett. B
	{\bf 371}, 34.
	
\item Kamionkowski M., Kosowsky A., Stebbins, A., 1996, ``A probe of
	primordial gravitational waves and vorticity'', Columbia
	preprint, astro-ph/9609132.

\item Khlopov, M. Yu., B. A. Malomed and Ya. B. Zel'dovich, 1985, 
	Mon. Not. R. astr. Soc. {\bf 215}, 575.

\item Knox, L. and M. S. Turner, 1994, Phys. Rev. Lett. {\bf 73}, 3347.

\item Knox, L., 1995, Phys. Rev. D {\bf 52}, 4307.

\item Kodama, H. and M. Sasaki, 1984, Prog. Theor. Phys. Supp. {\bf 78}, 1.

\item Kolb, E. W. and M. S. Turner, 1990, {\em The Early Universe}, 
	(Addison-Wesley, Redwood City, California).

\item Kolb, E. W. and S. L. Vadas, 1994, Phys. Rev. D {\bf 50}, 2479.

\item Koranda, S. and B. Allen, 1994, Phys. Rev. D {\bf 50}, 3713.

\item Kosowsky, A., 1996, Annals. Phys. {\bf 246}, 49.

\item Kosowsky, A. and M. S. Turner, 1995, Phys. Rev. D {\bf 52}, R1739.

\item Krauss, L. M. and M. White, 1992, Phys. Rev. Lett. {\bf 69}, 869.

\item Levin, J. J., 1995a, Phys. Rev. D {\bf 51}, 462.

\item Levin, J. J. 1995b, Phys. Rev. D {\bf 51}, 1536. 

\item Liddle, A. R., 1994a, Phys. Rev. D {\bf 49}, 739.

\item Liddle, A. R., 1994b, Phys. Rev. D {\bf 49}, 3805; Phys. Rev. 
	D {\bf 51}, 4603 (E).

\item Liddle, A. R. and D. H. Lyth, 1992, Phys. Lett. B {\bf 291}, 391.

\item Liddle, A. R. and D. H. Lyth, 1993a, Phys. Rept. {\bf 231}, 1.

\item Liddle, A. R. and D. H. Lyth, 1993b, Mon. Not. R. astr. Soc. 
	{\bf 265}, 379.

\item Liddle, A. R., D. H. Lyth and W. Sutherland, 1992, Phys. Lett. B
	{\bf 279}, 244.

\item Liddle, A. R. and M. S. Turner, 1994, Phys. Rev. D {\bf 50}, 758.

\item Liddle, A. R., P. Parsons and J. D. Barrow, 1994, Phys. Rev. 
	D {\bf 50}, 7222.
	
\item Liddle, A. R., D. H. Lyth, R. K. Schaefer, Q. Shafi and P. T. V.
	Viana, 1996, Mon. Not. R. astr. Soc. {\bf 281}, 531.

\item Lidsey, J. E., 1991a, Class. Quant. Grav. {\bf 8}, 923.

\item Lidsey, J. E., 1991b, Phys. Lett. B {\bf 273}, 42.

\item Lidsey, J. E., 1992, Class. Quant. Grav. {\bf 9}, 149.

\item Lidsey, J. E., 1993, Gen. Rel. Grav. {\bf 25}, 399.

\item Lidsey, J. E. and P. Coles, 1992, Mon. Not. R. astr. Soc. 
	{\bf 258}, 57P.
 
\item Lifshitz, E. M., 1946, Zh. Eksp. Teor. Phys. {\bf 16}, 587.

\item Linde, A. D., 1982a, Phys. Lett. B {\bf 108}, 389.

\item Linde, A. D., 1982b, Phys. Lett. B {\bf 116}, 335.

\item Linde, A. D., 1983, Phys. Lett. B {\bf 129}, 177.

\item Linde, A. D., 1990a, Phys. Lett. B {\bf 249}, 18.

\item Linde, A. D., 1990b, {\em Particle Physics and Inflationary
	Cosmology} (Harwood Academic, Chur, Switzerland).

\item Linde, A. D., 1991, Phys. Lett. B {\bf 259}, 38.

\item Linde, A. D., 1994, Phys. Rev. D {\bf 49}, 748.

\item Linde, A. D., 1995, Phys. Lett. B {\bf 351}, 99.

\item Linde, A. D. and A. Mezhlumian, 1995, Phys. Rev. D {\bf 52}, 6789.

\item Lindley, D., 1980, Mon. Not. R. astr. Soc. {\bf 193}, 593.

\item Lucchin, F. and S. Matarrese, 1985a, Phys. Rev. D {\bf 32}, 1316.

\item Lucchin, F. and S. Matarrese, 1985b, Phys. Lett. B {\bf 164}, 282.

\item Lucchin, F., S. Matarrese and S. Mollerach, 1992, Astrophys. J. 
	Lett. {\bf 401}, 49.

\item Lyth, D. H., 1985, Phys. Rev. D {\bf 31}, 1792.

\item Lyth, D. H. and E. D. Stewart, 1992, Phys. Lett. B {\bf 274}, 168.

\item Lyth, D. H., 1996, ``What would we learn by detecting a gravitational
	wave signal in the cosmic microwave background anisotropy?'',
	Lancaster preprint, hep-ph/9606387.

\item Ma, C.-P. and E. Bertschinger, 1995, Astrophys. J. {\bf 455}, 7.

\item MacGibbon, J. H., 1987, Nature 320, 308.

\item MacGibbon, J. H. and B. J. Carr, 1991, Astrophys. J. {\bf 371}, 447.

\item Maeda, K., 1989, Phys. Rev. D {\bf 39}, 3159.

\item Makino, N. and M. Sasaki, 1991, Prog. Theor. Phys. {\bf 86}, 103.

\item Mangano, G., G. Miele and C. Stornaiolo, 1995, Mod. Phys. Lett. A
	{\bf 10}, 1977.

\item Mather, J. C. et al., 1994, Astrophys. J. {\bf 420}, 439.

\item Mielke, E. W. and F. E. Schunck, 1995, Phys. Rev. D {\bf 52}, 672.

\item Misner, C. W., K. S. Thorne and J. A. Wheeler, 1973, 
	{\em Gravitation} (Freeman, San Francisco).

\item Mollerach, S., S. Matarrese and F. Lucchin, 1994, Phys. Rev. 
	D {\bf 50}, 4835.

\item Mukhanov, V. F., 1985, Pis'ma. Zh. Eksp. Teor. Fiz. {\bf 41},
	402 [JETP Lett. {\bf 41}, 493].

\item Mukhanov, V. F., 1988, Zh. Eksp. Teor. Fiz. {\bf 94}, 1 [Sov.
	Phys. JETP {\bf 41}, 493].

\item Mukhanov, V. F., 1989, Phys. Lett. B {\bf 218}, 17.

\item Mukhanov, V. F., H. A. Feldman and R. H. Brandenberger, 1992, 
	Phys. Rept. {\bf 215}, 203.

\item Muslimov, A. G., 1990, Class. Quant. Grav. {\bf 7}, 231.

\item Nakamura, T. T. and E. D. Stewart, 1996, Phys. Lett. B {\bf 381}, 
	413.

\item Nasel'skii, P. D. and A. G. Polnarev, 1985, Astron. Zh. {\bf 62}, 
	833 (Sov. Astron. {\bf 29}, 487). 

\item Olive, K. A., 1990, Phys. Rept. {\bf 190}, 307.

\item Olive, K. A., D. N. Schramm, G. Steigman and T. Walker, 1990, 
	Phys. Lett. B {\bf 236}, 454.

\item Peebles, P. J. E. and R. J. Dicke, 1979, in General Relativity, 
	eds S. W. Hawking and W. Israel (Cambridge University Press,
	Cambridge). 

\item Pogosyan, D. Yu. and A. A. Starobinsky, 1995, Astrophys. J. 
	{\bf 447}, 465.

\item Polarski D. and A. A. Starobinsky, 1995, Phys. Lett. B {\bf 356},
	196.
	
\item Pollock, M. D. and D. Sahdev, 1989, Phys. Lett. B {\bf 222}, 12. 

\item Randall, L., M. Solja\u{c}i\'{c} and A. H. Guth., 1996,
	Nucl. Phys. B {\bf 472}, 377. 

\item Sachs, R. K. and A. M. Wolfe, 1967, Astrophys. J. {\bf 147}, 73.

\item Sahni, V., 1990, Phys. Rev. D {\bf 42}, 453.

\item Salopek, D. S., 1992, Phys. Rev. Lett. {\bf 69}, 3602.

\item Salopek, D. S, J. R. Bond and J. M. Bardeen, 1989, Phys. Rev. 
	D {\bf 40}, 1753.

\item Salopek, D. S. and J. R. Bond, 1990, Phys. Rev. D {\bf 42}, 3936.

\item Salopek, D. S. and J. R. Bond, 1991, Phys. Rev. D {\bf 43}, 1005.

\item Sasaki, M., 1986, Prog. Theor. Phys. {\bf 76}, 1036.

\item Sasaki, M., T. Tanaka, K. Yamamoto and J. Yokoyama, 1993, 
	Phys. Lett. B {\bf 317}, 510.

\item Sasaki, M. and E. D. Stewart, 1996, Prog. Theor. Phys. {\bf 95}, 71. 

\item Sato, K., 1981, Mon. Not. R. astr. Soc. {\bf 195}, 467.

\item Schaefer, R. K. and Q. Shafi, 1994, Phys. Rev. D {\bf 49}, 4990.

\item Seljak, U. and E. Bertschinger, 1994, in ``Present and Future of
	the Cosmic Microwave Background'', eds. J.L. Sanz, E.
	Martinez-Gonzalez and L. Cajon. Springer Verlag (Berlin).

\item Seljak, U. and Zaldarriaga, M., 1996a, Astrophys. J. {\bf 469}, 437.

\item Seljak, U. and Zaldarriaga, M., 1996b, ``Signature of gravity
	waves in polarization of the microwave background'', CfA preprint,
	astro-ph/9609169.

\item Silk, J., 1967, Astrophys. J. {\bf 151}, 459.

\item Smoot, G. F. et al., 1992, Astrophys. J. Lett. {\bf 396}, L1.

\item Souradeep, T. and V. Sahni, 1992, Mod. Phys. Lett. A {\bf 7}, 3541.

\item Starobinsky, A. A., 1979, Pis'ma Zh. Eksp. Teor. Fiz. {\bf 30}, 
	719 (JETP Letters {\bf 30}, 682).

\item Starobinsky, A. A., 1980, Phys. Lett. B {\bf 91}, 99.

\item Starobinsky, A. A., 1982 Phys. Lett. B {\bf 117}, 175.

\item Starobinsky, A. A., 1985, Pis'ma Astron. Zh. {\bf 11}, 323 (Sov. 
	Astron. Lett. {\bf 11}, 133).

\item Starobinsky, A. A. and J. Yokoyama, 1995, ``Density 
        Perturbations in Brans--Dicke theory'', Kyoto preprint,
	gr-qc/9502002.

\item Stebbins, R. T., P. L. Bender, J. E. Faller, J. L. Hall, D. Hils, 
	and M. A. Vincent., 1989, in Fifth Marcel Grossman Meeting
	(World Scientific, Singapore).

\item Steinhardt, P. J. and M. S. Turner, 1984, Phys. Rev. D {\bf 29}, 2162.

\item Steinhardt, P. J., 1994, in ``Anisotropies two years after 
	COBE'', ed L. M. Krauss (World Scientific, Singapore).

\item Stewart, E. D. and D. H. Lyth, 1993, Phys. Lett. B {\bf 302}, 171.

\item Stompor, R., 1994, Astron. Astrophys. {\bf 287}, 693

\item Sunyaev, R. A. and Ya. B. Zel'dovich, 1970, Astrophys. Sp. Sci. 
	{\bf 7}, 20.

\item Sunyaev, R. A. and Ya. B. Zel'dovich, 1980, Ann. Rev. Astr. Astrophys. 
	{\bf 18}, 537.

\item Tegmark, M. and G. Efstathiou, 1996, Mon. Not. R. astr. Soc.
	{\bf 281}, 1297.

\item Thorne, K. S., 1987, in 300 years of Gravitation, eds 
	S. W. Hawking and W. Israel (Cambridge Univ. Press, Cambridge).

\item Thorne, K. S., 1995, ``Gravitational Waves'', CalTech preprint,
	gr-qc/9506086.

\item Turner, M. S., 1993a, Phys. Rev. D {\bf 48}, 3502.

\item Turner, M. S., 1993b, Phys. Rev. D {\bf 48}, 5539.

\item Turner, M. S., M. White and J. E. Lidsey, 1993, Phys. Rev. 
	D {\bf 48}, 4613.

\item Turner, M. S. and M. White, 1996, Phys. Rev. D {\bf 53}, 6822.

\item Vilenkin, A. and Shellard, E. P. S., 1994, {\em Cosmic Strings and
	other topological defects} (Cambridge University Press).

\item Wagoner, R. V., 1970, Phys. Rev. D {\bf 1}, 3204.

\item Wands, D., 1994, Class. Quant. Grav. {\bf 11}, 269.

\item Wang, Y., 1996, Phys. Rev. D {\bf 53}, 639.

\item White, M., 1992, Phys. Rev. D {\bf 42}, 4198.

\item White, M., D. Scott and J. Silk, 1994, Ann. Rev. Astron. Astrophys.
	{\bf 32}, 319.

\item White, M., D. Scott, J. Silk and M. Davis, 1995, Mon. Not. R.
	astr. Soc. {\bf 276}, L69.

\item Whitt, B., 1984, Phys. Lett. B {\bf 145}, 176.

\item Wright, E. et al., 1992, Astrophys. J. Lett. {\bf 396}, L13.

\item Yokoyama, J., 1995, ``Formation of MACHO-primordial black holes 
	in inflationary cosmology'', YITP preprint. 
	
\item Zaldarriaga, M. and Seljak, U.,  1996, ``An all-sky analysis of
	polarization in the microwave background'', CfA preprint,
	astro-ph/9609170

\item Zel'dovich, Ya. B., 1972, Mon. Not. R. astr. Soc. {\bf 160}, 1P.

\item Zel'dovich, Ya. B. and A. A. Starobinsky, 1976, Pis'ma Zh. Eksp. Teor.
	Fiz. {\bf 24}, 616 (1976) (JETP Lett. {\bf 24}, 571).

\end{description}
\newpage
\section*{Tables}

\begin{table}[h]
\begin{center}
\begin{tabular}{c|c|c}
\hline \hline
              & Gravitational Waves        & Gravitational Waves     \\
              & Important                  & Negligible              \\
\hline \hline
         &            &                         \\
$n<1$ & $\epsilon$ large, $\eta < 2\epsilon$ & $\epsilon$ small, $\eta < 
	-2\epsilon$ \\
         &  Power-Law Inflation     &   Natural Inflation               \\
\hline
$n \simeq 1$ & $\epsilon$ large, $\eta \simeq 2\epsilon$ & $\epsilon$, 
	$|\eta|$ small \\ 
         &  Intermediate Inflation  &   Hybrid Inflation                  \\
\hline
$n>1$ & $\epsilon$ large, $\eta > 2\epsilon$ & $\epsilon$ small, $\eta > 
	2\epsilon$ \\
         &   &   Hybrid Inflation\\
\end{tabular}
\end{center}
\footnotesize{\hspace*{.3in} Table 1: This table illustrates the different 
possible inflationary behaviors, and quotes a specific inflation
model which gives each (except the bottom left case, which while
possible in principle has not had any specific inflationary model
devised). The description `large' implies significantly larger than
zero (but still less than unity).}
\end{table}

\begin{table}
\begin{center}
\begin{tabular}{c|c|c}
\hline \hline 
        & lowest-order      & next-order (exact) \\
\hline \hline
 & & \\
$V(\phi_0)$     & $H(\phi_0)$
&  $H(\phi_0)$, $\epsilon(\phi_0)$  \\
$V'(\phi_0)$    & $H(\phi_0)$, $\epsilon(\phi_0)$ 
& $H(\phi_0)$, $\epsilon(\phi_0)$, $\eta(\phi_0)$ \\
$V''(\phi_0)$   & $H(\phi_0)$, $\epsilon(\phi_0)$, $\eta(\phi_0)$
& $H(\phi_0)$, $\epsilon(\phi_0)$, $\eta(\phi_0)$, $\xi(\phi_0)$  \\
$V'''(\phi_0)$  & $H$, $\epsilon(\phi_0)$, $\eta(\phi_0)$, $\xi(\phi_0)$ 
& -------- \\
\end{tabular}
\end{center}
\footnotesize {\hspace*{.3in} Table 2: A summary of the inflationary
parameters [$H$ and the slow-roll parameters $\epsilon$, $\eta$, and 
$\xi$ defined in  Eqs.\  (\ref{epsilon})--(\ref{xi})] needed to 
reconstruct a given  derivative of the potential to a certain order.  
See Eqs.\ (\ref{V})--(\ref{V''}).  Note that the next-order result is {\em 
exact}.}
\end{table}

\begin{table}
\begin{center}
\begin{tabular}{c|c|c}
\hline \hline
observable & lowest-order & next-order \\
\hline \hline
 & & \\
$A_T^2(k_0)$		& $H(\phi_0)$	
	& $H(\phi_0)$, $\epsilon(\phi_0)$ \\
$A_S^2(k_0)$		& $H(\phi_0)$, $\epsilon(\phi_0)$	& 
$H(\phi_0)$, $\epsilon(\phi_0)$, $\eta(\phi_0)$ \\
$n(k_0)$			& $\epsilon(\phi_0)$, $\eta(\phi_0)$	& 
$\epsilon(\phi_0)$, $\eta(\phi_0)$, $\xi(\phi_0)$\\
$dn/d\ln k|_{k_0}$	& $\epsilon(\phi_0)$, $\eta(\phi_0)$, $\xi(\phi_0)$ 	
& -------- \\
\end{tabular}
\end{center}
\footnotesize {\hspace*{.3in} Table 3: The observables, $A_T^2$, $A_S^2$,
$n$, and $dn/d\ln k$ at the point $k_0$ may be expressed in terms of $H$ and 
the slow-roll parameters at the point $\phi_0$.  Table 3 lists the inflation 
parameters
required to predict the observable to the indicated order.  (See Section 
5.1.)} 
\end{table}

\begin{table}
\begin{center}
\begin{tabular}{c|c|c}
\hline \hline
parameter & lowest-order & next-order  \\
\hline \hline
 & &  \\
$H$			
	& $A_T^2$
		& $A_T^2$, $A_S^2$    \\
$\epsilon$
	&  $A_T^2$, $A_S^2$
		& $A_T^2$, $A_S^2$, $n$   \\
$\eta$
	& $A_T^2$, $A_S^2$, $n$
		& $A_T^2$, $A_S^2$, $n$, $dn/d\ln k$  \\
$\xi$
	& $A_T^2$, $A_S^2$, $n$, $dn/d\ln k$ 
		&   --------                                \
\end{tabular}
\end{center}
\footnotesize {\hspace*{.3in} Table 4:  The inflation parameters
may be expressed in terms of the observables, $A_T^2$, $A_S^2$,
$n$, and $dn/d\ln k$ (see Section 5.1).  Through judicious use of the 
consistency relations one may employ different combinations of observables
than listed here, e.g., use of $n_T$ rather than $A_T^2/A_S^2$.}
\end{table}

\begin{table}
\begin{center}
\begin{tabular}{c|c|c|c}
\hline \hline
        & lowest-order      & next-order &  next-to-next-order   \\
\hline \hline
 & & &\\
$V$     
	& $A_T^2$ 						 
		& $A_T^2$, $A_S^2$
			& $A_T^2$, $A_S^2$, $n$ \\
$V'$
	& $A_T^2$, $A_S^2$ 				 
		& $A_T^2$, $A_S^2$, $n$ 
			&  $A_T^2$, $A_S^2$, $n$, $dn/d\ln k$ \\
$V''$   
	& $A_T^2$, $A_S^2$, $n$ 
		& $A_T^2$, $A_S^2$, $n$,  $dn/d\ln k$ 
			&  -------- \\
$V'''$  
	& $A_T^2$, $A_S^2$, $n$, $dn/d\ln k$ 
		&  -------- 
			&  -------- \\
\end{tabular}
\end{center}
\footnotesize {\hspace*{.3in} Table 5: A summary of the observables 
needed to reconstruct a given derivative of the potential to a certain
order.  The potential and its derivatives are given at a point $\phi_0$, and
$A_T^2$, $A_S^2$, $n$, and $dn/dk$ are to be evaluated at the point $k_0$.}
\end{table}

\begin{table}
\begin{center}
\begin{tabular}{c|c|c|c}
\hline \hline 
Model 1        & Input                  & Output (power-law fit) &  Output 
(including  $dn/d\ln k|_{k_0}$)   \\
\hline \hline
 & & &\\
$A_S^2$ & $2.5 \times 10^{-10}$ &  $(2.45 \pm 0.09) \times 10^{-10}$ &  
	$(2.45 \pm 0.10) \times 10^{-10}$ \\
$A_T^2$ & $0.12 \times 10^{-10}$ &  $(0.132 \pm 0.015) \times 10^{-10}$ & 
	$(0.132 \pm 0.015) \times 10^{-10}$ \\
$n-1$   & $-0.1$ & $-0.11 \pm 0.02$ & $-0.115 \pm 0.035$\\
$n_T$   & $-0.1$ & $-0.25 \pm 0.10$ & $-0.25 \pm 0.10$\\
$dn/d\ln k|_{k_0}$ & $0$ &   ---      & $0.003 \pm 0.018$
\end{tabular}

\vspace*{12pt}
\begin{tabular}{c|c|c|c}
\hline \hline
Model 2        & Input                  & Output (power-law fit) &  Output 
(including  $dn/d\ln k|_{k_0}$)   \\
\hline \hline
 & & &\\
$A_S^2$ & $1.34 \times 10^{-10}$ &  $(1.27 \pm 0.04) \times 10^{-10}$ &  
	$(1.28 \pm 0.04) \times 10^{-10}$ \\
$A_T^2$ & $0.094 \times 10^{-10}$ &  $(0.09 \pm 0.01) \times 10^{-10}$ & 
	$(0.09 \pm 0.01) \times 10^{-10}$ \\
$n-1$   & $0.00$ & $0.04 \pm 0.02$ & $0.06\pm 0.03$\\
$n_T$   & $-0.2$ & $-0.12 \pm 0.11$ & $-0.12 \pm 0.11$\\
$dn/d\ln k|_{k_0}$ & $0$ &   ---      & $-0.01 \pm 0.02$
\end{tabular}
\end{center}
\footnotesize {\hspace*{.3in} Table 6: Input and output values from the 
two simulated data sets. The amplitudes are given at the central $k$
value (in log units) for the scalars, notionally corresponding to the
20-th multipole.}
\end{table}

\begin{table}
\begin{center}
\begin{tabular}{c|c|c|c}
\hline \hline 
Model 1  &  Underlying      & Lowest-order      &   Next-order     \\
  &  potential       & reconstruction    &   reconstruction \\
\hline \hline
 & & &\\
$10^{12} V(\phi_0)/m_{{\rm Pl}}^4$ & 28.2 &  $31 \pm 4$ & $31 \pm 4$  \\
$10^{12} V'(\phi_0)/m_{{\rm Pl}}^3$ & -43.6  & $-51 \pm 9$ & 
$-52 \pm 9$ \\
$10^{12} V''(\phi_0)/m_{{\rm Pl}}^2$ & 67.5 &  $83 \pm 25$ & ---    
\end{tabular}

\vspace*{12pt}
\begin{tabular}{c|c|c|c}
\hline \hline
Model 2  &  Underlying      & Lowest-order      &   Next-order     \\
  &  potential       & reconstruction    &   reconstruction \\
\hline \hline
 & & &\\
$10^{12} V(\phi_0)/m_{{\rm Pl}}^4$ & 22.4 &  $21 \pm 2$ & $21\pm 2$  \\
$10^{12} V'(\phi_0)/m_{{\rm Pl}}^3$ & -38.9 & $-40 \pm 7$ & $-37 \pm 6$ \\
$10^{12} V''(\phi_0)/m_{{\rm Pl}}^2$ & 94.5 &  $123 \pm 27$ & ---    
\end{tabular}
\end{center}
\footnotesize {\hspace*{.3in} Table 7: Input potential compared with 
reconstructions for the two models.}
\end{table}

\newpage
\section*{Figure Captions}

\vspace*{24pt}
\noindent
{\em Figure 1}\\
A schematic illustration of the reconstruction strategy. The spectra $A_S$ 
of the density perturbations and $A_T$ of the gravitational waves are 
measured over some range of scales which corresponds to some interval of the 
underlying potential $V(\phi)$. 

\vspace*{24pt}
\noindent
{\em Figure 2}\\
The simulated data of Model 1, with error bars. The circles are $A_S^2$ and 
squares are $A_T^2$. The horizontal axis is in $h \, {\rm Mpc}^{-1}$. The 
lines show the best power-law fits to the simulated data, as given in Table 
2. Showing the data in the form of the spectra is schematic; an analysis of 
true observations would directly fit the amplitude and spectral index to 
measured quantities.

\vspace*{24pt}
\noindent
{\em Figure 3}\\
The reconstructed potentials compared to the underlying one, from the data 
in Model 1 in Table 4. The dashed line shows the true underlying exponential 
potential. The two solid lines, which nearly overlap, are Taylor series 
reconstructions, one using just lowest-order information and the other using 
the available next-order information. The length of these lines corresponds 
to the range of $k$ for which the simulated data is available. The 
observational errors (not shown) dominate the theoretical errors, and of 
course when taken into account the reconstructions are consistent with the 
true potential.

\vspace*{24pt}
\noindent
{\em Figure 4}\\
An illustration of the area law. Reconstruction finds $\epsilon$ and 
perhaps its derivative, between 60 and 50 $e$-foldings from the end of 
inflation, illustrated by the solid part of the curve which ends at a scalar 
field value indicated by $\phi_{50}$. After large-scale structure scales 
leave the horizon, $\epsilon$ (now shown as a dotted curve) must behave so 
that it reaches unity just as the shaded area under the curve of 
$\epsilon^{-1/2}$ against $\phi/m_{{\rm Pl}}$ reaches $50/\sqrt{4\pi}$. 

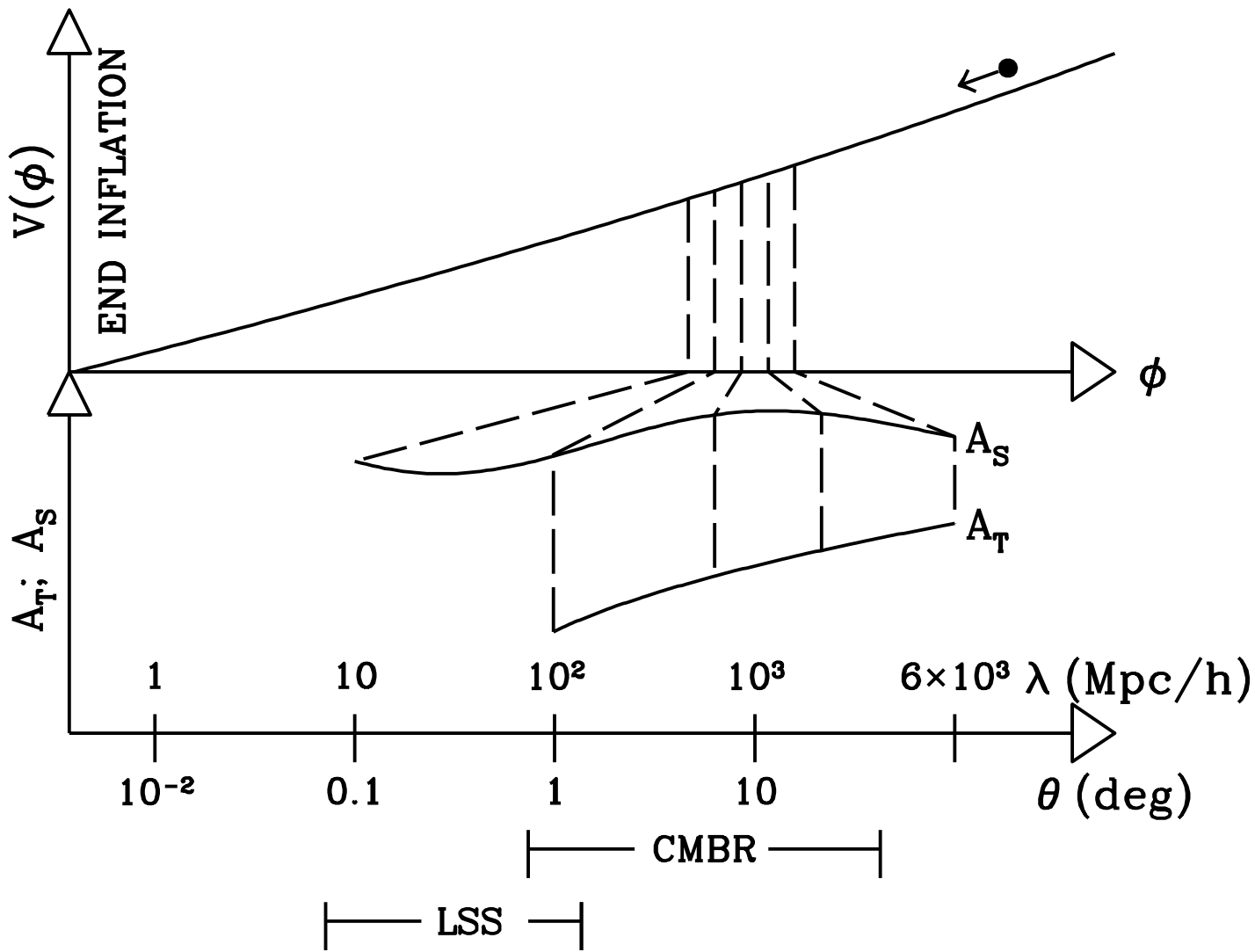
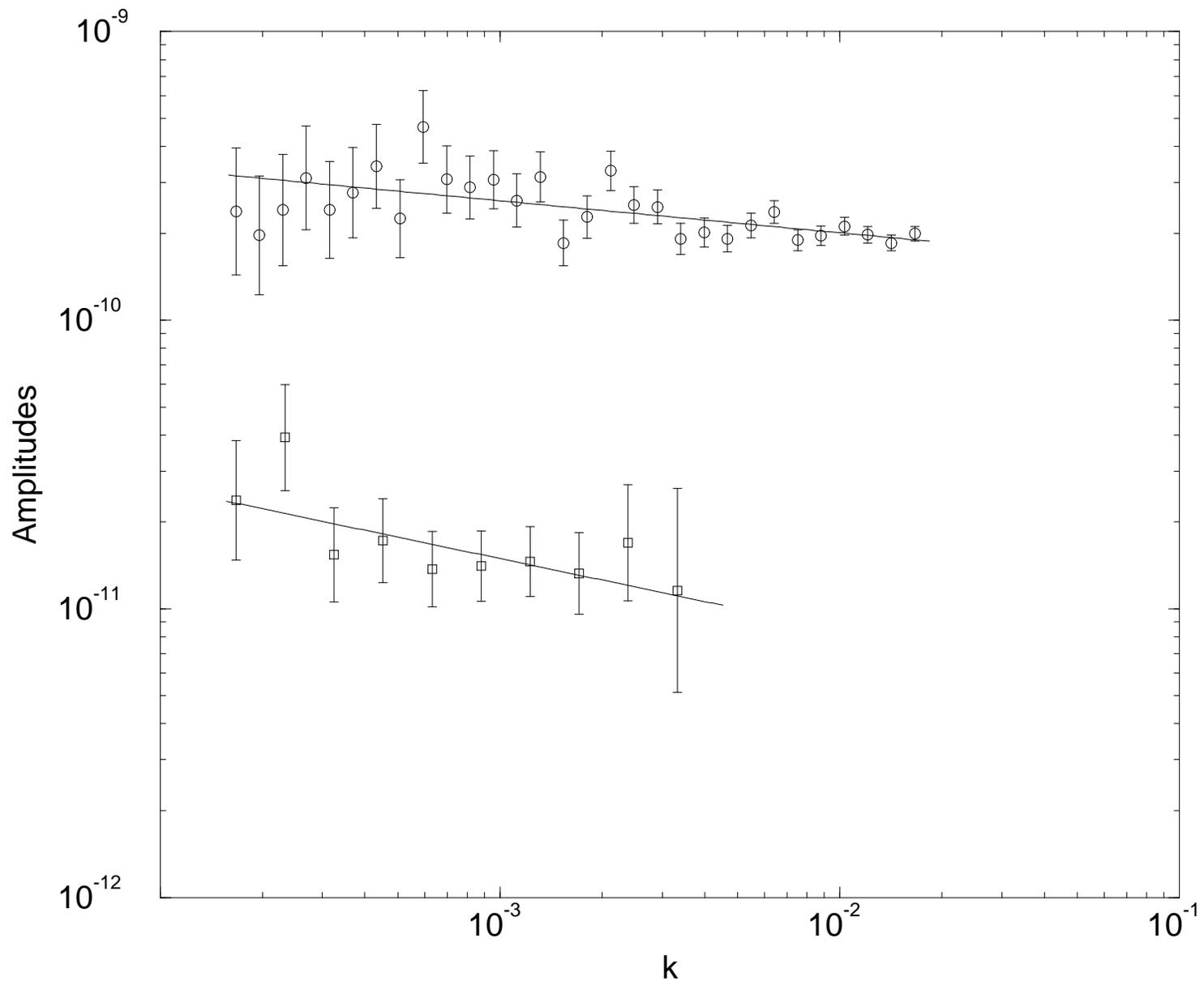
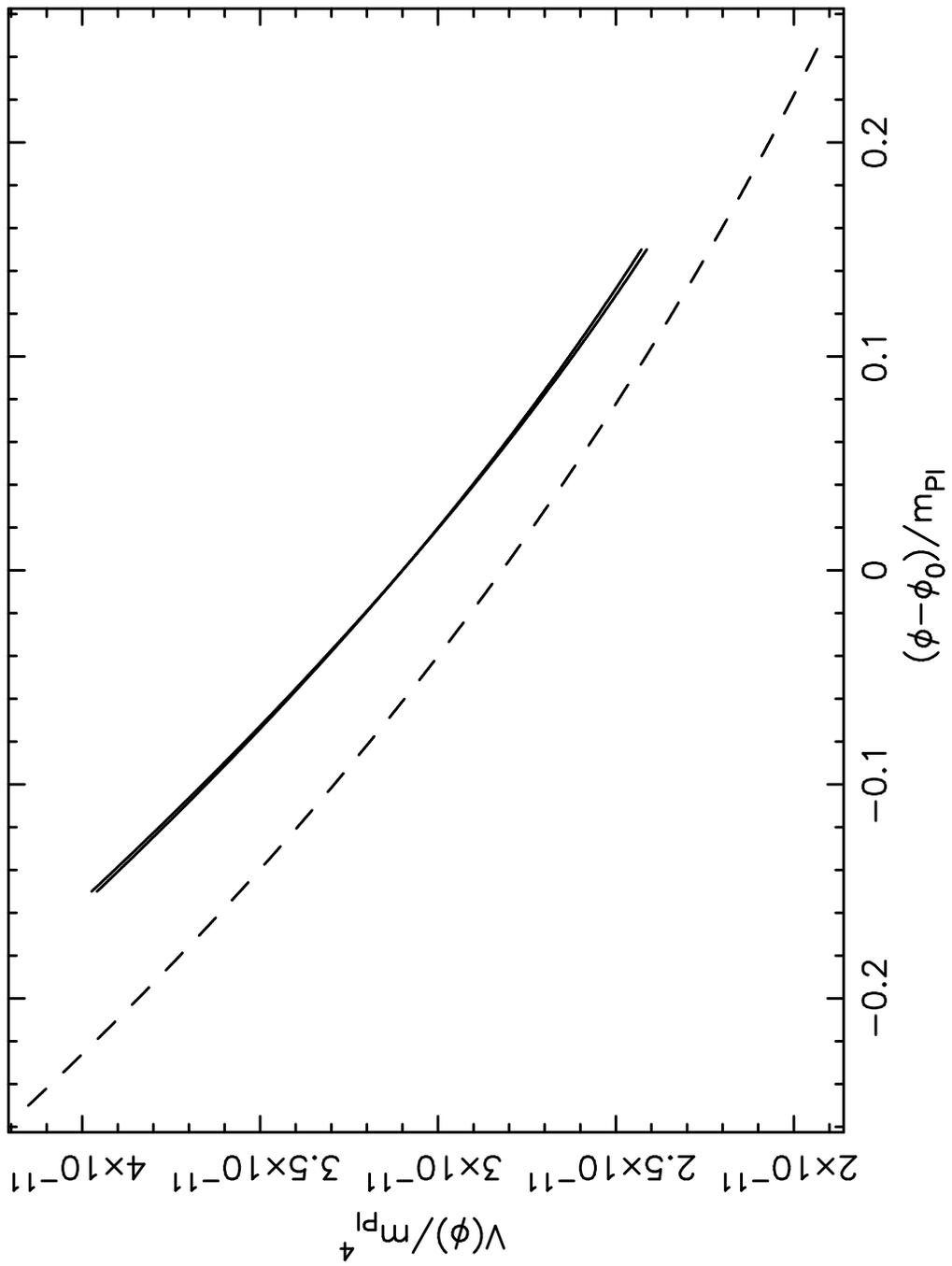
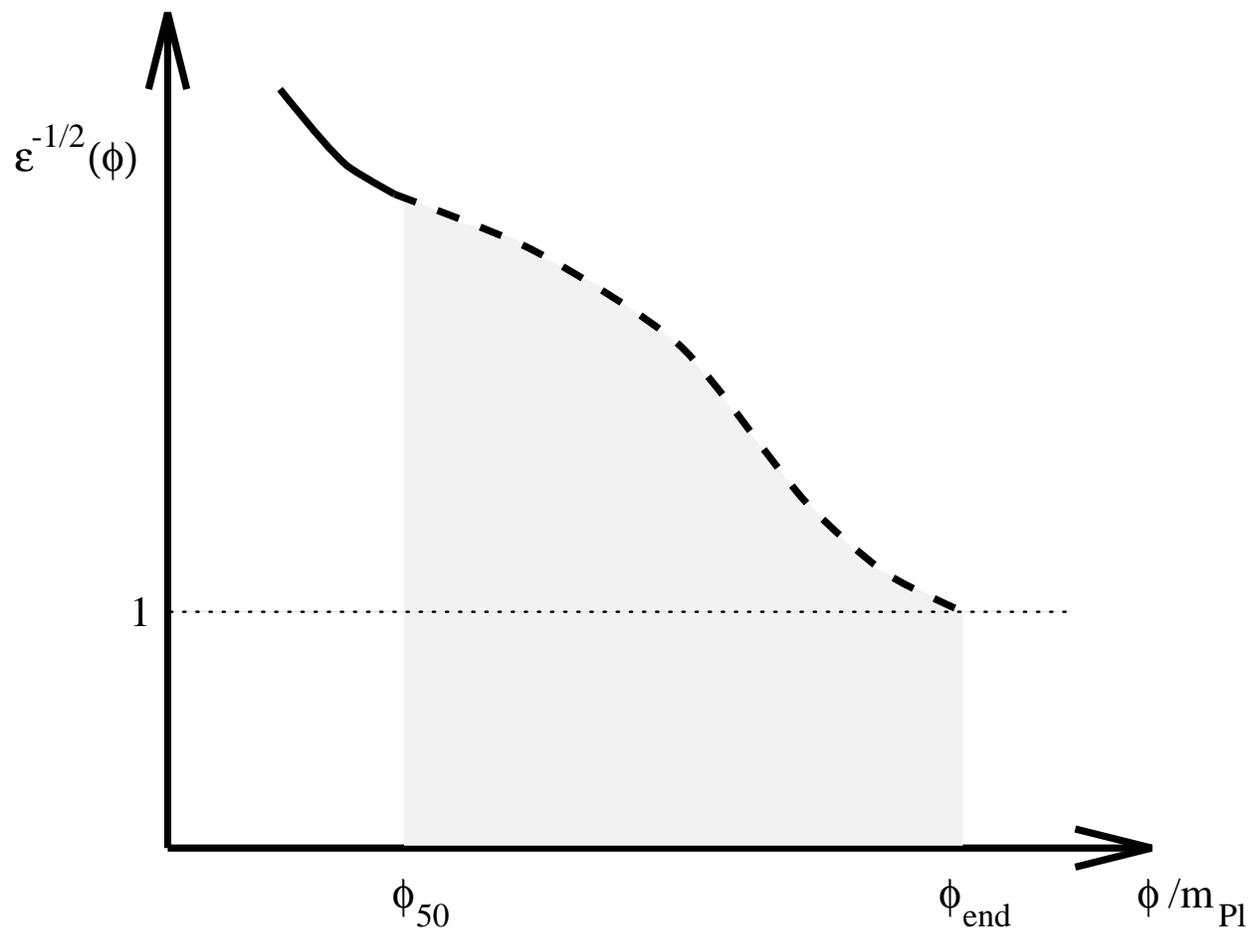
\end{document}